\documentclass[11pt]{article}

\usepackage{amsmath, amssymb,amsthm,bm,bbm,fullpage}
\usepackage{mathrsfs}
\usepackage{cite}
\usepackage{colortbl}
\usepackage[all]{xy}
\usepackage{epsfig}
\usepackage{subfigure}
\usepackage{graphics}
\usepackage{multirow}

\usepackage{lineno}

\usepackage{setspace}
\usepackage{caption}

\theoremstyle{remark} 
	
	\doublespace

\begin{document}

\begin{centering}
{\huge
\textbf{The Game Theory of Fake News}
}
\bigskip
\\
Alexander J. Stewart$^{1,*}$, Antonio A. Arechar$^{2,3}$, David G. Rand$^{3}$, Joshua B. Plotkin$^{4}$
\\
\bigskip
\end{centering}
\begin{flushleft}
{\footnotesize
$^1$ School of Mathematics and Statistics, University of St Andrews, St Andrews, UK
\\
$^2$ Center for Research and Teaching in Economics, CIDE, Aguascalientes, MX
\\
$^3$ Sloan School of Management, MIT, Cambridge, MA, USA
\\
$^4$ Department of Biology, University of Pennsylvania, Philadelphia, PA, USA
\\
$^*$ E-mail: ajs50@st-andrews.ac.uk
}
\end{flushleft}

\noindent\textbf{A great deal of empirical research has examined who falls for misinformation and why. Here, we introduce a formal game-theoretic model of engagement with news stories that captures the strategic interplay between (mis)information consumers and producers. A key insight from the model is that observed patterns of engagement do not necessarily reflect the preferences of consumers. This is because producers seeking to promote misinformation can use strategies that lead moderately inattentive readers to engage more with false stories than true ones --  even when readers prefer more accurate over less accurate information. We then empirically  test people's preferences for accuracy in the news.  In three studies, we find that people strongly prefer to click and share news they perceive as more accurate -- both in a general population sample, and in a sample of users recruited through Twitter who had actually shared links to misinformation sites online.   
Despite this preference for accurate news -- and consistent with the predictions of our model -- we find markedly different engagement patterns for articles from misinformation versus mainstream news sites. Using 1,000 headlines from 20 misinformation and 20 mainstream news sites, we compare Facebook engagement data with 20,000 accuracy ratings collected in a survey experiment. Engagement with a headline is negatively correlated with perceived accuracy for misinformation sites, but positively correlated with perceived accuracy for mainstream sites.  
Taken together, these theoretical and empirical results suggest that consumer preferences cannot be straightforwardly inferred from empirical patterns of engagement.}

\clearpage 

False or misleading information is a fundamental problem for people trying to form an accurate understanding of the world \cite{Lazer1094}.  
There has been widespread concern in recent years about the spread of misinformation -- for example during Brexit and the 2016 US Presidential Election \cite{Lazer1094}, the COVID-19 pandemic \cite{doi:10.1177/0956797620939054,Weste1912444117}, and the 2020 US Presidential election \cite{PennyMisinfoRev} -- among policy makers, scientists, and the general public, leading to an explosion of academic research on this topic. 

Most of this prior work has been empirical in nature, using observational data and experiments to ask questions such as: Who reads, shares, and believes misinformation \cite{Guesseaau4586,GuessNHB,Grinberg374,doi:10.1073/pnas.1804840115,osmundsen_bor_vahlstrup_bechmann_petersen_2021,Guilbeault9714}? Why do people fall for misinformation \cite{PENNYCOOK201939,BagoJEP,MosNatComm,luca_munger_nagler_tucker_2022}? What interventions can combat misinformation \cite{PennyNature,Guess15536,doi:10.1177/0956797617714579,10.1371/journal.pone.0175799}? In contrast to this large body of empirical research, however, there has been very little work using formal models to explore the spread of misinformation. To this end, here we develop a game-theoretic model for the evolution of news engagement. Considering this question from a game-theoretic perspective highlights the necessity of not simply modeling consumers as isolated agents, but rather of considering the dynamics of news engagement and \textit{dissemination}.
Inferences about the revealed preferences of consumers 
that fail to account for these dynamics are likely to be misleading \cite{Bak-Colemane2025764118,Stewart19,BakColComm}. To fully understand the spread of misinformation, therefore, we must examine the strategic considerations that arise on both sides of the news market \cite{Gravino,Gentzkow,10.1257/jep.31.2.211}.

To this end, 
we introduce an asymmetric ``misinformation game'' played between news outlets, who can choose to publish true or false information, and news consumers, who can choose whether or not to engage with stories from a given outlet.
We focus on the dissemination strategies of outlets that try to generate engagement with either true or false stories, and the news engagement strategies of 
readers who seek to consume either true or false stories, but operate under cognitive constraints.

We show that simple strategies, which vary the rate of production of true and false stories over time can, when adopted by publishers seeking to spread misinformation, can successfully induce a substantial level of engagement with false stories among truth-seeking readers who try to maximize their consumption of accurate information. In particular, our model predicts that publishers who seek to spread misinformation should adopt strategies that: i) include a mix of true and false  stories over time and ii) increase the rate of disseminating misinformation as the level of engagement they receive from readers increases. 

Our model predicts that, as a result of strategic dissemination, outlets seeking to spread misinformation can induce somewhat inattentive readers to have greater engagement with false stories than with true stories, creating an apparent preference for misinformation among their readers; whereas outlets seeking to spread accurate information should produce the opposite pattern. Importantly, we predict that this trend will hold even when all readers in fact value accuracy, and false headlines are no more ``attractive'' (e.g. due to novelty, emotional appeal, etc) than accurate headlines.

We then complement these theoretical results by collecting new empirical data. First, we measured the expressed preferences for engaging with accurate versus inaccurate information among 511 subjects sampled from the general population, as well as 119 Twitter users who had shared links to fake new sites. Both groups express a strong preference for sharing and engaging with accurate news stories. However, in a second empirical study integrating $20,000$ accuracy ratings for $1,000$ news stories published across 40 outlets  with Facebook engagement ratings for each story from Crowdtangle \cite{CrowTan}, we also find that -- in spite of the preference for accuracy demonstrated in the first studies -- 
misinformation sites tend to generate higher engagement with less accurate stories, while mainstream sites generate the opposite pattern. 
Taken together, our results call in to question the idea that patterns of engagement with misinformation reflect the preferences of news consumers. To properly account for the spread of misinformation, we must account for the strategies of those who \emph{produce} the news, rather than just the preferences 
of readers.
\\
\\
\noindent \textbf{The misinformation game.} 
We develop a theory of the production and consumption of misinformation using a game-theoretic approach. We consider an asymmetric, asynchronous, infinitely repeated coordination game, which we dub the ``misinformation game'', in which a news transmitter (i.e.~a news outlet or a platform promoting news stories) chooses whether to transmit pieces of true or false information (i.e. news stories) while a receiver (i.e.~news consumer) chooses whether or not to engage with each piece of transmitted news (Figure 1).

\begin{figure*}\centering \includegraphics[width=0.75\linewidth]{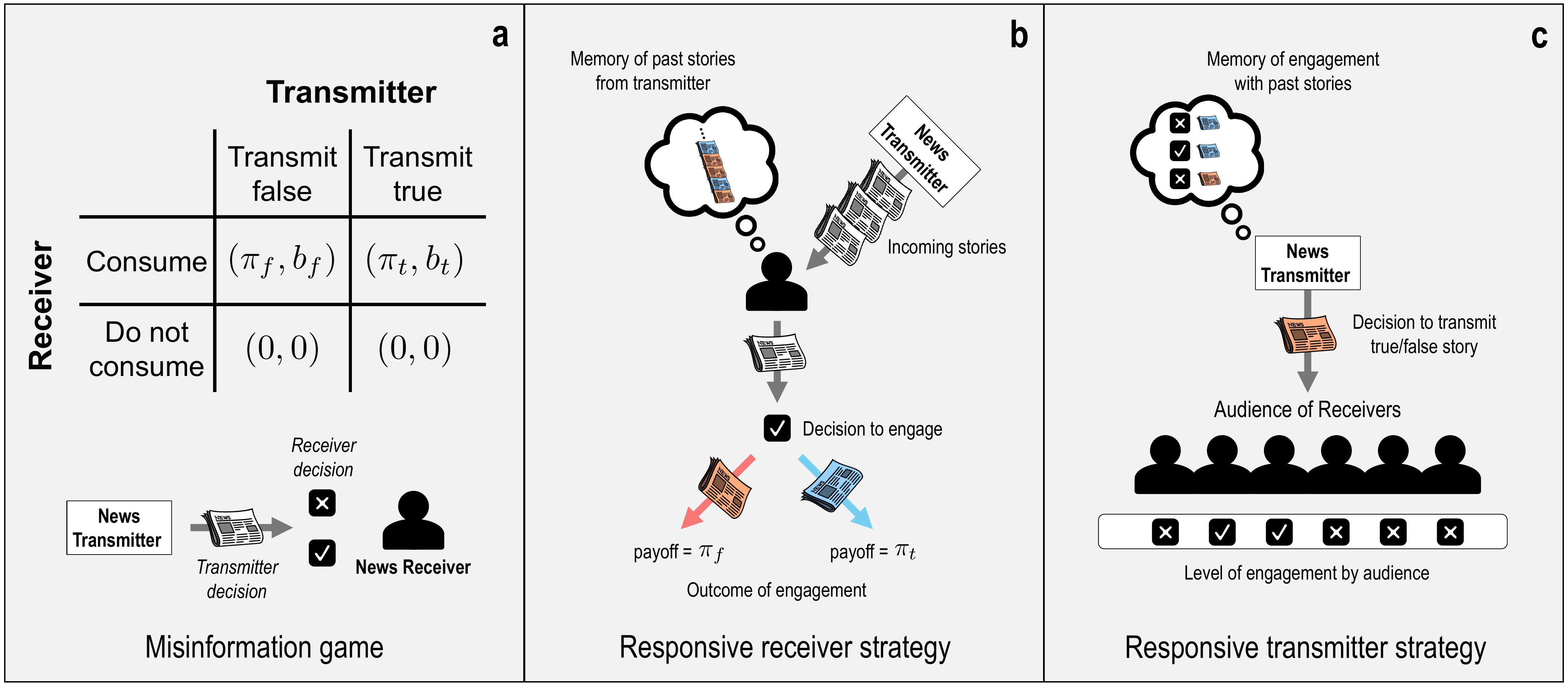}
\caption{\footnotesize \textbf{The misinformation game.} a) We developed the ``misinformation game'' in which transmitters choose whether to share true or false stories and receivers decide whether to engage with each story or not. We assume that receivers gain utility $\pi_t$ from engaging with accurate information and utility $\pi_f$ for engaging with misinformation. Receivers are incentivized to seek true news stories and avoid misinformation stories when $\pi_t>0$ and $\pi_f<0$.
b) We assume that, when deciding whether to engage with a given story, receivers can take account its perceived accuracy, and the accuracy of past stories shared by transmitters. Receivers update  their strategy using a myopic, noisy optimization process. c) We assume that a transmitter chooses each subsequent story to be true or false based on the veracity of their previous story and the level of engagement it received. The number of independent receivers whose engagement a transmitter considers determines how precisely the transmitter can target its stories. }
\end{figure*}

Receivers derive a utility  $\pi_t$ from engaging with true news stories, and utility $\pi_f$ from engaging with false stories (Figure 1); they derive no utility if they do not engage. A payoff structure $\pi_t>0$ and $\pi_f<0$ reflects the preferences of readers who prefer accurate information and seek to avoid misinformation; a payoff structure $\pi_t<0$ and $\pi_f>0$ reflects receivers who prefer misinformation and avoid accurate information. In the SI (Section 1 and Section 3.9) we also provide results for receivers who are indifferent to the veracity of news stories, i.e. for a payoff structure $\pi_t>0$ and $\pi_f>0$. 

Similarly, transmitters derive benefit $b_t$ or $b_f$ from engagement by a receiver with a true or false story, and no payoff if the receiver does not engage. The differences in these benefits may reflect differences in the ease of producing such content, or some underlying ideological preference of the transmitter.
We consider two types of news transmitters: a misinformation transmitter that seeks to promote misinformation stories, so that $b_t=0$ and $b_f>0$, and a mainstream transmitter that seeks to promote accurate stories, so that $b_t>0$ and $b_f=0$. In the SI (Section 3.2), we also consider a ``clickbait'' transmitter who simply aims to maximize engagement regardless of accuracy ($b_t=b_f>0$).  
\\
\\
\noindent \textbf{Transmitter strategies.} 
We model the dynamics of news dissemination and engagement as an optimization process, first specifying the strategy spaces available to transmitters and receivers.
A transmitter can choose to share either a true or false story with a probability that depends on the level of engagement their previous story generated. We first introduce a transmitter strategy space, in which a given transmitter strategy has a baseline accuracy, along with a feedback term described by 
\begin{eqnarray}
\nonumber r_{kt}=&\alpha_0&+\ \ \ \ \ \ \sum_{l}\gamma_l\left(\tfrac{k}{N}\right)^i\\
r_{kf}=&\underbrace{\beta_0}_\text{baseline accuracy}&+\ \ \ \ \ \ \sum_l\underbrace{\theta_l\left(\tfrac{k}{N}\right)^l}_\text{ feedback}.
\end{eqnarray}

Here $r_{kj}$ defines the probability of sharing a true story given that the previous story was true or false ($j\in\{t,f\}$) and that $k$ out of $N$ targeted receivers chose to engage with the story. The parameters $\gamma_l$ and $\theta_l$ describe the shape of the (polynomial) feedback function. When there is no feedback ($\gamma_l=\theta_l=0$ for all $l$) transmitters do not take account of past engagement. For convenience we define $\gamma=\sum_l\gamma_l$ and $\theta=\sum_l\theta_l$. Note that in any give round of the iterated game, a transmitter must either share a true or false story, and so the overall rate of news production is constant.
\\
\\
\noindent \textbf{Receiver strategies.} 
We assume that receivers can pay different levels, and different types, of attention when deciding whether to engage with a news story. They can attend to the likely veracity of the headline, and to their recent experience of engagement with the source that transmitted the story. In particular, a receiver engages with a story with probability $q^l_{ij}$ where

\begin{align}
    \nonumber &q^l_{ij}=
   \underbrace{a_0\delta_{lt}}_\text{$P(\text{engage}|\text{true \& attentive})$}+\underbrace{(1-a_0)(1-a_1)p_0}_\text{$P(\text{engage}|\text{no memory \& inattentive})$}+\\
   &\underbrace{(1-a_0)a_1p_{ij}}_\text{$P(\text{engage}|\text{memory \& inattentive})$}
\end{align}

Indices $i$ and $j$ describe the outcome of the previous news story shared by the transmitter: index $j\in\{t,f\}$ indicates whether the transmitter's \emph{previous} story was true or false and index $i\in\{c,n\}$ indicates whether the receiver either engaged with ($c$) or did not engage with ($n$) the previous story. The index $l\in\{t,f\}$ indicates whether the transmitter's \emph{current} story is true or false. With probability $a_0$ the receiver 
is able to directly assess the veracity of the current story \cite{luca_munger_nagler_tucker_2022}, where $\delta_{lt}$ is the Kronecker delta, equal to 1 if the newly shared story is true, and 0 otherwise. If the receiver is unable to assess 
the veracity of the current story (e.g.~due to lack of background knowledge, or failure to pay attention), then with probability $a_1$ they make their decision based on their previous experience of the news source, according to a behavioral strategy $\mathbf{p}=\{p_{ct},p_{cf},p_{nt},p_{nf}\}$, where $p_{ct}$ is the probability of engaging given that they chose to engage in the previous round and the story was true, and so on. Finally, if the reader attends to neither veracity nor past experience, they engage with the story with fixed probability $p_0$.
\\
\\
\noindent \textbf{Dynamics of reader engagement.}
The dynamics of transmission and engagement in our model occur across two different timescales. First, on short timescales, news is produced and consumed (or not) by transmitters and receivers using \emph{fixed} strategies for a repeated game. Second, over longer timescales, transmitters and receivers update their strategies via a noisy optimization process (see SI section 1 for full details). Thus, strategy optimization occurs with respect to the utility derived from the equilibrium payoffs across rounds of news production/consumption generated under the (infinitely) repeated misinformation game.

To understand the dynamics of reader engagement under this model, we begin by analyzing the dynamics of the repeated misinformation game for a pair of \emph{fixed} transmitter and receiver strategies (see SI Section 1, where we also give an equivalent expression for the case of multiple receivers). We show that a transmitter using a strategy of the type described by Eq.~1 enforces a specific relationship between the proportion of transmitted stories that are false, $v_f$, and the overall probability the receiver engages with true and false stories, $v_{tc}$ and $v_{fc}$:

\begin{equation}
    v_f=\frac{1-\alpha}{1-\alpha+\beta}-\frac{\theta}{1-\alpha+\beta} v_{fc}-\frac{\gamma}{1-\alpha+\beta} v_{tc}
\end{equation}
\\
Remarkably, this relationship between the misinformation transmission probability per story $v_f$ and the receiver's overall engagement with accurate stories, $v_{tc}$ and misinformation $v_{fc}$ holds regardless of the receiver's strategy for engagement -- and therefore regardless of the receiver's preference for consuming true versus false stories. This kind of unilaterally enforced constraint, imposed by one player in a repeated game, has been extensively studied in the context of so-called ``extortion strategies'' in the Iterated Prisoner's Dilemma [REFs]. Here we adapt this idea for asymmetric games related to news production and consumption.

Eq.~3 holds regardless of the strategy of the receiver -- and so it can be understood as a constraint enforced by the (fixed) transmitter strategy on the optimization process of any receiver. That is, Eq.~3 shows that a news site can unilaterally shape the dynamics of consumer engagement by constraining how engagement and misinformation transmission covary, as the consumer updates his strategy. 
For example, a transmitter can choose their strategy to ensure that the probability of production of false stores ($v_f$) is positively correlated with consumer engagement ($v_{tc}$ and $v_{fc}$, Eq.~3), regardless of what strategies the consumer explores. When a transmitter uses this strategy, consumers who increase their overall engagement with a news source \emph{for whatever reason} will always encounter greater levels of misinformation as they do so.

This type of transmitter strategy is particularly important when we consider receivers who seek to increase their engagement with true stories, interacting with a transmitter who seeks to spread misinformation. 
When such receivers engage more with such a transmitter, they will receive increasing amounts misinformation. If, in response, they then decrease their engagement with the transmitter’s stories they will be targeted with more true stories, and thus they are incentivized to increase their engagement once again. The result of this dynamic is a \emph{negative} correlation between story accuracy and receiver engagement, even though the receiver is seeking to increase their engagement with accurate information. An illustrative example of this type of dynamic is shown in Figure 2 and in SI Figure S2.

Critically, these results mean that the empirical observation of false stories receiving more engagement than true stories (e.g.~Vosoughi \emph{et al.} 2018 \cite{Vosoughi1146}) does \textit{not} necessarily imply that false stories are actually more attractive to readers -- since this same pattern can be generated by a clever transmitter responding to truth-seeking consumer behavior. 
(Whereas in the absence of any feedback between transmitter and receiver behavior, i.e.~when $\theta=\gamma=0$ in Eq.~3, there will be no correlation between the probability of producing misinformation $v_f$ and the probability of engagement among receivers).  
\\
\begin{figure*}\centering \includegraphics[width=0.75\linewidth]{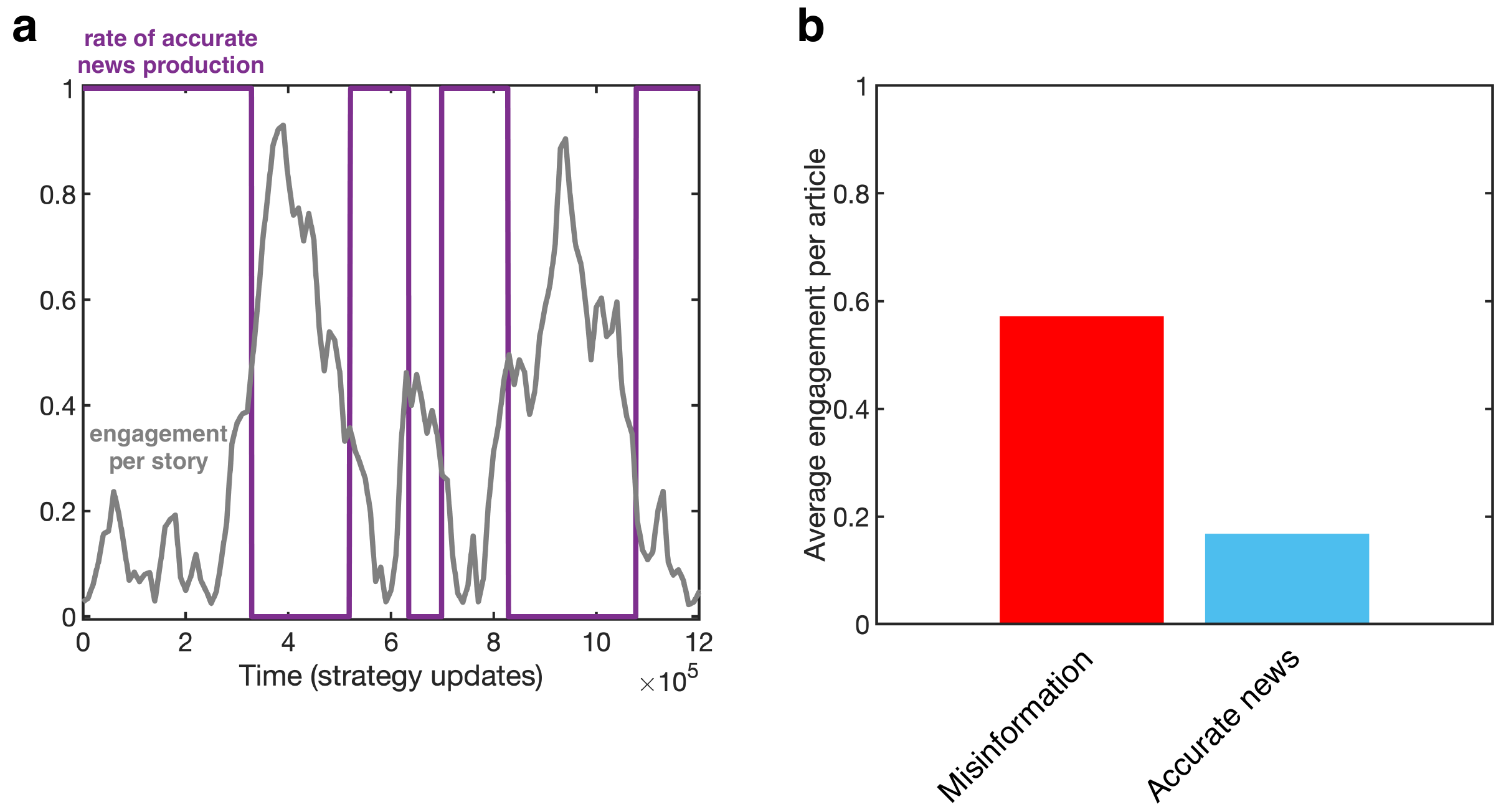}
\caption{\footnotesize \textbf{A transmitter strategy can unilaterally drive engagement with misinformation} An illustrative example of a transmitter strategy that drives engagement with misinformation. We selected a transmitter strategy that employs non-linear feedback, in the form of a sigmoidal function $r_{kt}=1/(1+\exp[\lambda(k/N-0.5)])$ and $r_{kf}=1/(1+\exp[\lambda(k/N-0.25)])$, where we have set $\lambda=100$ and the population of receivers to be $N=100$ (see SI section 1.6). a) Dynamics of production of accurate news (purple) and engagement (grey) for a receiver using a noisy optimization process with low attention to accuracy ($a_0=0$), experience ($a_1=0$) and to payoff ($\sigma=1$). The transmitter strategy produces a pattern of low levels of misinformation when engagement is low, switching to high levels of misinformation production when engagement is high. b) As a result of these dynamics, the average engagement probability per article is higher for misinformation (red) than for accurate news stories (blue), even though the consumer was seeking to increase engagement with true stories.}
\end{figure*}

\noindent\textbf{Finding successful transmitting strategies.}Having defined a strategy space for transmitters and receivers, and having described how a transmitter can unilaterally constrain the dynamics of the misinformation game (Eq.~3), we now examine the dynamics of a receiver's engagement strategy in response to a strategic transmitter (Figure S3). We study a single receiver who employs a noisy optimization process in an attempt to increase engagement with true stories (that is, to increase payoff given truth-seeking preferences, see SI section 1.5). We focus on a myopic optimization process in which the receiver compares the payoff they gain under their current strategy to that under a randomly selected alternate strategy, and tends to adopt the alternate strategy if it will increase their payoff, such that they adopt the new strategy, and discard the old, with probability $\pi_{i\to j}$ determined by a Fermi function $\pi_{i\to j}=\frac{1}{1+\exp[\sigma(w_i-w_j)]}$
where $w_i$ is the payoff received under strategy $i$, and $\sigma$ determines the level of attention the player pays to their payoffs (see SI section 1.5). The case of a single transmitter and receiver corresponds to perfect microtargeting by the transmitter (i.e.~the transmitter is able to promote different types of news to specific receivers in direct response to their engagement habits). However our results also hold in the more realistic case where a transmitter is able to target their stories only at the level of a group of receivers (see SI Section 3.6), and in the case where receivers interact with multiple transmitters (see SI Section 3.8). 
In the SI (Section 1.4) we also consider the case where receivers are perfectly rational \cite{8563,10.2307/2296617}, and the case where receiver behavior is the product of social learning \cite{Traulsen:2006zr,Guilbeault9714} (SI Section 3.3), where we find similar results as for the local optimization process. 

We simulated strategy optimization for inattentive receivers ($a_0=0$ and $\sigma=1$) against $10^8$ randomly selected transmitter strategies $\mathbf{r}=\{\alpha,\beta,\gamma,\theta\}$ who use feedback (Eq. 3) as well as $10^8$ randomly selected transmitter strategies $\mathbf{r}=\{\alpha,\beta,0,0\}$ who do not use feedback. 
We measured the equilibrium probability with which the transmitter shares false stories with the receiver, $v_f$, and the probability with which the receiver engages with false stories (probability of engaging per false story) $v_{fc}/v_f$. We also measured the probability with which the transmitter shares true stories with the receiver, $v_t=1-v_f$, and the probability with which the receiver engages with true stories (probability of engaging per true story) $v_{tc}/v_t$ (see Methods). 

We define the most successful misinformation transmission strategies as those that i) shared misinformation more often than not, $v_f>0.5$ and ii) produced a misinformation engagement probability within the top 10\% of all transmitter strategies considered (see Methods and SI Figure S3). We consider alternate definitions, in which the minimum amount of misinformation shared varies, in the SI (Figure S4). Likewise, we define successful accurate news transmission strategies as those that produce $v_t>0.5$ along with true news engagement within the top 10\% of all transmitter strategies considered. 
Note that accurate transmitters, when faced with inattentive and myopic readers whose incentives are unknown, may also share a mix of true and false stories as they attempt to induce engagement, just as misinformation sites may share accurate stories to draw readers in.
\\
\\
\noindent \textbf{Characterizing successful producer strategies.} To quantify how transmission strategies shape engagement, we calculated the resulting correlation between story accuracy and receiver engagement. 
When feedback is present, the most successful accurate and misinformation dissemination strategies induce characteristic, and opposite, patterns of engagement among readers in our model: successful misinformation sites induce higher reader engagement with each false story, as well as greater overall engagement with false than true stories (Figure S3, Table S2). This pattern arises even when we assume that receivers prefer accurate news over misinformation, so there is no inherent appeal of false stories.

To understand this phenomenon we inspected the strategies of successful misinformation sites under our model. We find that all successful misinformation sites indeed use  strategies that employ feedback in such a way as to enforce a positive correlation between engagement and misinformation transmission. A significant proportion ($72\%$, see Methods) use a type of responsive strategy
that enforces $v_f\geq v_{fc}+v_{tc}$, i.e.~successful misinformation site strategies tend to increase their false stories output rapidly in response to increased engagement. However, if engagement drops, they tend to increase their output of true stories (to draw the user back in). As a result, they ``mash up'' true and fake stories as engagement fluctuates over time.

The behavior of successful accurate news sites, which seek to generate engagement with true stories, shows the opposite pattern to successful misinformation sites. All successful accurate strategies enforce a \textit{negative} correlation between engagement and misinformation transmission.  
A substantial proportion ($56\%$, see Methods) use a strategy 
that enforces $1-v_f\geq v_{fc}+v_{tc}$, i.e.~successful strategies of sites seeking to promote accurate information tend to decrease their output of false stories rapidly in response to increased engagement, but may share more false stories when engagement is low (in a misguided attempt to draw readers back in). We also show  that transmitter strategies generated through a process of co-optimization by transmitters and receivers, produces similar patterns (see SI Section 3.3).
\\
\\
\noindent \textbf{Co-optimization of transmitter and receiver strategies.} Next we explored the dynamics of reader engagement under different reader preferences, $\pi_t$ and $\pi_f$, and different levels of reader attention to payoff, $\sigma$. We allowed readers  to optimize their engagement strategies when interacting with misinformation and accurate transmitters who also seek to optimize their transmission strategy (see Methods). We find three distinct regions for engagement patterns against both misinformation and accurate transmitters (Figure 3). We begin by considering the case of receivers interacting with misinformation transmitters. When receivers prefer accurate information and are highly attentive (region F1), they engage more with accurate news stories. When  receivers who prefer accurate news  are at least somewhat inattentive (region F2), however the strategic behavior of the transmitter causes the receivers to engage more with misinformation despite their preference for true news. When receivers who prefer inaccurate information interact with misinformation transmitters, there is no mis-match between transmitter and receiver preferences and receivers always engage more with misinformation than with accurate stores (region F3). When considering the case of receivers interacting with accurate transmitters, we find a symmetric set of outcomes (regions T1-T3). 

\begin{figure*}[th!] \centering \includegraphics[width=0.75\linewidth]{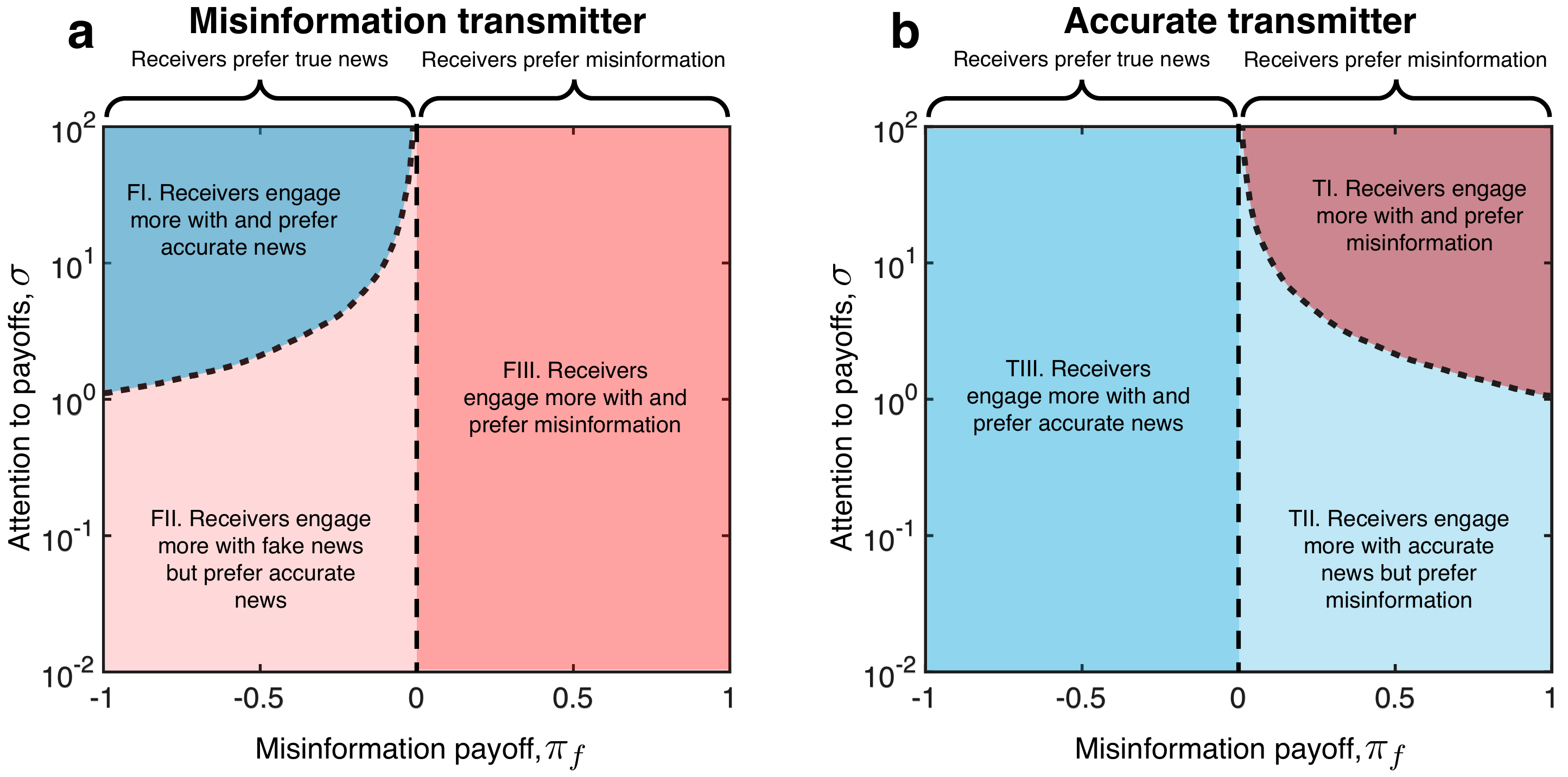}
\caption{\footnotesize \textbf{Transmitter strategy interacts with receiver preferences and attentiveness to determine engagement --} 
a) We calculated the engagement patterns among receivers optimizing under different preferences (where we have set $\pi_t=-\pi_f$), and different levels of attention to payoff, $\sigma$. We allow a misinformation transmitter ($b_f=1$ and $b_t=0$) and a receiver to co-optimize and identified the regions in which receivers have a higher probability of engaging with accurate vs fake stories. In region F1, receivers prefer true news and attention to payoff is high, so that receivers engage more with accurate than with fake stories. In region F2, receivers prefer accurate news but attention to payoff is low, and receivers engage more with fake stories. In region F3, receivers prefer fake stories and engage more with fake stories regardless of their level of attention to payoffs. b) We carried out the same procedure as in panel a, for an accurate transmitter ($b_f=0$ and $b_t=1$).  In region TI, receivers prefer misinformation and attention to payoffs is high, so that receivers engage more with fake than with accurate stories. In region T2, receivers prefer misinformation but attention to payoffs is low, and receivers engage more with accurate stories. In region T3, receivers prefer accurate stories and engage more with accurate stories regardless of their level of attention to payoffs. In all cases receiver strategic exploration was local, with $a_0=a_1=0$ (equivalent results when $a_0>0$ and $a_1>0$ are shown in SU Figures S13 and S17). Transmitter attention to payoffs set at $\sigma=100$. Optimization occurred over $10^4$ time-steps using ensembles of $10^3$ replicates for each value of $\{\pi_f,\sigma\}$ (see Methods).}
\end{figure*}

Critically, there are two large regions (F2 and T2), which produce patterns of receiver engagement that do not reflect the receiver's own preferences. In these regions -- where receivers are not highly attentive and have preferences that are misaligned with those of the transmitter -- it is the transmitter's preferences, and their use of responsive strategies to implement those preferences, that determine engagement among receivers. Our predictions for these regions are distinct from the predictions of previously proposed theories for explaining engagement with misinformation \cite{10.1257/jep.31.2.211,Vosoughi1146,PennyNature}, which allow only for regions F1/T1 and F3/T3 (in which receivers' preferences are aligned with their engagement patterns; SI Section 2).

We demonstrate the robustness of our theoretical results in the SI (Section 3). We show that the existence of region F2/T2 holds when receivers can choose between multiple different news sources (SI Section 3.8).  We also explore the impact of microtargetting and receiver attention to accuracy. We find that when microtargeting is low, the difference in engagement between false versus true stories declines. Thus the ability to target news stories at specific receivers, either by news sites directly or by social media algorithms, can exacerbate the ability of misinformation sources to  drive engagement with false stories despite reader preferences for accuracy (Figure S11). We also explored the impact of two other forms of receiver attention: i) attention to and/or prior knowledge of story accuracy, $a_0$, ii) memory of past interactions when deciding to engage with a news source, $a_1$ (see Methods and SI Section 2). 
We find that increasing prior knowledge of or attention to headline accuracy reduces both the engagement probability per false story and the overall engagement with false stories under an  optimization process (Figure S15-S17).

We also show that transmitters who seek only to maximize engagement among receivers ($b_f=b_t$), but make different assumptions about the type of news those receivers prefer, can inadvertently shape the patterns of engagement among at least somewhat inattentive readers. Transmitters who assume receivers prefer misinformation (e.g.~because it is more novel) employ strategies that seek to reinforce increased engagement by increasing the amount of false stories they share in response (see SI Section 1). As a result, false stories will get more engagement than true stories, which will then reinforce the transmitter's initial assumption (even if it is wrong), creating a self-fulfilling prophecy. Reciprocally, transmitters who assume receivers prefer true news employ strategies that seek to reinforce increased engagement by increasing the amount of true news they share in response -- and generate a pattern of increased engagement with true news. Transmitters who make no assumption about receivers preferences (and hence try out all possible strategies without bias) and transmitters who assume receivers prefer true news, 
both produce greater engagement with true than with false stories (Figure S5). 
\\
\\
\noindent \textbf{Empirical patterns of misinformation engagement.}
Our analysis of the misinformation game presented above identifies three classes of  engagement dynamics with misinformation: The patterns of engagement among those who read misinformation sites fall into either region F1, F2 or F3 of Figure 3a. Only if readers prefer accurate information \emph{and} are highly attentive to their patterns of news consumption will they engage more with accurate stories than with false stories from misinformation sites (region FI). However, if readers prefer accurate information, but are not highly attentive to their patterns of news consumption, they will fall into region F2 -- where misinformation sites can induce high engagement with fake stories by using feedback, despite readers' preference for veracity (Eq. 2). Finally, if readers prefer inaccurate information, they will fall into region F3, where their preference  straightforwardly leads them to engage more with inaccurate information. As shown in Figure 3b, the converse pattern holds for mainstream news sites that aim to spread accurate information. We now present two sets of experiments designed to determine which region(s) reflect the actual dynamics of misinformation engagement observed empirically.

First, we ask whether inaccurate articles generate more or less engagement than accurate articles -- and whether this correlation between engagement and perceived accuracy differs for content published by misinformation versus mainstream sites.
To do so, we used data on Facebook engagement with news from publishers that prior work has determined to be either misinformation or mainstream outlets \cite{Pennycook2521}. We sampled a total of $1,000$ articles from 40 outlets (20 misinformation, 20 mainstream). 
We selected the 25 most recently available news stories for each site from the Crowdtangle database (see Methods), providing a snapshot of the output from each site (and avoiding selecting on the dependent variable by only analyzing high-engagement articles \cite{Allen21}). For each story, we retrieved  Facebook engagement ratings from Crowdtangle \cite{CrowTan} (see Methods). We also assessed each story's perceived accuracy by recruiting $1,000$ American participants from Amazon Mechanical Turk to rate the accuracy of $20$ headlines (yielding a total of $20,000$ accuracy ratings), which has been shown to produce good agreement with the ratings of professional fact-checkers via the wisdom of crowds \cite{doi:10.1126/sciadv.abf4393}. 

\clearpage

\begin{figure*}[th!] \centering \includegraphics[width=0.75\linewidth]{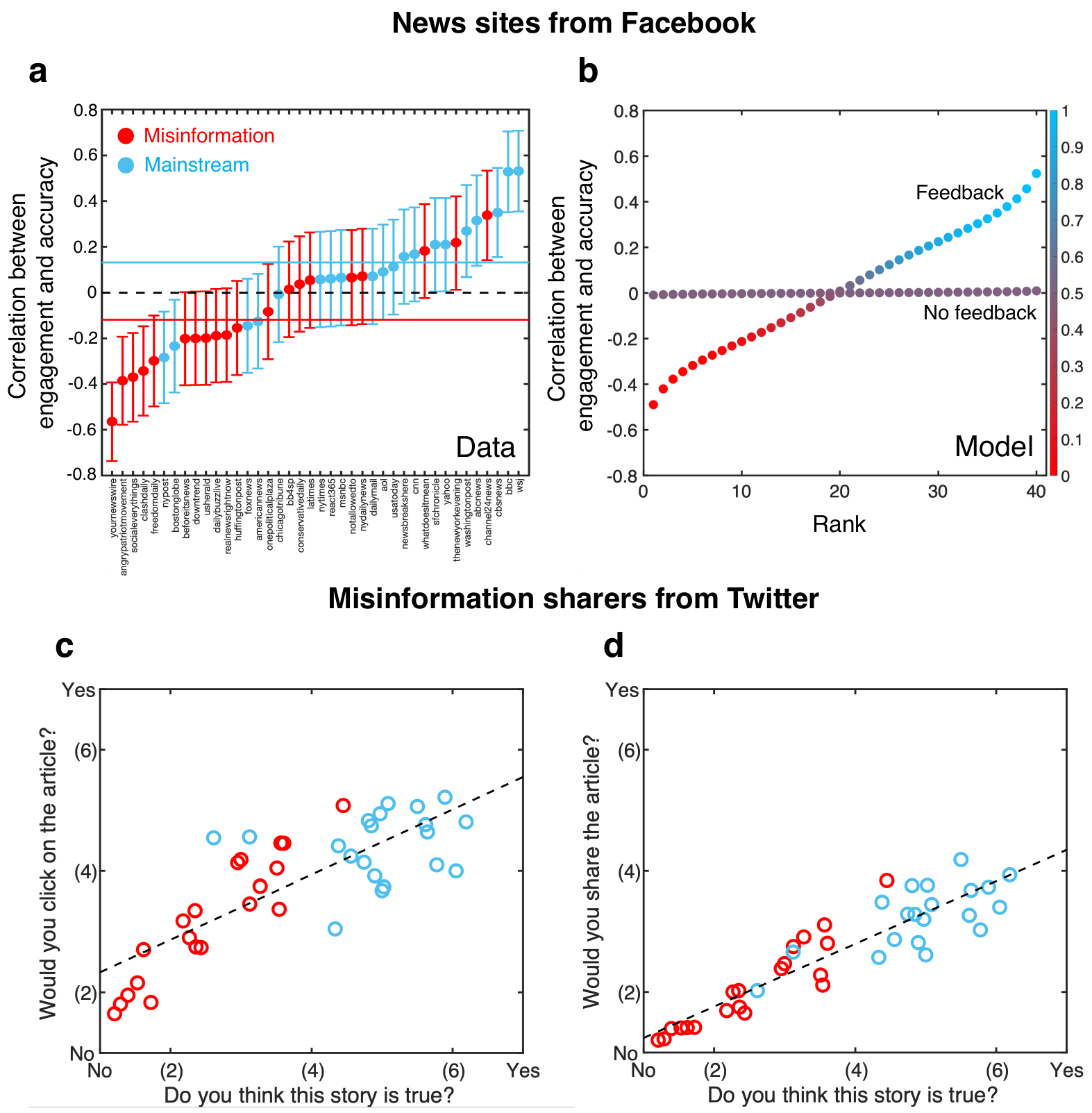}
\caption{\footnotesize \textbf{Less plausible stories receive more engagement on misinformation sites  -- even though misinformation sharers prefer to share accurate news} We selected 20 mainstream and 20 misinformation sites identified in previous studies of misinformation \cite{Pennycook2521} (see Table S5 for list). Using Crowdtangle we selected the most recent 25 news stories for which engagement data on Facebook was available (see Methods). We then recruited American subjects from Amazon Mechanical Turk to assess headline accuracy ($20$ ratings per headline).  a) Regression coefficient between accuracy and $\log_{10}$-engagement for mainstream news sites (blue) and misinformation sites (red)). There is a significant positive correlation between accuracy and engagement across mainstream sites (Fisher's combined test, $p=0.004$; Random effects meta-analysis \cite{DERSIMONIAN1986177,https://doi.org/10.1002/sim.1186}, $p=0.005$, standardized average correlation coefficient -- blue line -- $\mu=0.13$; See Methods and SI Section 3). However there is a significant negative correlation between accuracy and engagement across misinformation sites (Fisher's combined test, $p=0.005$; Random effects meta-analysis, $p=0.013$; standardized average correlation coefficient -- red line -- $\mu=-0.11$; See Methods and SI Section 3). b) Computational experiment, selecting $10^4$ replicates of 20 successful accurate transmitters and 20 successful misinformation transmitters as described in Figure S3. For each replicate we rank ordered the regression coefficients from lowest to highest. Shown is the average regression coefficient in each rank with and without feedback. Colors indicate the proportion of sites in that position that are accurate (blue) or misinformation (red). When feedback is present, the qualitative pattern observed empirically is reproduced by the model. As in Figure 3, computational results are for local receiver strategic exploration (see SI section 1.5) with transmitter error rates of $0.3$ (see Methods).c) We selected 20 mainstream (blue) and 20 misinformation (red) stories identified in previous studies of misinformation \cite{Pennycook2521}. We then recruited a sample of 119 participants directly from Twitter  who had previously shared misinformation to assess the accuracy of 10 headlines of each type. Each participant was then asked to rate their willingness to click on and d) share the article associated with the headline. We observe a significant positive association between accuracy and willingness to click ($\beta=.491, z=11.37, p<0.001$) or share ($\beta=.640, z=14.01, p<0.001$).  As per our pre-registered analysis plans, all analyses use linear regression conducted at the rating level, with all variables z-scored and using robust standard errors clustered on subject and headline. For visualization, we average ratings across subjects for each headline and plot headline-level associations.}
\end{figure*}

\clearpage

Before turning to the key question of the correlation between engagement and perceived accuracy, we begin by noting an expected basic descriptive pattern: mainstream news sites, which we assume seek to promote accurate information, tend to share headlines with higher accuracy ratings than misinformation sites ($p<0.001$, Figure 4a). Importantly, however, both mainstream and misinformation sites show wide variation in perceived headline accuracy. Thus, there is substantial overlap in plausibility between the content produced by the two kinds of sites, with many articles from misinformation sites being rated as more accurate than many articles from mainstream sites.

How, then, does engagement on Facebook vary across this range of perceived accuracy? The answer is strikingly different for mainstream versus misinformation sites. 
We find a significant negative correlation between engagement and perceived accuracy for articles from misinformation sites ($p=0.005$), whereas we find a significant positive correlation for articles from  mainstream sites ($p=0.004$). 

This pattern rules out the case in which all readers are highly attentive and prefer accurate news, which would lead to more engagement for more accurate articles regardless of publisher type; as well as the case in which all readers are highly attentive and prefer inaccurate news, which would lead to less engagement for more accurate articles regardless of publisher type.  Instead, the observed pattern is consistent with readers who belong in either region F2/T2 or region F3/T3 for both misinformation and mainstream sites. That is, readers of misinformation sites either prefer less accurate information (region F3), or they prefer accurate information but are sufficiently inattentive that their engagement habits can be shaped by misinformation transmitters (region F2); and readers of mainstream sites either prefer accurate information (region T3), or they prefer inaccurate information but are sufficiently inattentive that their engagement habits can be shaped by mainstream transmitters (region T2) . In order to differentiate between these possibilities, we conducted a second set of experiments to measure  users' preferences for engaging with accurate versus inaccurate news.
\\
\\
\noindent \textbf{Empirical patterns of receiver engagement preference.}
We empirically determined the preferences of readers who engage with misinformation sites versus mainstream news sites through a set of preregistered survey experiments (see Methods). First, we recruited 511 members of the general population via Lucid (see Methods), of whom 124 indicated regularly using one or more fake sites; and second, we recruited 119 participants directly from Twitter who had actually shared links to misinformation sites previously (behaviorally demonstrating their engagement with such sites).  Each participant was shown a series of true and false headlines, and for each was asked to rate the headline's accuracy, and indicate how likely they would be to click on and to share the story if they saw it online. If we find that subjects who use misinformation sites prefer to engage with  content they perceived as inaccurate, this would suggest that reader preference is driving the greater engagement rates for more inaccurate headlines observed in Figure 4a (i.e. suggests we are in region F3 in Figure 3a). If, on the other hand, those who use misinformation sites actually prefer to engage with content they perceive as accurate, this is consistent with the interpretation that reader preferences alone are not driving the pattern of higher engagement for more inaccurate content observed in Figure 4a -- rather the strategies of producers may cause inaccurate content to receive more engagement (region F2 in Figure 3a).

The observed preferences are strikingly consistent across the different types of users. In all cases, there is a significant positive correlation between perceived accuracy and willingness of readers to click or share an article ($p<0.001$ for both outcomes across all three groups, Figure 4 and Figure S21-23). This holds for subjects recruited from Twitter who had previously shared misinformation, as well as those recruited from the general population regardless of their self-reported use of misinformation sites. The results are qualitatively equivalent when using objective accuracy (as measured by professional fact-checkers; $p<0.001$ for both outcomes across all three groups), and our findings continue to hold under replication in a Supplemental Experiment conducted with a general population sample in which participants were only asked their willingness to click and share each headline, without first rating the headline's accuracy; see SI section 4).

Thus, our empirical results indicate that the pattern of reader engagement with misinformation sites falls in region F2: the observed pattern reflects the transmitter's desire to generate high engagement with misinformation, and not a consumer preference for engaging with inaccurate news. For mainstream sites, the pattern of reader engagement with mainstream sites falls into region T3, in which the preferences of the reader and transmitter are aligned (both seeking high engagement with accurate news).

Taken together these empirical results imply that reader preferences cannot be reliably inferred from patterns of reader engagement. Rather they suggest a more complicated interaction, in which reader engagement with news may result
from a feedback loop generated by misinformation sites acting on not fully attentive readers who prefer accurate news -- but nonetheless engage more with inaccurate news due to the transmitter's behavioral strategy.

\section*{Discussion}

Understanding why people engage with misinformation has become increasingly important. Widespread belief in false information can destabilize democratic institutions, fuel populist movements and polarization, and undermine public health efforts, as seen during the COVID-19 pandemic. 
Here, we develop formal game theoretic models to shed new light on why misinformation spreads.
We developed a framework to study the 
 strategic interaction between readers and news sources, and combined this with empirical analyses of patterns of engagement with misinformation and mainstream news sites, as well as the expressed preferences of their readers.
Our theoretical analysis reveals two key features of the dynamics of consumer engagement with misinformation.
First, misinformation sites that are likely to be successful in producing high overall levels of engagement among 
readers will tend to produce patterns of engagement in which the false stories
receive more engagement than true stories receive, even if true stories are preferred over false stories by their readers (region F2 of Figure 3a). In contrast, mainstream sites that seek to promote engagement with accurate stories should tend to produce the opposite engagement pattern among their readers, unless those readers are highly attentive and actively prefer to engage with misinformation (Figure 3b). Second, a misinformation site committed to pushing as much false information as possible into the information ecosystem should  
-- when faced with readers who prefer accurate stories -- use responsive strategy and, as a result, publish a substantial amount of true news alongside the false. 


While our model is a highly simplified description of the interactions that occur in complex online information ecosystems, it illustrates the dangers of drawing conclusions about consumer preferences without accounting for the supply side of news. Our results cast a new light on previous findings that false stories receive more engagement than true stories, which has been interpreted as reflecting a consumer preference for novel but false information \cite{Vosoughi1146}. This observation stands in contrast to other studies, which found that true stories receive as much or more engagement than false stories \cite{Grinberg374,PENNYCOOK201939}. Our results suggest that these different patterns may be explained by the different \textit{sources} of news examined in the different studies. Among claims that have been fact-checked by Snopes (largely coming from misinformation sources) -- as focused on in  \cite{Vosoughi1146} -- we might expect less accurate news to get more engagement due to the behaviour of its suppliers. Conversely, when news come from a more balanced set of sources, as in \cite{Grinberg374,PENNYCOOK201939}, we would not expect the same pattern to emerge.
Of course, our findings do not rule out the possibility that novelty may drive engagement with misinformation \cite{Vosoughi1146}, but our work shows that high levels of misinformation engagement can arise even in the absence of intrinsic novelty appeal or motivated reasoning, and especially when transmitters behave in a highly responsive manner by microtargeting their content \cite{tappin_wittenberg_hewitt_berinsky_rand_2022}. Our results also help to explain why misinformation sites publish so much content that is actually quite plausible (Figure S21). To garner engagement, responsive misinformation-spreading strategies must provide enough accurate-seeming content to avoid alienating their readers, who prefer to engage with accurate news.

The idea that misinformation sites can successfully induce readers into engaging with false stories 
has important practical implications. First, our findings suggest a new approach for identifying outlets that are seeking to spread misinformation: examining the relationship between articles' engagement and perceived accuracy. Outlets from which particularly implausible articles generate particularly high levels of engagement may be aiding the spread of misinformation by employing responsive transmission strategies. Observing such a relationship could be used as a signal that an outlet is intent on spreading misinformation, or making faulty assumptions about the preferences of readers.
Similarly, if such a relationship is observed in the patterns of engagement with inaccurate stories on a particular social media platform, this may suggest the platform uses responsive strategies to promote misinformation to users.
Second, the tendency to share misinformation can become self-reinforcing on the part of transmitters: sites that seek to spread misinformation using responsive strategies can create the incorrect impression that readers prefer misinformation. This pattern, if taken at face value, may then lead to increased misinformation production by  sites that do not have any particular agenda beyond simply maximizing engagement -- even when such sites would actually maximize engagement by publishing true articles. Third, we find that this apparent preference for misinformation can be reversed by encouraging readers to pay attention to key aspects of their news consumption habits. This observation reinforces the importance of (in)attention in combating misinformation.

In sum, we have demonstrated the importance of the feedback loop between news publishers and readers. We have shown that focusing only on reader behavior can lead to incorrect conclusions about readers' preferences, while considering publisher strategies can resolve apparent contradictions in the literature, and can highlight new directions for combating misinformation. News consumption does not occur in a vacuum, and to understand the dynamics of misinformation it is essential to explore supply as well as demand.    

\section*{Methods}
\noindent \textbf{Simulations.}
We performed simulations to determine the most successful transmitter strategies for fake and mainstream sites. We selected a transmitter strategy $\mathbf{r}=\{\alpha,\beta,\gamma,\theta\}$ (Eq. 2) with each parameter drawn uniformly from the interval required to produce a viable strategy. We initialized a receiver with a strategy that does not engage with any news from the transmitter, i.e. $\mathbf{q}=\{0,0,0,0\}$ (Eq. 1) which was then allowed to update under a local optimization process (see SI Section 1). For each set of receiver and transmitter strategies, we  calculated the stationary distribution for transmission and engagement of news stories $\mathbf{v}=\{v_{tc},v_{tn},v_{fc},v_{fn}\}$ under the iterated misinformation game. For a single receiver the stationary distribution can be found explicitly. For a larger population of receivers we simulated $10^4$ rounds of the game, which allowed us to estimate the stationary distribution numerically. In all simulations we assume that both transmitters and receivers experience execution errors \cite{Stewart17558} with probability $\varepsilon=10^{-3}$. Finally we assume a ``perception error'' among transmitters, in which they incorrectly label false stories as true, or vice versa, of $\eta=0.3$.
Transmitter strategies without feedback were produced in the same way, with the constraint $\theta=\gamma=0$.

We used the stationary distribution $\mathbf{v}$ to calculate receiver payoffs at equilibrium. We then allowed receiver strategies to update under a local optimization process (see SI Section 1.5 for further details). After  a burn in period of $10^4$ update events we measured the average engagement and engagement probability of receivers over an additional $10^4$ update events.

Simulations in which transmitters and/or multiple receivers co-optimize followed the same procure, with all players engaging in the local optimization process. Region plots (Figure 3 and SI Section 3) were produced by breaking the parameter space into a $100\times 100$ grid and co-optimizing $10^3$ replicates at each point.

Transmitters were identified as employing extortion strategies, $r_+^*$ or $r_-^*$, if their strategy lay within a $\Delta$-neighborhood of a ``true'' extortion strategy \cite{Stewart:2013fk,Hilbe:2013aa}, with  $\Delta=0.05$. Significant over-representation of extortion strategies among the most successful fake and mainstream sites (i.e those strategies that successfully produce engagement with fake and true stories respectively) was then determined by comparing the prevalence of the strategy in our data to the null distribution, i.e. the probability with which such strategies are randomly drawn. Significance level quoted is $p<0.01$.
\\
\\
\noindent \textbf{Modelling experiments.}
In order to assess the role of transmitter feedback in producing empirical patterns of engagement with different types of news site (Figure S3 and Figure 4b), we ran simulation experiments. We randomly selected successful transmitter strategies and their corresponding receiver strategy. For each transmitter we generated a sequence of 20 news stories according to their strategy, and calculated the probability of engagement among a population of $10^5$ receivers for each story based on the associated receiver strategy. This produced a single simulated experiment, and we calculated the regression coefficient for the standardized engagement rate against the perceived accuracy of the story under perception error $\eta$ (Figure S3). This process was repeated $4\times10^5$ times to produce a distribution of accuracy and regression coefficients (Figure S3), for both accurate and misinformation transmitters, with and without transmitter feedback.

Next we simulated a rank ordering of regression coefficients (main text Figure 4b) by randomly drawing 20 accurate transmitter strategies and 20 misinformation transmitter strategies, and calculating their simulated regression coefficient as described above. We then ordered all 40 strategies from lowest to highest to produce a ranking. We repeated this procedure $10^4$ times to produce an average regression coefficient in each ranked position, for transmitter strategies with and without feedback, as shown in Figure 4b.
\\
\\
\textbf{Experiment A: Empirical patterns of misinformation engagement.} We asked participants to evaluate the accuracy of 1,000 articles across 40 mainstream and misinformation sites, gathered from Crowdtangle, a tool for monitoring engagement on Social Media). The study was approved by MIT institutional review board (IRB) (protocol 1806400195). Informed consent was provided at the beginning of the study.
\\
\textbf{Participants.} From 22 to 29 November 2020, we recruited 1,000 participants from Amazon Mechanical Turk who met the following three criteria: located in the US, more than 100 studies completed on the platform, and more than 95 per cent of them approved. A total of 1,027 participants initiated the study but 27 did not complete the evaluation task and were excluded. The sample included 576 males and 424 females, with a mean age of 38.49 years (min. 19; max. 96). Median completion time was 4 minutes and 55 seconds.
\\
\textbf{Materials.} We used Crowdtangle to gather the 25 most recently available news stories from 40 media outlets (i.e. 1,000 articles; 500 posted by 20 misinformation sites and 500 posted by 20 mainstream sites), along with the headline, lede, date of publication, link and level of engagement. See Supplementary Information section 3 for additional details. We then used this data to present the participants outlined above with 20 article headlines and ledes, drawn randomly from within one of the two media outlet subsets, and asked them to assess the accuracy of the information they were faced with. Specifically, they were asked ``Do you think this story is true?'', to which they responded on a seven-point scale from ``Definitely NO'' to ``Definitely YES''. The study concluded with seven demographic questions (age, gender, education, political conservativeness on social and economic issues, political position, and political preference) and a section to leave comments. 
\\
\\
\textbf{Experiment B: Empirical patterns of receiver preference (Lucid).} We asked participants to assess the accuracy, likelihood of sharing, and likelihood of clicking on 40 headlines. We then asked which domains they regularly use for news (see Figure S22 and S23). The study was approved by MIT institutional review board (protocol 1806400195). Informed consent was provided at the beginning of the study.
\\
\textbf{Participants.} We recruited American participants via Lucid (a widely used platform for recruiting participants for online experiments) from  March 15 to 25, 2022, quota-matched to the national distribution on age, gender, ethnicity, and geographic region. In accordance with our preregistration (90974) we excluded: 1 participant located outside the US, 188 who failed one or two of the trivial attention checks at the study outset, and 79 who reported not having at least one social media account. We also excluded 21 participants who did not declare using at least one of the 60 listed domains for news (as these participants cannot be classified as users of either misinformation sites  or non-misinformation sites), leaving a final sample of 511 subjects. The sample included 237 males and 259 females, with a mean age of 47.01 years (min. 18; max. 87). Median completion time was 10 minutes and 39 seconds.
\\
\textbf{Materials.} We identified a pool of 40 news "cards" (i.e. representations of Facebook posts with an image, headline, and a source; balanced on veracity and partisan lean) and a list of 60 domains regularly used for news (20 identified as mainstream, 20 as hyper-partisan, and as 20 fake by Pennycook and Rand, 2019  \cite{Pennycook2521}). We then asked participants to : i) evaluate 20 cards, drawn randomly from the set of 40, on the accuracy (i.e. ``Do you think this story is true?'', seven-point scale from ``Definitely NO'' to ``Definitely YES''), likelihood of sharing (i.e. ``If you were to see the above headline online, would you share it?'', seven-point scale from ``Definitely NO'' to ``Definitely YES''), and likelihood of clicking (i.e. ``If you were to see the above headline online, would you click on it to read the article?'', seven-point scale from ``Definitely NO'' to ``Definitely YES'') of the information presented; ii) select which of the 60 domains they regularly use for news (with this information, we classified participants as misinformation media users if they selected at least one domain pre-identified as a misinformation outlet). The study concluded with a list of 20 exploratory items.
\\
\\
\textbf{Experiment C: Empirical patterns of receiver preference (Twitter).} We recruited participants for Experiment C through an advertisement campaign on Twitter conducted between April 22 and April 29, 2022.  We created a set of 24 ad creatives that paired different images with the text "We want to know your opinion! Take a 5 minute survey about the news. Just click here" and targeted a custom audience of  all the Twitter users who had previously engaged with content from any of the 20 misinformation sites identified by Pennycook and Rand, 2019  \cite{Pennycook2521}. Upon clicking on the advertisement, participants were redirected to a survey asking them  to assess the accuracy, likelihood of sharing, and likelihood of clicking on the same 40 headlines used in Experiment B above. 
The study was approved by MIT institutional review board (protocol 1806400195). Informed consent was provided at the beginning of the study.
\\
\textbf{Participants.} From April 22 to 29, 2022, 206 participants entered the survey. Of these, we excluded 87 who failed at least one of the two attention checks of our study. The final sample of 119 participants included 60 males, 53 females and 6 participants who preferred not to answer, with a mean age of 57.81 years (min. 28; max. 79). Median completion time was 6 minutes and 14 seconds.
\\
\textbf{Materials.} We use the same pool of 40 politically- and veracity-balanced news "cards" from Experiment B. We also asked participants to evaluate cards based on accuracy, likelihood of sharing, and likelihood of clicking, just like in Experiment B. The only difference between the questionnaires of the two experiments was that we asked participants to evaluate 10 cards instead of 20, that we did not asked them to select what domains they use for news, and  that we concluded with a list of 10 exploratory items instead of 20.

\clearpage

\section*{Detailed analysis of the misinformation game}

The stage game payoff matrix for the misinformation game is described in Table S1 for a single receiver and a single transmitter. We consider an infinitely iterated version of this game, which in general is played by a single transmitter and $N$ receivers, as described in the main text. This iterated game describes the long term dynamics of a receiver being served news stories by a transmitter.

\begin{figure}[th!] \centering \includegraphics[scale=0.2]{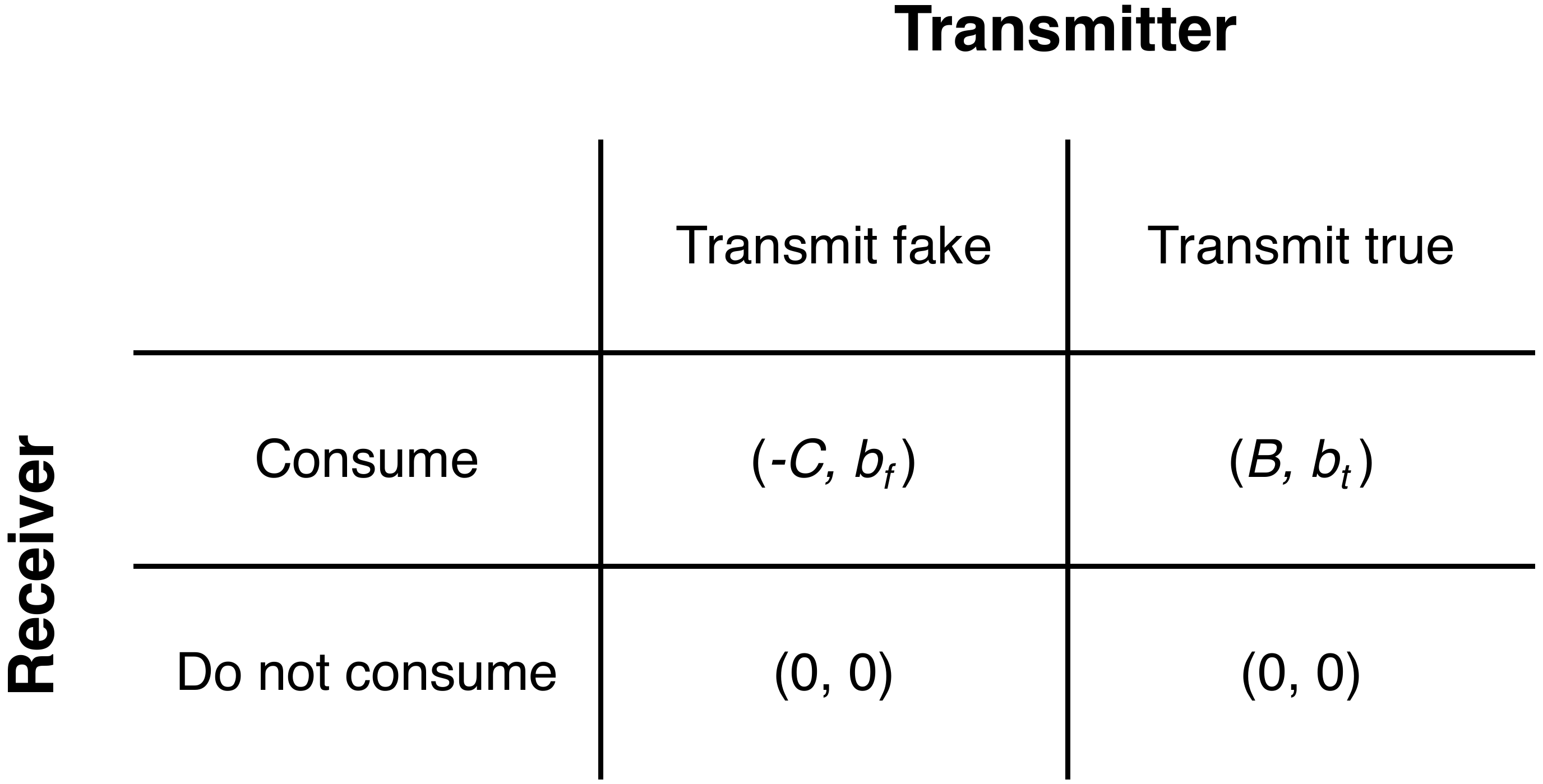}
\caption*{\small Table S1: \textbf{ Stage game payoff matrix} }
\end{figure}

We consider a transmitter strategy $\mathbf{r}$ as described by Eq. 1, such that the probability of a transmitter sharing true news is $r_{ki}$, where  the last piece of news shared was consumed by $k$ receivers and was of type $i$. The possible types of news are true ($t$) or fake ($f$), so that $i\in\{t,f\}$, and in a population of $N$ receivers the amount of engagement is $k\in[0,N]$. The transmitter strategy space we consider is thus memory-1. When using a strategy of this type, a transmitter unilaterally enforces a  linear relationship between receiver engagement and the amount of misinformation shared by the transmitter (as described in the main text). This relationship is given by Eq. 3 and holds regardless of the receiver strategy $\mathbf{p}$, i.e. it holds for other types of receiver strategy than given by Eq. 2, including cases where the receiver strategy employs memory of their full history of interactions with the transmitter.

To see this, encode the state of the repeated game in round $j$ as $v_{ki}^j$ where $k$ is the number  of receivers who engage in round $j$ and $i$ is the type of news shared by the transmitter. The probability that the game is in state $v_{ki}^{j+1}$ at round $j+1$ is then given by\footnote{Note that equation numbers in the supplement follow on from the main text}

\begin{equation}\tag{4}
v_{kt}^{j+1}=\sum^N_{m=0}\sum_{l\in\{t,f\}}v_{ml}^jr_{ml}\sum_{A\in F_k}\prod_{a\in A}p_a\prod_{b\in A^C}(1-p_b)
\end{equation}
\\
where $\text{Pr}(K=k)=\sum_{A\in F_k}\prod_{a\in A}p^a\prod_{b\in A^C}(1-p^b)$ is the Poisson Binomial distribution describing the probability that $k$ receivers engage, where $F_k$ encodes the possible subsets of $k$ integers that can be selected from $N$, i.e. the possible combinations of receivers who can engage with a piece of news to produce overall engagement level $k$. The set $A$ is just a particular such subset. Note that we allow each receiver strategy $p_a$ and $p_b$ to be different and depend on the full history of play of the receivers and transmitter i.e. we place no particular constraint on the receiver strategy used. The master equation for $v_{kf}$ is defined in the equivalent way with $(1-r_{ml})$ in place of $r_{ml}$ in Eq. 4.

At equilibrium the solution to Eq. 4 must satisfy $v_{ki}^{j+1}=v_{ki}^j=v_{ki}$. Below we give the conditions for a transmitter strategy $\mathbf{r}$ of the form Eq. 1 to generate an equilibrium solution satisfying Eq. 3 for a single transmitter and receiver. Note that we assume that receiver and transmitter actions are subject to a small execution error at rate $\varepsilon$, which ensures that the Markov process described by Eq. 4 has no absorbing states, and a unique stationary distribution. And so, since we assume an infinitely repeated game, the dynamics are guaranteed to converge to this equilibrium.

\subsection*{Stationary distribution}

The stationary distribution for the infinitely repeated game is encoded by the vector of probabilities $\mathbf{v}$. Each entry $v_{ki}\in [0,1]$ describes the stationary probability that the game is in state $(k,i)$, where $k \in \{0, 1, 2, \ldots N\}$ gives the number of receivers who choose to engage with the piece of news shared by the transmitted, and $i\in \{nt,f\}$ encodes the type of news shared. The master equation for the Markov process associated with the infinitely repeated game is given by Eq. 4. At equilibrium, for a true news item $i=t$, this equation has the form
m
\begin{equation}\tag{5}
v_{kt}=\sum^N_{m=0}\sum_{l\in\{t,f\}}v_{ml}r_{ml}\sum_{A\in F_k}\prod_{a\in A}p_a\prod_{b\in A^C}(1-p_b).
\end{equation}
An analogous equation holds for $i=f$ by replacing $r_{ml}$ with $(1-r_{ml})$.

In order to solve for the stationary distribution of this process, marginalized over all possible overall engagement values $k$, we first note that the overall probability of sharing misinformation by the transmitter is

\begin{equation}\tag{6}
v_{f}=\sum_{k=0}^N v_{kf}
\end{equation}
\\
and $v_t=1-v_f$ is the overall probability of production of true news by the transmitter. Applying this definition to Eq. 5 we recover

\begin{equation}\tag{7}
v_{t}=\sum^N_{m=0}\sum_{l\in\{t,f\}}v_{ml}r_{ml}
\end{equation}
\\
since in summing over all possible engagement levels $k$, we sum over the probabilities of all possible combinations of receiver overall engagement which necessarily total 1. If we now set $r_{mt}=\alpha_0+\sum_{i} \gamma_i \left(\frac{m}{N}\right)^i$ and $r_{mf}=\beta_0+\sum_{i} \theta_i \left(\frac{m}{N}\right)^i$ in Eq. 7 
we recover

\begin{equation}\tag{8}
v_{t}=\alpha_0 v_t+\beta_0 v_f+\sum^N_{m=0}\sum^N_{i=1}\left(v_{mt}\gamma_i \left(\frac{m}{N}\right)^i+v_{mf}\theta_i \left(\frac{m}{N}\right)^i\right).
\end{equation}
\\
where the terms $\gamma_i$ and $\theta_i$ are polynomial coefficients describing the response function of the transmitter to receiver engagement, given they previously shared true or false stories respectively.
Now we can write 
\begin{equation}\tag{9}
(v_{tc})^i=\sum^N_{m=0}v_{mt}\left(\frac{m}{N}\right)^i
\end{equation}
\\
where $(v_{tc})^i$, is the $i$th raw moment of the distribution of transmitter engagement with accurate news stories. i.e.  $(v_{tc})^1=v_{tc}$ is the average overall probability of receivers engaging with true news and so on. And so Eq. 8 can be written as

\begin{equation*}
v_{t}=\alpha_0 v_t+\beta_0 v_f+\sum^N_{i=1}\gamma_i (v_{tc})^i + \theta_i (v_{fc})^i
\end{equation*}
\\
and we recover an expression for the receiver engagement with true and false news in terms of the transmitter strategy. Next we note that when there is only one receiver and one transmitter, $(v_{tc})^i=v_{tc}$ and $(v_{fc})^i=v_{fc}$ for all $i>1$ and so we can write  $\gamma=\sum_{l}\gamma_l$ and $\theta=\sum_{l}\theta_l$ to give 

\begin{equation*}
v_{t}=\alpha_0 v_t+\gamma v_{tc}+\beta_0 v_f+\theta v_{fc}
\end{equation*}
\\
Replacing $v_f=1-v_t$ and rearranging gives us the relationship between the proportion of transmitted stories that are false and the overall probability with which receivers engage with true and false stories, presented in Eq 3 of the main text. Note that this expression holds for any polynomial transmitter strategy and an arbitrary receiver strategy when interactions are pairwise.

This equation, (i.e. Eq. 3 of the main text) holds when there is a single transmitter and receiver, or when there are multiple receivers and the transmitter uses only a linear response function. In the more general case of multiple transmitters and a non-linear response function, we recover a more general relationship in terms of the moments of the stationary distribution of the iterated game, i.e.

\begin{equation}\tag{10}
v_{t}=\alpha_0 v_t+\beta_0 v_f+\sum_i\gamma_i (v_{tc})^i+\sum_i\theta_i (v_{fc})^i
\end{equation}
\\
and so a transmitter can induce positive or negative correlations in the same way as for the simpler case of Eq. 3, by choosing a response function with positive or negative coefficients as desired.

\subsection*{Definition of responsive strategies}

We define a responsive strategy as a strategy which assuredly increases the amount of news served to receivers of a given type as the receivers increase their engagement, in a manner that reflects the preferences of the transmitter. Thus a responsive misinformation strategy increases the probability of sharing misinformation stories as receivers increase their engagement. For this to be true in the case of a single receiver (or in the case of multiple receivers and a linear transmitter strategy), the coefficients $\frac{\theta}{1-\alpha_0+\beta_0}$ and $\frac{\gamma}{1-\alpha_0+\beta_0}$ in Eq. 3 must be negative. The more general case of a non-linear transmitter strategy and multiple receivers is discussed in Section 1.6 below.

In order for a strategy to be viable Eq.1 must produce viable probabilities, i.e. $r_{ki}\in[0,1]$. For a single receiver (or a linear strategy and multiple receivers), this in turn requires $0\leq \alpha_0\leq 1$, $0\leq \beta_0 \leq 1$, $-\alpha_0\leq \gamma \leq 1-\alpha_0$ and $-\beta_0\leq \theta \leq 1-\beta_0$. A viable responsive misinformation strategy thus requires $-\alpha_0\leq\gamma<0$ and $-\beta_0\leq\theta<0$. Equivalently, a responsive mainstream news (i.e. accurate news spreading) strategy requires $0<\gamma\leq1-\alpha_0$ and $0<\theta\leq 1-\beta_0$.

We have defined a responsive transmitter strategy as one that always serves more news of their preferred type to receivers who consume more news. Thus a responsive misinformation strategy serves more fake stories to receivers who engage more with the transmitter's stories. As a result truth-seeking receivers who try to increase their benefit from true stories by engaging more, in fact receive increasing amounts misinformation. In contrast, receivers who engage little with the transmitter's stories are targeted with more true stories and thus are incentivized to engage more. The conditions for a strategy to be responsive are $\gamma>0$ and $\theta>0$. 

We define an extortion strategy as a particular type of responsive strategy that additionally enforces $v_f>v_{c}$ (for misinformation sites), where $v_c=v_{tc}+v_{fc}$ is the overall probability of engagement by receivers. Similarly, for mainstream sites extortion strategies must enforce $v_t>v_{c}$ . The term ``extortion'' was previously introduced in reference to analogous strategies studied in the Prisoner's Dilemma \cite{Press:2012fk,Stewart:2013fk,Stewart:2012ys,Hilbe:2013aa}. The extortion strategies studied here ensure that the rate of increase in fake (or true) news exceeds the rate of increase of overall engagement by receivers.

\subsection*{Necessary and sufficient conditions for extortion under linear transmitter strategies}

The definition of a extortionate misinformation site is that $v_f>v_{c}$ for all receiver strategies. In order to find necessary and sufficient conditions for a linear transmitter strategy to enforce such a relationship, first define the following parameters \cite{Stewart:2013fk,Akin2}

\begin{equation*}
\kappa=\frac{1 -  \alpha_0 - (\gamma - \theta)/2}{1 - \alpha_0 + \beta_0 + \theta}
\end{equation*}

\begin{equation*}
\lambda=-\frac{(\gamma - \theta)/2}{1 - \alpha_0 + \beta_0 + \theta}
\end{equation*}

\begin{equation}\tag{11}
\chi=-\frac{\gamma + \theta}{2(1 -  \alpha_0 + \beta_0 - (\gamma - \theta)/2)}
\end{equation}
where for simplicity  we have dropped the $0$ subscripts from $\alpha_0$ and $\beta_0$.
Here $|\chi|\leq1$ is required to produce a viable strategy. Note that the denominator in Eq 11, namely $(1 -  \alpha_0 + \beta_0 - (\gamma - \theta)/2)$, is required to be positive to produce a viable strategy and thus any misinformation promoting responsive strategy necessarily has $\chi>0$.

Substituting Eq. 11 into Eq. 3 produces the following expression

\begin{equation}\tag{12}
v_f-\kappa+\lambda(v_{tn}+v_{fc})=\chi\left[v_c-\kappa+\lambda(v_{tn}+v_{fc})\right]
\end{equation}
\\
where $v_{tn}$ is the probability with which true news is shared but not consumed. Eq. 12 is convenient because it allows us to state the conditions for extortion as i) $\chi>0$ (which is necessary for the strategy to be responsive) and ii) $v_c\leq\kappa+\lambda(v_{tn}+v_{fc})$ for all possible receiver strategies (which is necessary to ensure $v_{f}>v_c$, given that $|\chi|\leq 1$). If these two conditions are met, the strategy is responsive and $v_f>v_c$ for all receiver strategies, and so the strategy is an extortioner.

The first condition is met provided $\gamma + \theta<0$ (since $\gamma<1-\alpha_0$ is required to produce a viable strategy). To asses the second condition note that ii) implies $v_f+v_c\leq 2\kappa-2\lambda(v_{tn}+v_{fc})$. Also note that $v_f+v_c=1+v_{fc}-v_{tn}$ and so condition ii) is equivalent to

\begin{equation}\tag{13}
1+v_{fc}-v_{tn}\leq 2\kappa-2\lambda(v_{tn}+v_{fc})
\end{equation}
\\
All that remains is to inspect the limiting cases, where this conditions is most stringent. A receiver who never consumes enforces $v_{fc}=0$ and Eq. 13 becomes  $1-v_{tn}\leq 2\kappa-2\lambda v_{tn}$ which is guaranteed to hold provided 

\begin{equation*}
\kappa \geq 1/2
\end{equation*}

and

\begin{equation}\tag{14a}
\kappa\geq \lambda
\end{equation}
\\
The other limiting case of a receiver who always consumes enforces $v_{tn}=0$ and generates conditions 

\begin{equation*}
\kappa \geq 1/2
\end{equation*}
and
\begin{equation}\tag{14b}
\kappa\geq 1+\lambda.
\end{equation}

Eq. 14b is the more stringent of these two conditions and thus it provides necessary and sufficient conditions for extortion. Substituting back from Eq. 11 these conditions become

\begin{equation*}
\theta=-\beta_0
\end{equation*}
and
\begin{equation}\tag{15}
\alpha_0 + \beta_0+\gamma \leq 1.
\end{equation}

The equivalent condition for a mainstream transmitter is
\begin{equation*}
\gamma=1-\alpha_0
\end{equation*}
and
\begin{equation}\tag{16}
\alpha_0 + \beta_0+\theta \leq 1. 
\end{equation}
\\
And so we can define an extortionate misinformation strategy \cite{Press:2012fk} as one that ensures the amount of misinformation shared always exceeds the amount of news consumed i.e. $v_f>v_{fc}+v_{tc}$. For a misinformation transmitter using a linear strategy, this has the form $\mathbf{r}_-^*=\big\{\theta=-\beta_0,1-\alpha_0-\beta_0<\gamma\leq0\big\}$ (see below). Conversely, an extortion strategy for a transmitter seeking to spread accurate information ensures that the amount of true news shared always exceeds the amount of news consumed i.e. $v_t>v_{fc}+v_{tc}$. For such a transmitter a linear extortion strategy has the form $\mathbf{r}^*_+=\big\{\gamma=1-\alpha_0, 0\leq\theta<1-\alpha_0-\beta_0\big\}$ .

\subsection*{Nash equilibria in the misinformation game}

Our model assumes inattentive receivers who noisily optimization their engagement strategies as they seek to consume true stories. Such a process does not necessarily lead to optimal receivers behavior, or even to behaviors predicted by standard solution concepts such as Nash equilibria. However, it is useful to characterize Nash equilibria for the misinformation game.

For a mainstream site, which seeks to promote accurate information, there is no conflict between the incentives of transmitter and receiver, and so they both should adopt a Nash equilibrium in which the receiver always engages and the transmitter only shares true stories.
For a misinformation site, which seeks to promote false news, there is a conflict between the incentives of transmitter and receiver. 

The Nash equilibria for the two-player misinformation game can be easily characterized using the Folk Theorem for infinitely iterated games \cite{8563}. The minimax value for receivers in the misinformation game is 0, which is achieved by never consuming. If the receiver does consume, the transmitter can always hold them to a lower payoff by transmitting misinformation, assuming that the cost of consuming misinformation is positive, $\pi_f<0$. The minimax value for the transmitter is also 0, since we assume the payoff to the transmitter of misinformation being consumed is $b_f\geq0$ and also the payoff from true news being consumed is $b_t\geq 0$, whereas the payoff when no news is consumed is 0, and so the transmitter's minimax profile is uniquely determined by a receiver strategy that never consumes (Table S1). Thus any feasible payoff profile that provides a positive payoff to both players is individually rational.

\begin{figure}[th!] \centering \includegraphics[scale=0.2]{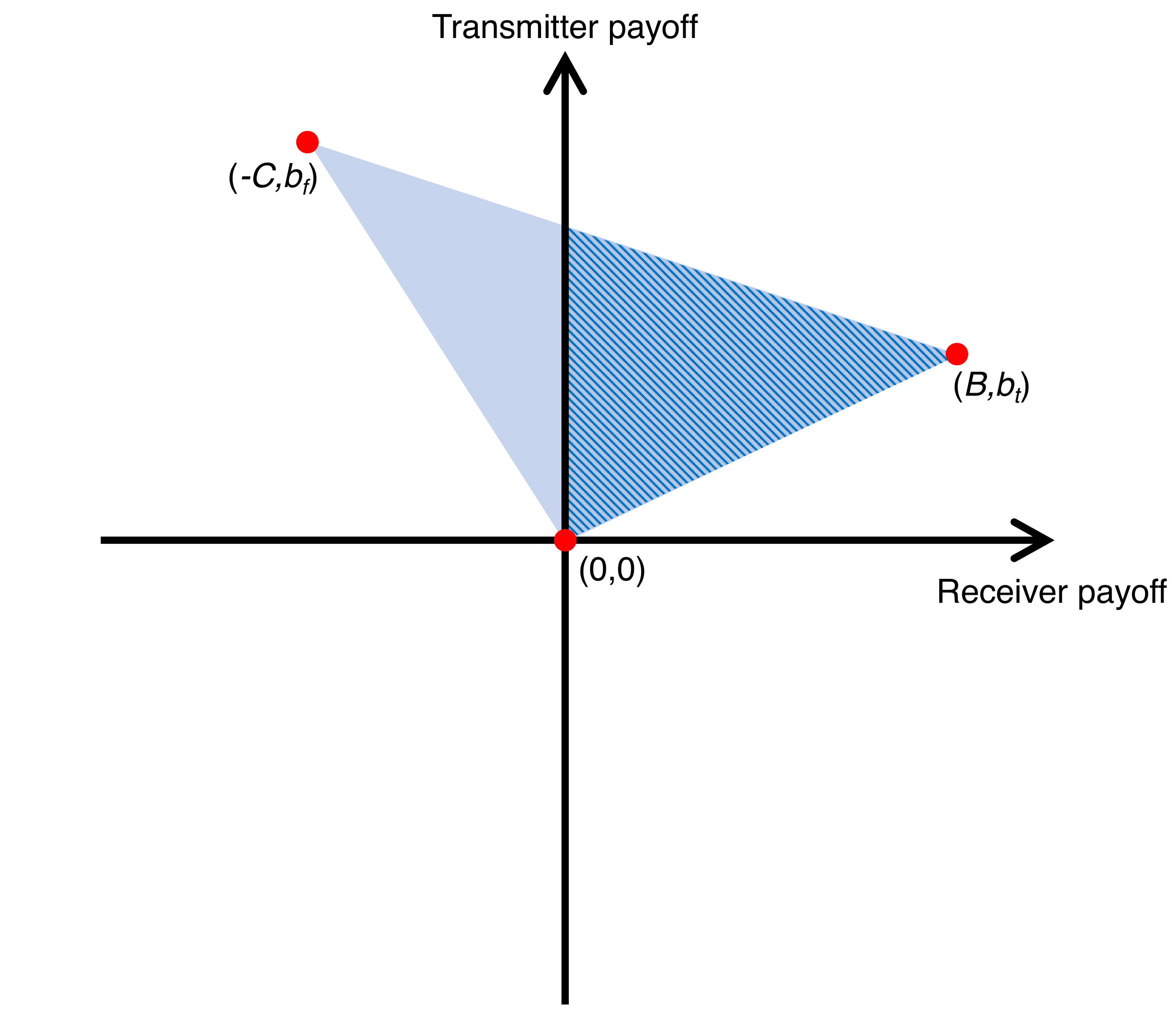}
\caption*{\small Figure S1: \textbf{Feasible, individually rational payoffs for the misinformation game --} The set of feasible payoffs (blue region) and the subset of individually rational payoffs for both players (striped region) for a misinformation game with  $b_f>b_t$ meaning that the transmitter prefers to generate engagement with misinformation. Individually rational payoffs for both players only arise when both players receive a positive payoff. An equilibrium in which the receiver always consumes and the transmitter always shares misinformation, $(-C,b_f)$ is feasible but not individually rational.}
\end{figure}

Any individually rational, feasible payoff is a Nash equilibrium of the game (Figure S1), and, since the game is assumed infinite, can always be enforced by players employing a grim trigger strategy. That is, if the receiver deviates from the prescribed course of play, the transmitter switches to only sharing misinformation. Any benefit gained by the receiver from deviating is lost in the long run, since the best they can do is to cease consuming and receive zero payoff in every subsequent round. Conversely, if the transmitter deviates from the equilibrium, the receiver can switch to never consuming news, with the same result that any benefit to the transmitter gained by deviating eventually lost. As discussed above, the set of individually rational, feasible payoffs (Figure S1) in the misinformation game corresponds to cases in which both transmitter and receiver have positive payoffs. Although a misinformation site receives  payoffs  $b_t=0$ and $b_f>0$, this nonetheless means that they must share a mixture of both true and misinformation, since the payoff that results from only sharing misinformation produces a negative payoff to the receiver.

To explore this point further, consider a misinformation site who simply transmits true stories with fixed probability $r$ each round. The payoff to a receiver who consumes news from the receiver at equilibrium probability $v_c$ is simply $w_t=v_c(Br-C(1-r))$, which is positive provided  $r>\frac{C}{B+C}$. If this condition is met, the receiver's payoff increases with overall engagement, and so an equilibrium at which the receiver always consumes and the transmitter shares true stories with probability $r>\frac{C}{B+C}$ can be enforced as a Nash equilibrium of the misinformation game (e.g. by employing a grim trigger, as discussed above). This illustrates the need, in a general sense, for a misinformation site to ``mash up'' true and fake stories to produce engagement. Finally, note that if $b_t>0$, $r=1$ and $v_c=1$ produces positive payoffs for both players, and can also be enforced as a Nash equilibrium -- i.e. mainstream news sites do not need to mash up true and fake stories when faced with rational receivers who value true news. 

Thus our model predicts that misinformation sites should always share a mixture of true and fake stories if their readers value accuracy, regardless of whether those readers are rational (as per this Nash equilibrium calculation), or lazy and irrational (as per our simulations above).

\subsection*{Modelling local optimization}

For the most part we do not assume that receivers in the misinformation game are perfectly rational, but rather update their behavior according to a local optimization process \cite{Traulsen:2006zr}. Under this process a receiver temporarily adopts an alternate strategy $j$ to the one they are currently employing, $i$, and compares the payoff they receive with the new strategy to that received under the old strategy. They then adopt the new strategy, and discard the old, with probability $\pi_{i\to j}$ determined by a Fermi function.

\begin{equation}\tag{17}
\pi_{i\to j}=\frac{1}{1+\exp[\sigma(w_i-w_j)]}
\end{equation}
\\
where $w_i$ is the payoff received under strategy $i$, and $\sigma$ determines the level of attention the player pays to their payoffs. If $\sigma<1$ little attention is paid to payoffs and the optimization process is very noisy, with receivers frequently adopting suboptimal engagement strategies. If $\sigma\gg1$ a high level of attention is paid to payoffs, and only strategies that improve payoffs are adopted.

When choosing the alternate strategy we assume that $j$ is small perturbation $\Delta_{\mu}\in[0,0.05]$ to the current strategy, such that each element of the receiver strategy  $\mathbf{p}=\{p_0,p_{ct},p_{cf},p_{nt},p_{nf}\}$ is increased or decrease by an independently drawn $\Delta_{\mu}$, with the constraint that each element must remain a viable probability (i.e. $p_{ij}\in[0,1]$).

\subsection*{Time series for linear and non-linear transmitter strategies}

In the main text we present the time series for co-optimization (Figure 2) for a transmitter strategy that employs non-linear feedback, defined by $r_{kt}=1/(1+\exp[\lambda(k/N-0.5)])$ and $r_{kf}=1/(1+\exp[\lambda(k/N-0.25)])$, where we set $\lambda=100$ and the population of receivers to be $N=100$. In this example the non-linear feedback can be expressed as a Taylor series as

$$
r_{kt}=1+\sum_{n=0}^\infty\frac{(-1)^{n+1}\lambda^n E_n(0)}{2n!}\left(\frac{k}{N}-\frac{1}{2}\right)^n
$$
similarly

$$
r_{kf}=1+\sum_{n=0}^\infty\frac{(-1)^{n+1}\lambda^n E_n(0)}{2n!}\left(\frac{k}{N}-\frac{1}{4}\right)^n
$$

where $E_n(0)$ is the $n$th Euler polynomial evaluated at 0. This allows us to compute the coefficients $\gamma_i$ and $\theta_i$ in Eq. 10. In general this must be done numerically. Here we have $\alpha_0=\beta_0=1/2$ and

$$
\gamma_1=\sum_{n=1}^\infty\frac{(-1)^{n+1}\lambda^n E_n(0)}{2n!}\left(\frac{1}{2}\right)^{n-1}=\frac{1-\exp[\lambda/2]}{1+\exp[\lambda/2]}
$$
and

$$
\theta_1=\sum_{n=1}^\infty\frac{(-1)^{n+1}\lambda^n E_n(0)}{2n!}\left(\frac{1}{4}\right)^{n-1}=2\times\frac{1-\exp[\lambda/4]}{1+\exp[\lambda/4]}
$$
and so for this strategy the first-order feedback terms satisfy $\gamma_1<0$ and $\theta_1<0$, i.e. they tend to generate a negative correlation between misinformation and engagement.
This is illustrated in the time-series in Figure 2 shows how a transmitter using this strategy  induces engagement with misinformation in a population of optimizing transmitters.

The same effect can be produced for simple linear strategies, as shown in Figure S2 below. Here a transmitter using a fixed linear strategy $\mathbf{r}=\{1.0,0.1,-0.9,-0.1\}$ interacts with an optimizing receiver. While the time-series produced do not show a visibly obvious correlation, when we plot the rate of misinformation production against the rate of engagement, we see an unambiguous correlation, as predicted by Eq. 3, and analogous to the effects shown for the nonlinear transmitter strategy in Figure 2.

\begin{figure}[th!] \centering \includegraphics[scale=0.33]{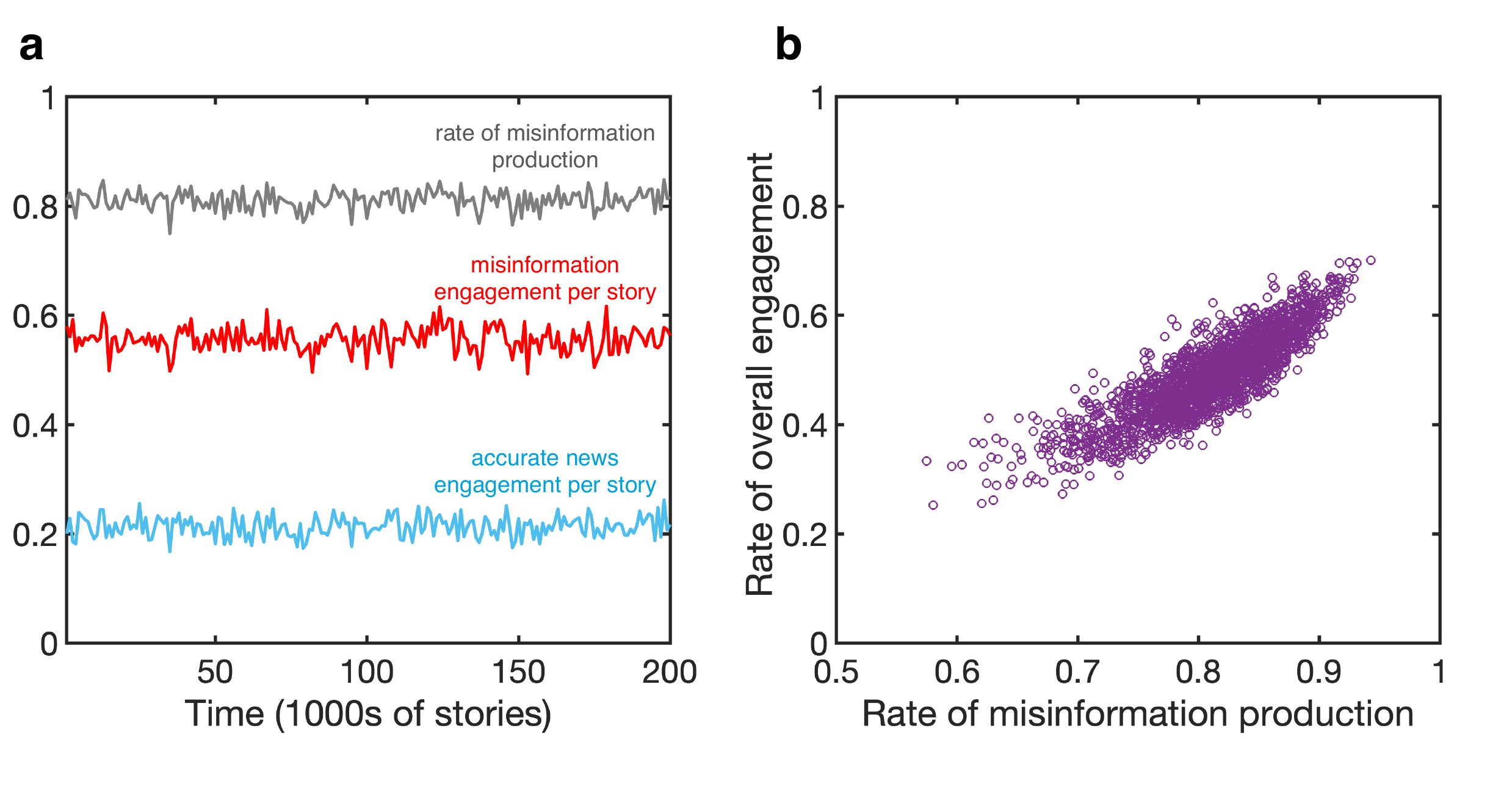}
\caption*{\small Figure S2: \textbf{Linear transmitter driven engagement with misinformation.} An illustrative example of transmitter-driven engagement with misinformation for a linear transmitter strategy. We selected a transmitter strategy which employs linear feedback, $\mathbf{r}=\{1.0,0.1,-0.9,-0.1\}$, which induces a positive correlation between rate of misinformation production and engagement (Eq. 3). a) Dynamics of engagement with accurate news (blue), misinformation (red) by the receiver, and overall production of misinformation (grey) by the transmitter, for a receiver employing a noisy optimization process with low attention ($a_0=a_1=0$ and $\sigma=1$). b) The result of these dynamics are a positive correlation between engagement rate per article and rate of misinformation production. In all cases receiver mutations were local (see SI section 1.5) and we set receiver payoffs $\pi_t=2$ and $\pi_f=-1$.}
\end{figure}

\subsection*{Modelling optimization through social learning}

In addition to local optimization, receivers whose strategies are updated through a process of cultural evolution via payoff-biased imitation \cite{Traulsen:2006zr}. Under this model, a pair of receivers, $i$ and $j$, are selected at random from the population of receivers. Receiver $i$ then adopts the strategy of receiver $j$ with probability

\begin{equation}\tag{18}
\phi_{i\to j}=\frac{1}{1+\exp[\sigma(w_i-w_j)]}
\end{equation}
\\
In addition, mutations are introduced to the receiver population at rate $\mu$, under which a player spontaneously adopts a news strategy drawn uniformly from the space of possible receiver strategies. Note that Eq. 17 and Eq. 18 are identical, but here $j$ is drawn from the strategies present in the population, while in Eq. 17, $j$ is chosen as a perturbation to the current strategy. As such, the process of copying other players' strategies tends to result in many different players using the same strategy, compared to the case where each player independently optimizes their own strategy. 

\subsection*{Feedback and receiver engagement}

Finally, we show that in order to produce the region F2 or T2 of main text Figure 3, feedback is required in the strategies used by transmitters. To see this, consider a transmitter strategy without feedback, such that $r_{mt}=r_{mf}=\alpha_0$, i.e. misinformation is produced at a constant rate $1-\alpha_0$. The probability of engagement with misinformation by an arbitrary (but inattentive, $a_0=0)$) receiver strategy $\mathbf{p}$ is then

\begin{eqnarray*}
v^{j+1}_{ct}=\alpha_0( v_{ct}^j p_{ct,H}+ v^j_{nt}p_{nt,H}+ v^j_{cf}p_{cf,H}+ v^j_{nf}p_{nf,H})\\
v^{j+1}_{cf}=(1-\alpha_0) (v_{ct}^j p_{ct,H}+v^j_{nt}p_{nt,H}+v^j_{cf}p_{cf,H}+v^j_{nf}p_{nf,H})
\end{eqnarray*}
\\
where $j$ is the round of play, and $p_{ab,H}$ is the probability of engaging given that the previous round had outcome $ab$ and $H$ is the full history of interaction by the receiver with the transmitter. Under this model 
\[
v_t=\alpha_0
\]
and so at equilibrium we have a difference in engagement probability by the receiver of

\[
\frac{v_{ct}}{v_t}-\frac{v_{ft}}{v_f}=\frac{1}{\alpha_0} v_{ct}-\frac{1}{1-\alpha_0}v_{cf}=0
\]
\\
and so no difference in engagement is produced when transmitter strategies do not use feedback, provided receiver strategies are inattentive (i.e. are not making use of prior knowledge of veracity when deciding to engage). If, in contrast, receivers are attentive and see true news we have

\begin{eqnarray*}
v^{j+1}_{ct}=(1-a_0)\alpha_0( v_{ct}^j p_{ct,H}+ v^j_{nt}p_{nt,H}+ v^j_{cf}p_{cf,H}+ v^j_{nf}p_{nf,H})+a_0\\
v^{j+1}_{cf}=(1-a_0)(1-\alpha_0) (v_{ct}^j p_{ct,H}+v^j_{nt}p_{nt,H}+v^j_{cf}p_{cf,H}+v^j_{nf}p_{nf,H})
\end{eqnarray*}
\\
and the engagement difference at equilibrium is

\[
\frac{v_{ct}}{v_t}-\frac{v_{ft}}{v_f}=\frac{1}{\alpha_0} v_{ct}-\frac{1}{1-\alpha_0}v_{cf}=a_0/\alpha_0
\]
\\
i.e. the engagement difference reflects the preferences of the receiver. Since region F2 and T2 represent scenarios in which engagement patterns are at odds with receiver preference, we see that absent feedback, region F2 and region T2 cannot be produced.

\clearpage

\section*{Comparison to previous models}

In this section we compare our model to three previously proposed theories of misinformation engagement and/or production \cite{10.1257/jep.31.2.211,Vosoughi1146,PennyNature}. We ask whether the theories presented in these papers can produce region F2 and T2 of main text Figure 3, and thereby explain  our empirical results (main text Figure 4). That is, we ask whether our proposed theory is uniquely predictive of the observed data. We find that, indeed, the other theories cannot reproduce the observed patterns since they use transmitter strategies that do not employ feedback (see Section 1.7 above).

\subsection*{Vosoughi et. al. 2018}

Vosoughi et. al. \cite{Vosoughi1146} present empirical results showing that, among news stories that were questionable enough to get fact-checked -- by snopes.com, politifact.com, factcheck.org, truthor-fiction.com, hoax-slayer.com, or urbanlegends.about.com -- false news spread further on social networks and persisted longer. They also find that false news was more novel, in an information-theoretic sense, than true news. The authors are careful to note that ``Although we cannot claim that novelty causes retweets or that novelty is the only reason why false news is retweeted more often, we do find that false news is more novel and that novel information is more likely to be retweeted.''

This suggests that readers derive benefit from engaging with misinformation due to its novelty. However such an explanation, absent feedback between receiver and transmitter sharing strategy, would predict greater engagement with false stories among both fake and mainstream sites 
, which is not consistent with our empirical findings (main text Figure 4).

\subsection*{Pennycook et. al 2021}

Pennycook et. al \cite{PennyNature} investigate the role of inattention in shaping consumer engagement with misinformation. The authors present a model in which receiver utility is derived from a consumer's  preferences and the level of attention they pay to a topic. Fitting the model parameters to experimental data they find
the best-fit preference parameters indicate that participants value accuracy as much as or more than partisanship. Thus, they would be unlikely to share false but politically concordant content if they were attending to accuracy and partisanship''
suggesting a model in which consumers prefer true news to false news, 
and occasionally share  misinformation accidentally, due to inattention. This account would therefore predict
greater engagement with true stories than false stories, regardless of whether to producer is a misinformation or mainstream site
, which is inconsistent with our empirical findings (main text Figure 4).

\subsection*{Allcott and Gentzkow 2017}
Allcott and Gentzkow \cite{10.1257/jep.31.2.211} develop a model of supply and demand of misinformation. Under this model  ``misinformation arises in equilibrium because it is cheaper to provide than precise signals, because consumers cannot costlessly infer accuracy, and because consumers may enjoy partisan news.''

This model considers consumer decisions at the level of the news outlet (e.g. ``shall I subscribe to this newspaper or not?'') rather than at the level of the individual story, and thus does not make direct predictions about our key empirical outcome, the relative probability of engagement for true versus false news within a given outlet. However, the model's implications can be extrapolated to our setting. 
On the demand side, this means that consumers/receivers either (i) engage with misinformation accidentally due to inattention or (ii) derive benefit from engaging with misinformation due to e.g. partisan preference. Inattention alone, absent feedback between transmitter and receiver sharing strategy, leads in the limit to equal levels of engagement with true and false news. Similar to the case of a consumer preference for novelty discussed above, a partisan preference for false news, absent feedback between receiver and transmitter sharing strategy, would predict greater engagement with false stories among both fake and mainstream sites, which is not consistent with our empirical findings (main text Figure 4).

On the supply side, this theory posits that transmitters broadcast false news because it is cheaper to produce than true news (i.e.~the net payoff for transmitters from receivers engaging with false news is greater than it is for true news). By this logic, it is payoff maximizing for transmitters to produce false news so long as the engagement advantage of true news over false news is sufficiently small (or non-existent) -- i.e.~due to receivers being inattentive and/or preference for falsehoods. This explanation for the production of misinformation, while entirely plausible, makes no prediction for the different patterns of engagement among consumers with mainstream versus misinformation sources, which is our key empirical result (main text Figure 4).

\clearpage

\section*{Further analysis and extensions of the model}

In this section we perform robustness checks on our model, and show that our results hold as we alter our model specification. We show that i) our results are robust to different definitions of ``successful'' transmitters ii) our results are robust to different models of receiver learning, iii) our results are robust to different receiver group sizes but require a moderate to high degree of transmitter microtargetting, iv) our results require low levels of receiver attention, v) our results are robust to competition between transmitters and vi) our results are robust to changes to consumer demand for misinformation.

\subsection*{Identifying successful transmitter strategies}

In order to find successful mainstream and misinformation transmission strategies we generated $10^8$ receiver strategies as described in SI Section 1. We then defined successful misinformation transmitters as those which achieved i) in the 90th percentile for misinformation engagement, and ii) a higher probability of transmitting fake than true news, $v_f>0.5$. Similarly, we defined successful mainstream transmitters as those in the 90th percentile for true news engagement who transmit a higher probability of true news than fake, $v_t>0.5$. These definitions were used to select the strategies used to produce Figure S3 below and discussed in the main text. 

The most successful mainstream and misinformation dissemination strategies induce characteristic, and opposite, patterns of engagement among readers in our model (Figure~3). In particular, successful misinformation sites induce higher reader engagement with each false news story, as well as greater overall engagement of false news than true news (Table S2). This phenomenon arises even though we assume that receivers strictly prefer true news over false news, so there is no inherent appeal of false news stories under our model. 

To understand this phenomenon we inspected the strategies of successful misinformation sites under our model. We find that all successful misinformation sites indeed use responsive strategies that enforce a positive correlation between engagement and false news transmission. A significant proportion ($72\%$, see SI Section 1) use a type of extortion strategy \cite{Press:2012fk} 
that enforces $v_f\geq v_{fc}+v_{tc}$, i.e.~successful misinformation site strategies tend to increase their false news output rapidly in response to increased engagement by a given user. However, if engagement drops, they tend to increase their output of true stories (to draw the user back in). As a result, they ``mash up'' true and fake stories as engagement fluctuates over time.

\begin{figure}[th!] \centering \includegraphics[scale=0.25]{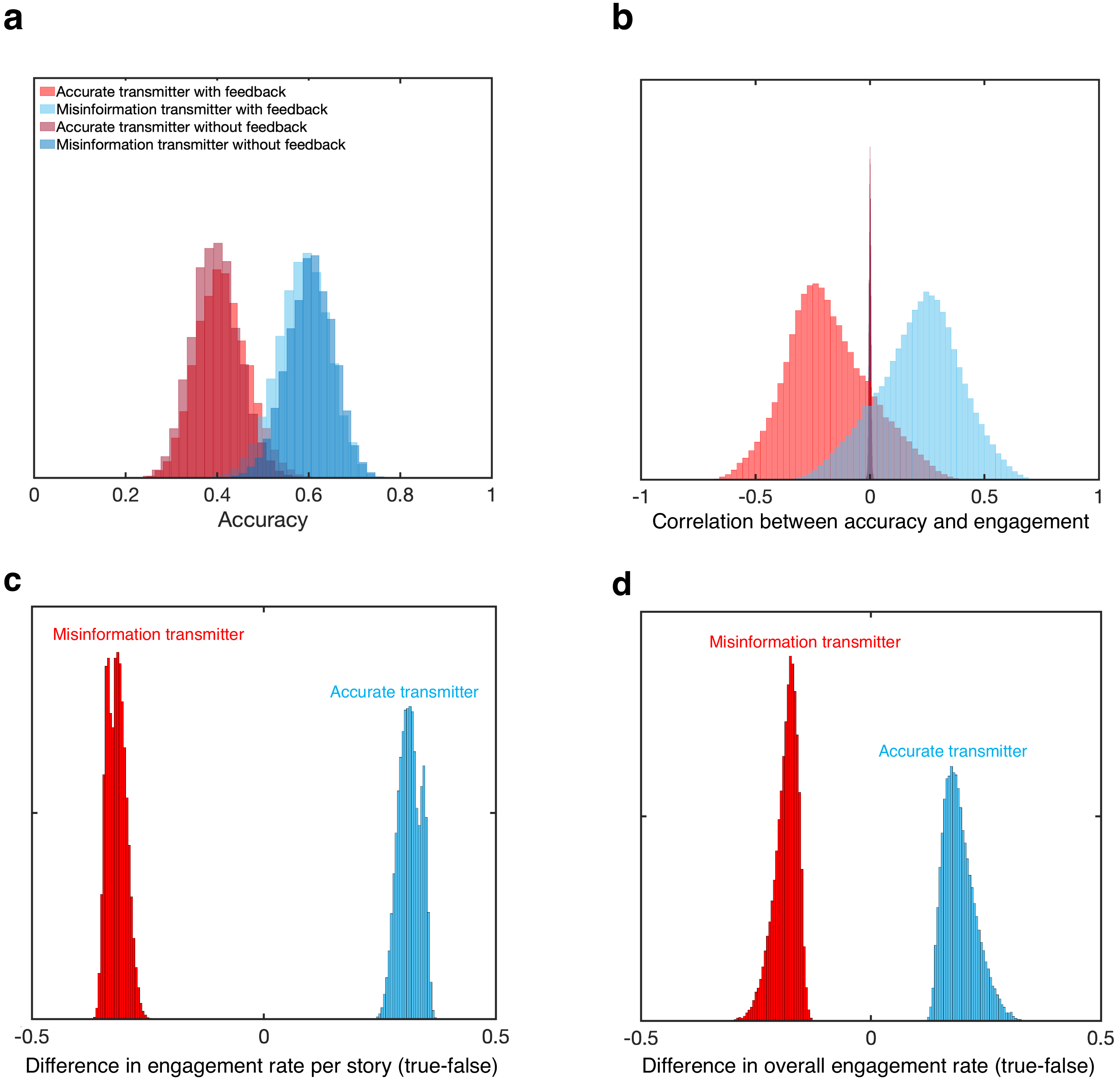}
\caption*{\small Figure S3: \textbf{Feedback dramatically changes predicted patterns of news engagement --} 
a) We identified strategies that are successful at promoting engagement with true news stories (accurate sites) or with misinformation stories (misinformation sites). To do this we randomly drew $10^8$ transmitter strategies (Eq.~2) and allowed a single receiver, incentivized to engage with true news ($\pi_t=2$ and $\pi_f=-1$), to optimize their engagement strategy (see SI Section 1) over the course of $10^4$ interactions. The receiver was assumed to be inattentive ($a_0=a_1=0$ and $\sigma=1$). We identified the transmitter strategies that  successfully promote engagement with true news stories (mainstream sites, blue), i.e those that produce a true news engagement probability within the 90th percentile of the $10^8$ transmitter strategies considered, as well as $v_t>0.5$. Similarly, we identified the transmitter strategies that successfully promote engagement with misinformation (misinformation sites, red), i.e those that produce a misinformation engagement probability within the 90th percentile, as well as $v_f>0.5$.  We plot the distribution of accuracy among the most successful transmitters, for strategies that make use of feedback ($\gamma\neq 0$ and $\theta\neq 0$, light colors) and strategies that do not make use of feedback ($\gamma=\theta=0$). We see in both cases that accurate transmitters tend to share accurate stories and misinformation transmitters tend to share fake stories. b) We calculated the regression coefficient between engagement and accuracy for each transmitter strategy, for a group of $10^3$ receivers engaging with 20 stories from each source. In the absence of feedback (dark colors) there is no correlation between accuracy and engagement, whereas in the presence of feedback (light colors) accurate transmitters generate a positive correlation while misinformation transmitters tend to generate a negative correlation. In all cases receiver strategic exploration was local (see SI section 1.5) and we assumed transmitter error rates of $0.3$ (see SI Section 1). Results for other parameter choices are shown in the SI Section 3. c)  We  plot the difference between engagement with true and fake stories, $v_{tc}/v_t-v_{fc}/v_f$, for all mainstream and misinformation sites. Mainstream site strategies induce engagement with accurate stories while misinformation site strategies induce the opposite effect. d) We also report the difference in overall engagement with true stories, $v_{tc}$, and overall engagement with fake stories $v_{fc}$. In all cases receiver mutations were local (see SI section 1.5) and we set receiver payoffs $\pi_t=2$ and $\pi_f=-1$.}
\end{figure}

The behavior of successful mainstream sites, that seek to generate engagement with true stories, shows the opposite pattern from of successful misinformation sites. All successful mainstream strategies enforce a \textit{negative} correlation between engagement and false news transmission.

\begin{figure}[th!] \centering \includegraphics[scale=0.2]{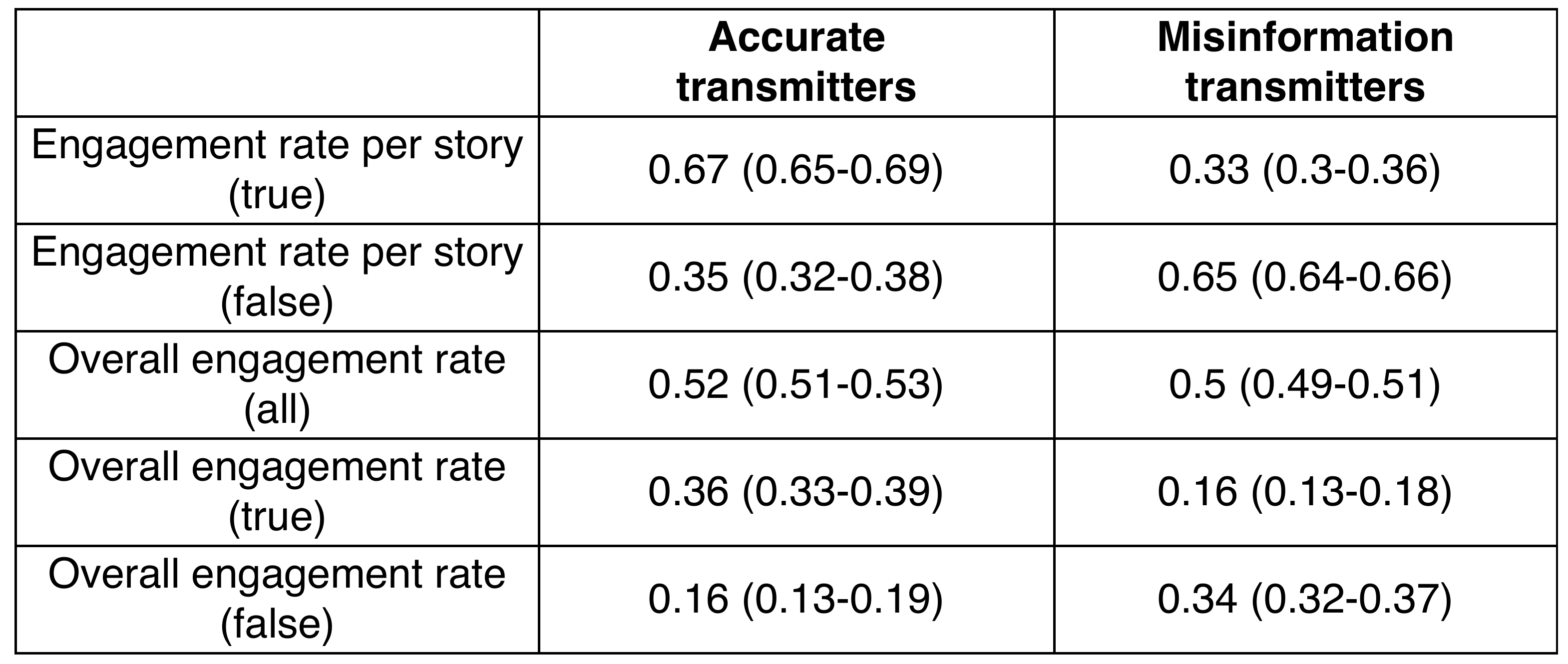}
\caption*{\small Table S2: \textbf{News engagement patterns.}  Summary of the engagement rates per story and overall engagement rates for mainstream and misinformation site transmission strategies shown in Figure S3.}
\end{figure}

A significant proportion ($56\%$, see SI Section 1) use a strategy 
that enforces $1-v_f\geq v_{fc}+v_{tc}$, i.e.~successful mainstream site strategies tend to decrease their output of false stories rapidly in response to increased engagement, but may share more false stories when engagement is low (in a misguided attempt to draw readers back in).

The full, unfiltered distribution of engagement per story and overall engagement for the generated strategies is shown in Figure S4a-b. The distributions have a weak bias towards true news consumption ($p<0.001$) reflecting the noisy optimization process of the receivers, who prefer to consume true and avoid misinformation.

\begin{figure}[th!] \centering \includegraphics[scale=0.2]{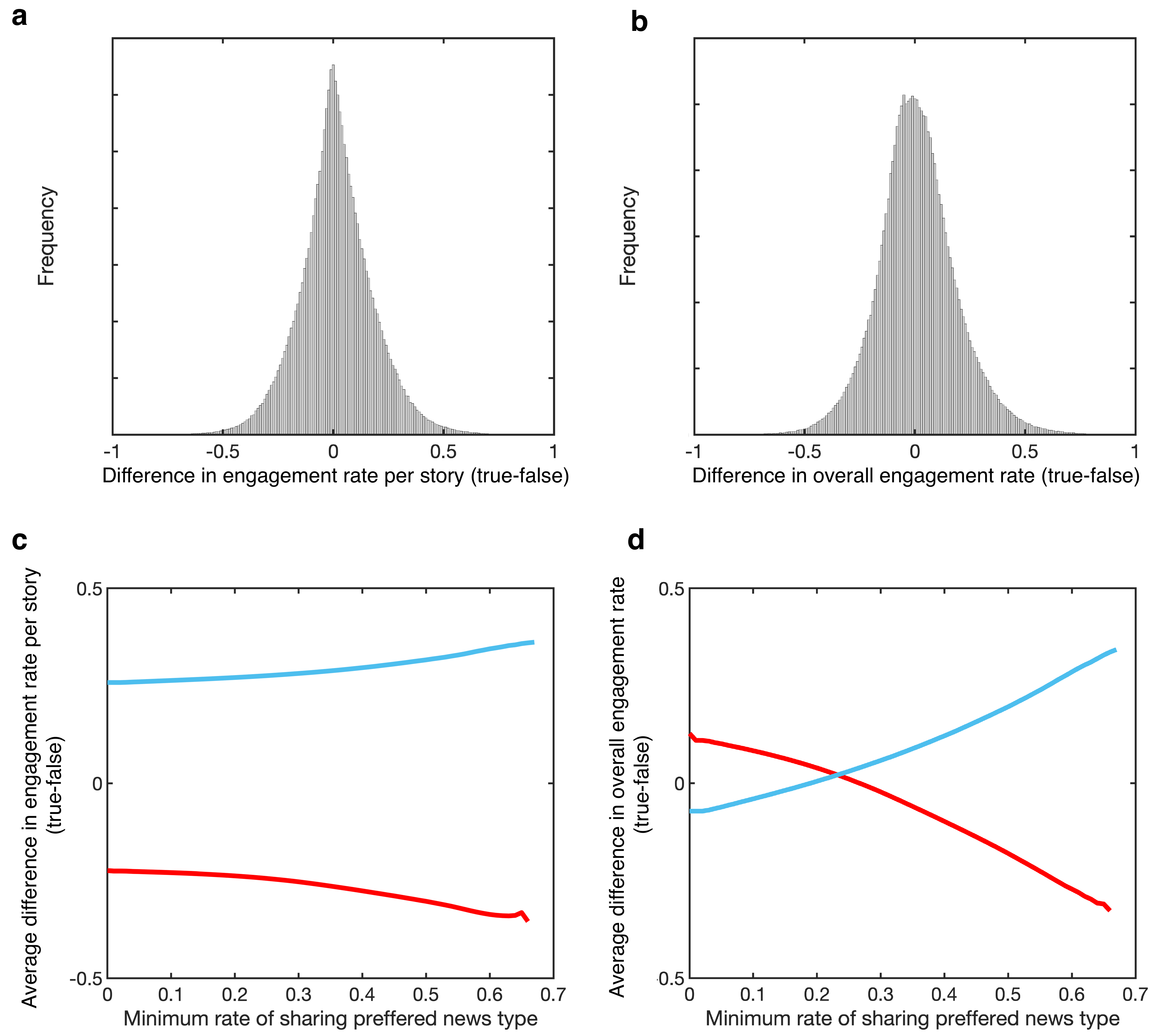}
\caption*{\small Figure S4: \textbf{Full distribution of engagement per story and overall engagement --} Shown are the distributions of the difference in engagement per story (a) and overall engagement (b) with true vs fake stories for $10^8$ transmitter strategies. We define successful misinformation strategies as those with $v_f>0.5$ and misinformation engagement, $v_{fc}/v_f\geq0.704$ which is the 90th percentile among all observed strategies.  Similarly we define successful mainstream strategies as those with $v_t>0.5$ and misinformation engagement, $v_{tc}/v_t\geq0.758$ which is the 90th percentile among all observed strategies. c) We calculated the average difference in engagement as a function of the threshold probability of sharing the transmitter's proffered news type used when to determine which transmitter strategies are successful. We see that both mainstream (blue) and misinformation sites (red) produce an engagement difference that is insensitive to this choice of threshold. d) However the overall engagement probability is highly sensitive to the threshold, so that when the threshold is low the preferred news type is consumed less frequently than the non-preferred type. This arises because many successful sites produce high levels of engagement with their preferred news type while sharing it rarely.}
\end{figure}

We also explored the effect of changing the definition of ``successful'' transmitters of true and misinformation. We look at sites in the top 10\% of engagement, who share their preferred news type greater than 50\% of the time. In Figure S4c-d we vary this threshold and look at the average difference between engagement per story and overall engagement. We see that engagement per story difference is insensitive to the threshold choice, whereas overall engagement difference reverses direction when the threshold is low, indicating that there are many news sites who are good at producing engagement with true or misinformation while also sharing such stories rarely.

\clearpage

\subsection*{Co-optimizing transmitters and receivers}

We explored the behaviors of transmitters and receivers when both players co-optimize their strategies (main text Figure 3 and Figure S5). 
In order to identify regions F1-F3 and T1-T3 (Figure 3) we allowed ensembles of transmitter and receiver strategies to co-optimize over $10^4$ time steps (see SI Section 1). Next we then expanded this analysis to consider transmitters who seek only to maximize engagement among receivers ($b_f=b_t$), but make different assumptions about the type of news those receivers prefer. Transmitters who assume receivers prefer misinformation (e.g.~because it is more novel) employ strategies that seek to reinforce increased engagement by increasing the amount of false stories they share in response (see SI Section 1). Similarly transmitters who assume receivers prefer true news employ strategies that seek to reinforce increased engagement by increasing the amount of true news they share in response. Transmitters who make no assumption about receivers preferences (and hence try out all possible strategies without bias) and transmitters who assume receivers prefer true news, 
both produce greater engagement with true than with false stories (Figure S5a). 
In contrast, transmitters who assume receivers prefer misinformation, produce greater engagement with false than with true stories.
 
 \begin{figure}[th!] \centering \includegraphics[scale=0.25]{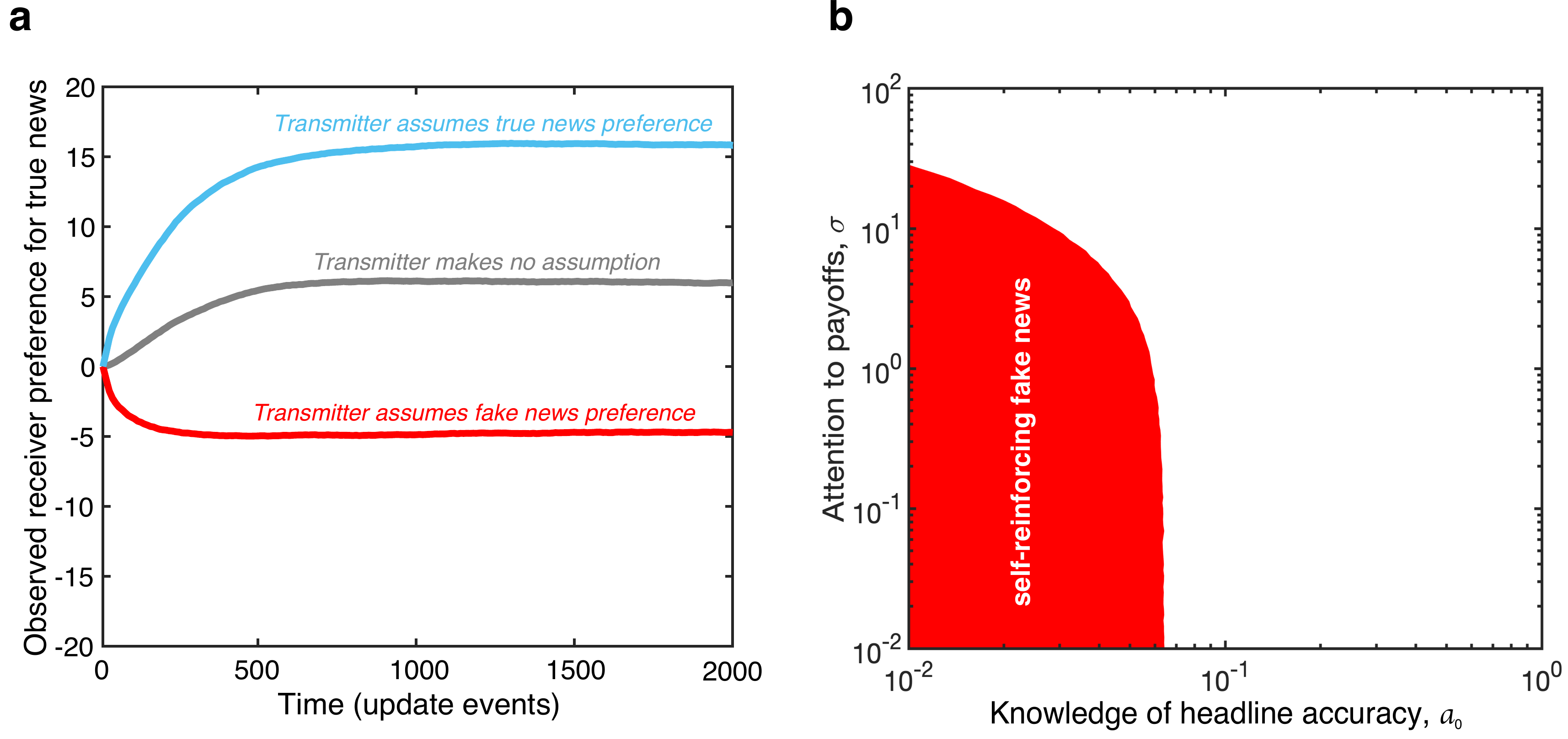}
\caption*{\small Figure S5: \textbf{Self-reinforcing misinformation} We allowed both receivers and transmitters to optimize under the process described in main text Figure 3. Here we assume that transmitters seek only engagement with news, regardless of veracity ($b_f=b_t=1$). But transmitters draw strategies that reflect their assumptions about the receivers' preferences (see SI Section 1). a) Time series of receiver engagement as receivers and transmitters co-optimize. We initialized with a transmitter that always shares true news and a receiver that never engages. We calculated the percentage difference in probability of engagement with a true versus a fake story, averaged across $10^4$ replicate simulations. When transmitters make no assumption (gray), or assume receivers prefer true news (blue), true news receives greater engagement, which reflects the underlying receiver preferences. When transmitters incorrectly assume receivers prefer misinformation (red), however, false stories receive greater engagement -- that is, the assumption that receivers prefer misinformation is self-reinforcing. b) We calculated the average engagement difference 
for different levels of receiver attention to payoffs, $\sigma$, and for different probabilities that receivers have prior knowledge about headline accuracy, $a_0$. It is only when attention to accuracy and attention to payoffs are low that misinformation is self-reinforcing (red region).
In all cases receiver mutations were local (see SI Section 1.5) and receiver payoffs are set to $\pi_t=2$ and $\pi_f=-1$. Transmitter mutations were global. Receiver attention to payoffs was set at $\sigma=1$, attention to accuracy at $a_0=0$ in panel a) with memory $a_1=0$. Transmitter attention to payoffs was set at $\sigma=100$.}
\end{figure}

As these results show, transmitters tend to elicit behavior among receivers that reinforce their assumptions about receiver preferences. We explored more broadly (Figure S5b) the conditions under which an assumed preference for misinformation is self-reinforcing in this way (i.e.~leads receivers to engage more with false stories than with true). We find that misinformation is self-reinforcing when attention to accuracy is low (i.e.~when receivers either cannot assess or do not pay attention to the accuracy of a headline) and when attention to payoffs is low. This suggests that increasing reader attention to the news they are consuming can reduce not only their own consumption of misinformation but also the production of misinformation by transmitters who make faulty assumptions about receiver preferences.

\clearpage

\subsection*{Comparison of most successful transmitters with co-evolved transmitters}

In our characterization of successful misinformation transmitter strategies above we selected strategies in the 90th percentile of engagement with misinformation that also produce misinformation at a probability $v_f>0.5$, with an analogous selection process for successful transmitters of accurate information (Figure S3). A natural alternative method for selecting transmitter strategies is to look at those that emerge via an optimization process, such as that used to produce Figure 3. Figure S6 below shows the patterns of engagement that arise for misinformation and accurate transmitter selected via this process, as compared to the top 10\% of transmitters. We see that such transmitters produce the same clear pattern of engagement with receivers, i.e. misinformation transmitters generate engagement with misinformation, while accurate transmitters generate engagement with accurate stories.

\begin{figure}[th!] \centering \includegraphics[scale=0.25]{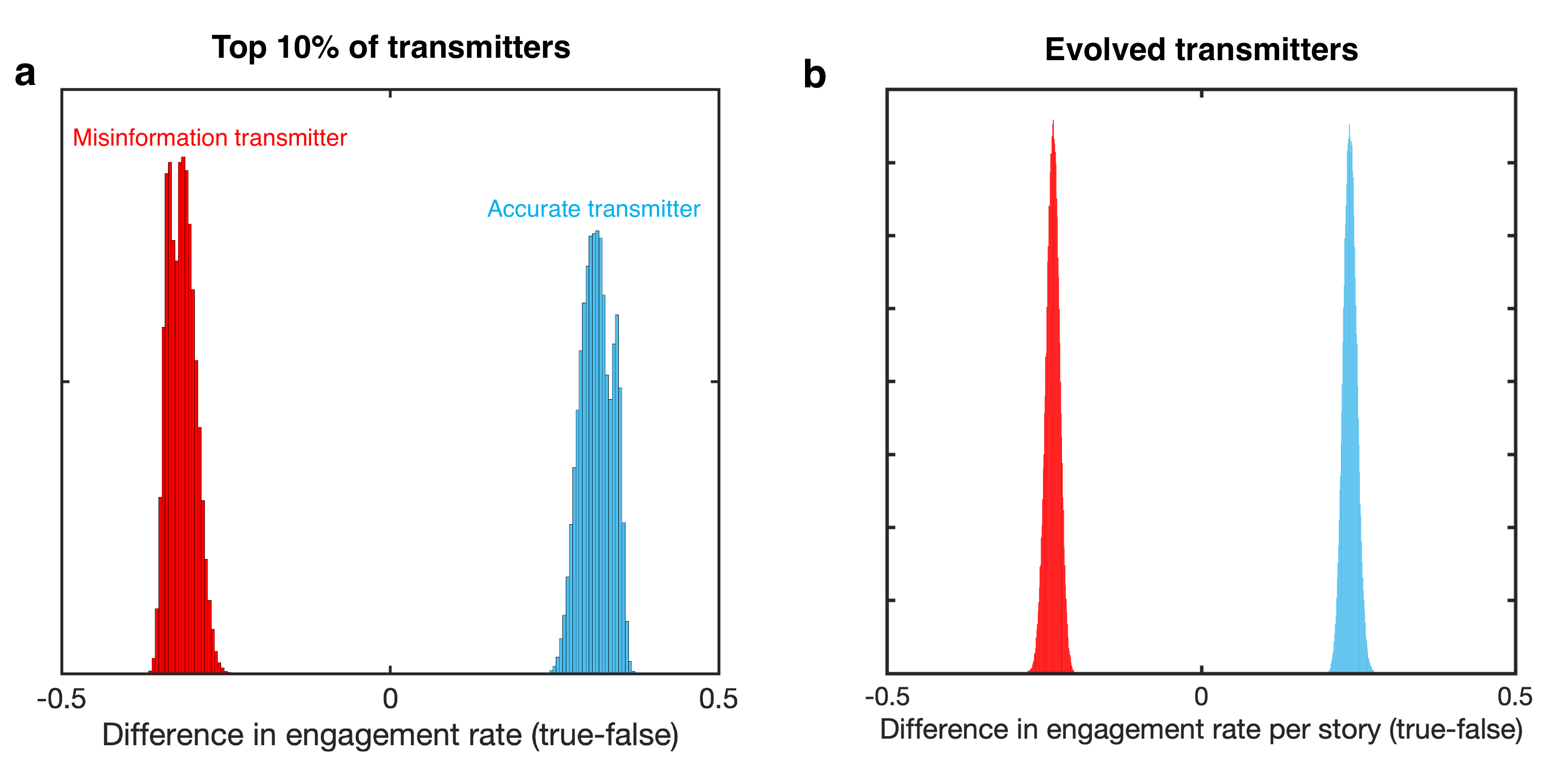}
\caption*{\small Figure S6: \textbf{Comparison of evolved transmitters to the top 10\% --} 
a) We identified strategies that are successful at promoting engagement with true news stories (accurate sites) or with misinformation stories (misinformation sites) in the same way as in Figure S3. We  plot the difference between engagement with true and fake stories, $v_{tc}/v_t-v_{fc}/v_f$, for all mainstream and misinformation sites. Mainstream site strategies induce engagement with accurate stories while misinformation site strategies induce the opposite effect. b) We plot the equivalent distribution for strategies evolved via a co-optimization process. We see a qualitatively similar pattern of engagement in both cases. In all cases receiver mutations were local (see SI section 1.5) and we set receiver payoffs $\pi_t=2$ and $\pi_f=-1$ and transmitter payoffs $b_t=0$ and $b_f=1$. For the optimization process, transmitter mutations were global. Receiver attention to payoffs was set at $\sigma=1$, attention to accuracy at $a_0=0$ with memory $a_1=0$. Transmitter attention to payoffs was set at $\sigma=100$.}
\end{figure}

\clearpage

\subsection*{Impact of absolute transmitter payoff}

In our analysis and simulations we have typically assumed that transmitters who prefer misinformation receive payoffs $b_f=1$ and $b_t=0$, while transmitters that prefer accurate stories receive payoffs $b_f=0$ and $b_t=1$. This is justified since the Nash equilibria for the game are unchanged under the addition of a constant to a player's payoffs. However such a translation can impact the dynamics of an optimization process. To check the robustness of our results we reproduced Figure S6 and Figure 3 with payoffs $b_f=2$ and $b_t=1$ for misinformation transmitters, and payoffs $b_f=1$ and $b_t=2$ for accurate transmitters (Figure S7 and S8). As the figures show, both the distribution of engagement (Figure S7) and the size of the region FII in Figure 3 (Figure S8) are unchanged under these alternate payoffs.

\begin{figure}[th!] \centering \includegraphics[scale=0.25]{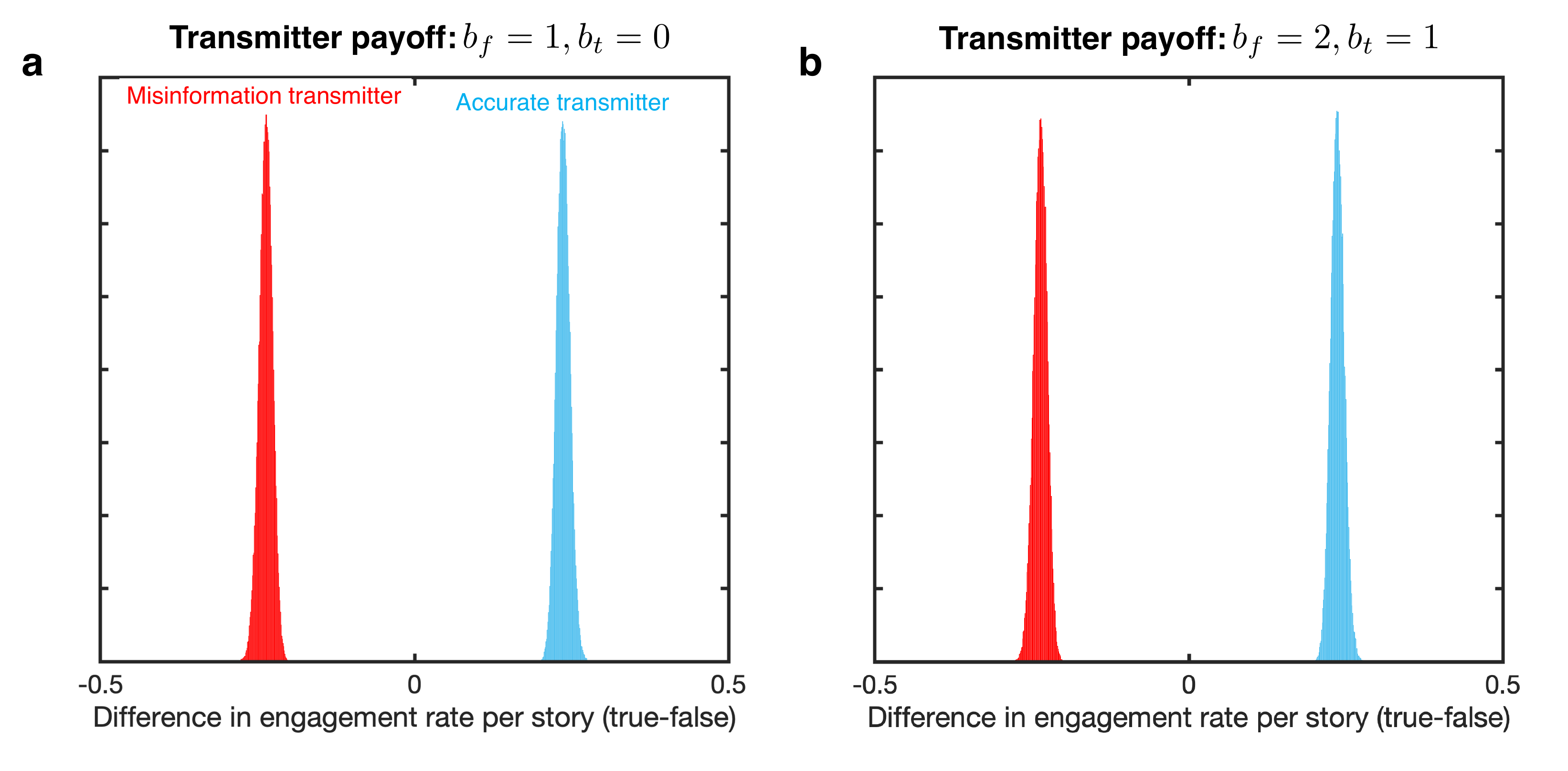}
\caption*{\small Figure S7: \textbf{Effect of transmitter payoffs on distribution of engagement under optimization --} 
a) We plot the distribution of engagement for strategies evolved via a co-optimization process using the same method as Figure S6, with transmitter payoffs $b_t=0$ and $b_f=1$. b) We compare this to the distribution of engagement for strategies evolved via a co-optimization process using the same method as Figure S6, with transmitter payoffs $b_t=1$ and $b_f=2$.
We see no difference between the cases. In all cases receiver mutations were local (see SI section 1.5) and we set receiver payoffs $\pi_t=2$ and $\pi_f=-1$. For the optimization process, transmitter mutations were global. Receiver attention to payoffs was set at $\sigma=1$, attention to accuracy at $a_0=0$ with memory $a_1=0$. Transmitter attention to payoffs was set at $\sigma=100$.} 
\end{figure}

\begin{figure}[th!] \centering \includegraphics[scale=0.25]{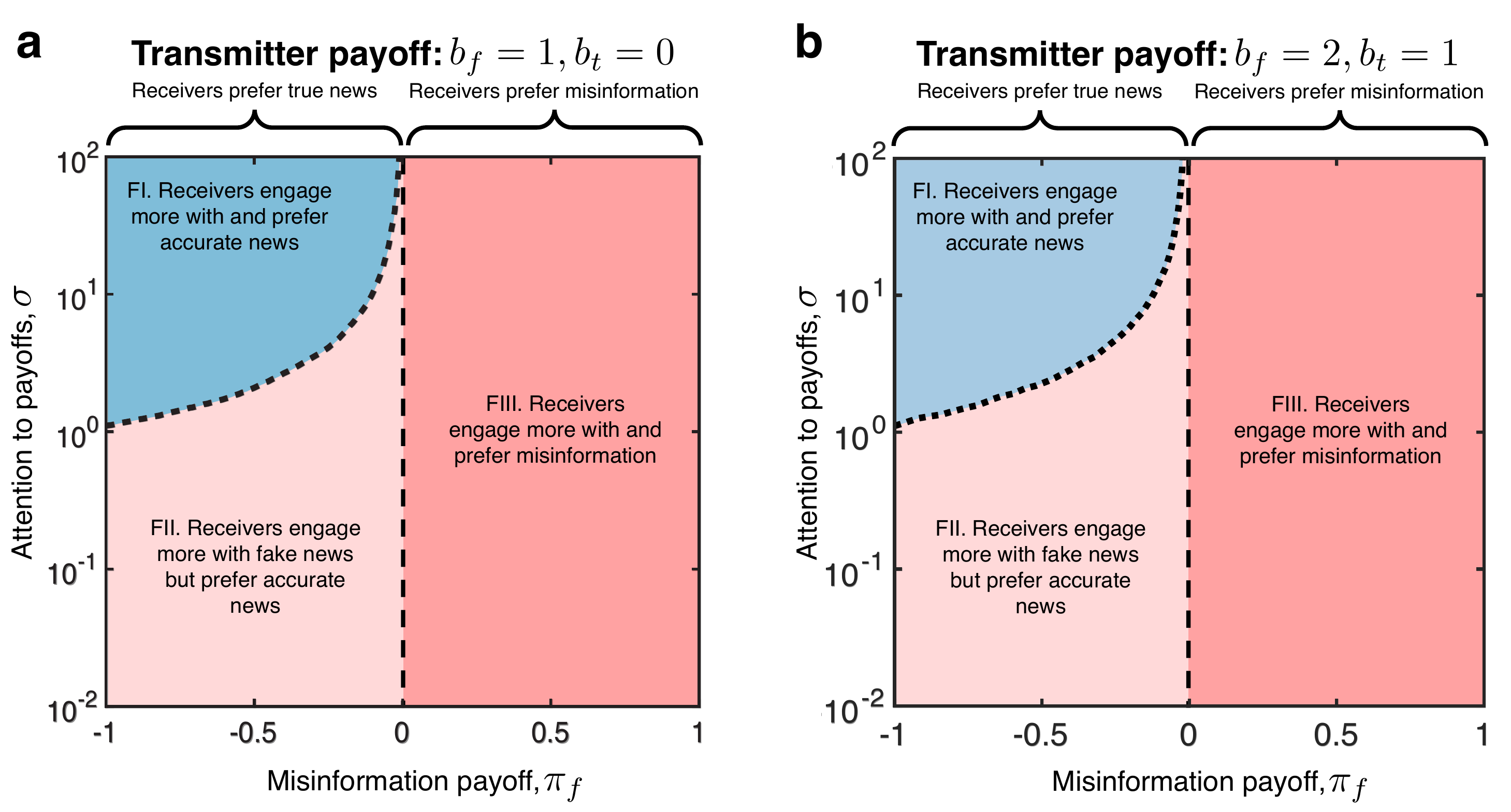}
\caption*{\small Figure S8: \textbf{Effect of transmitter payoffs, receiver preferences and attentiveness on engagement --} 
a) We calculated the engagement patterns among receivers optimizing under different preferences in the same way as for main text Figure 3 with transmitter payoffs set at $b_t=0$ and $b_f=1$. b) We compared this to a transmitter with payoffs $b_t=1$ and $b_f=2$. We see no difference between the patterns of engagement among receivers that emerges, as captured by the size of regions FI, FII and FIII. In all cases receiver strategic exploration was local, with $a_0=a_1=0$, with transmitter attention set at $\sigma=100$. Optimization occurred over $10^4$ time-steps using ensembles of $10^3$ replicates for each value of $\{\pi_f,\sigma\}$.}
\end{figure}

\clearpage

\subsection*{The effect of receiver learning model choice}

We studied the effect of different receiver learning model choice. In particular we explored social learning via a model of cultural evolution, as described in Section 1.6 of this Supplement, on the level of engagement per story and overall engagement in a population of receivers. Figure S9 shows the impact of group size on engagement per story and overall engagement with fake and true news with a fixed single misinformation transmitter as in Figure S3. Assuming that the rate of strategy mutations is $\mu=1/N$ where $N$ is the size of the group of receivers, we see that increasing group size has no effect on overall engagement and only a modest effect on engagement per story, tending to increase true news and reduce misinformation engagement per story, without reversing the pattern of higher engagement per story with fake stories than with true.

\begin{figure}[th!] \centering \includegraphics[scale=0.5]{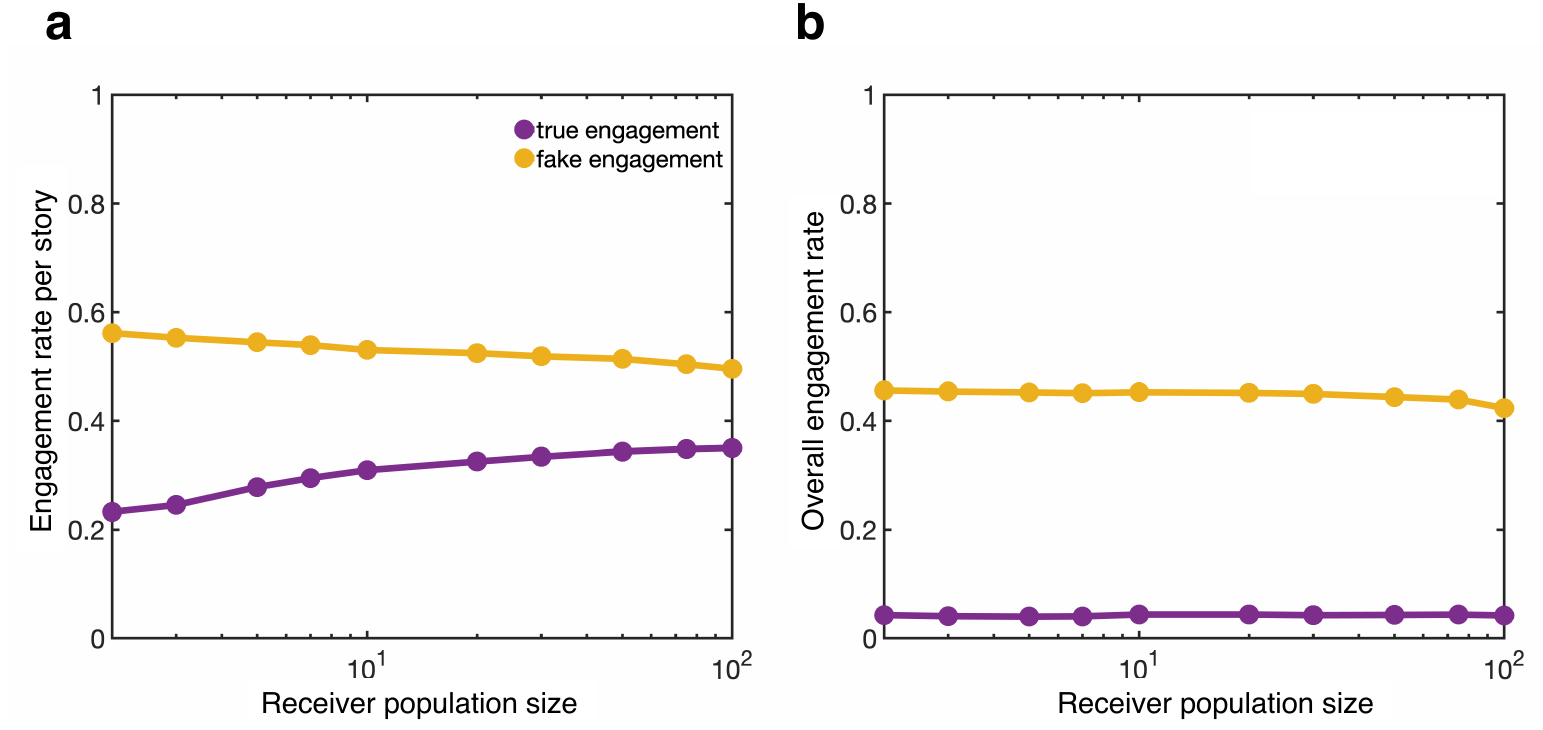}
\caption*{\small Figure S9: \textbf{Cultural evolution of receiver strategy --} a) A group of receivers evolving in response to a single transmitter, under the same process as described is Section 1.6 of the Supplement. We vary the size of the receiver group (x-axis) and calculate the equilibrium level of engagement with fake (orange) and true (purple) stories. b) We also calculate the equilibrium level of overall engagement. In all cases receiver mutations were local (see SI Section 1) and we set receiver payoffs $\pi_t=2$ and $\pi_f=-1$. Here misinformation sites use a fixed linear responsive strategy $\mathbf{r}=\{1.0,0.1,-0.9,-0.1\}$ see SI Section 1}
\end{figure}

\clearpage

\subsection*{The effect of receiver group size and transmitter microtargetting}

We studied the effect of relaxing the assumption, employed in the main text (Figure 2-3) and in preceding sections of the SI, that a transmitter can directly target each receiver, in response to his or her engagement with the previous news article. We show that the effects observed -- in particular the apparent attractiveness of misinformation -- hold under a wider range of conditions in which a transmitter targets groups of individuals. 

First we studied the effect of varying the size of the group of receivers engaging with a single transmitter, when all receivers use the same strategy to engage with the transmitter, with optimization taking place at the level of the group rather than at the level of the individual, as in Figure S3. Figure S10 shows that, in such a scenario, group size has no impact on either engagement per story or overall engagement even as group size varies across two orders of magnitude.

\begin{figure}[th!] \centering \includegraphics[scale=0.5]{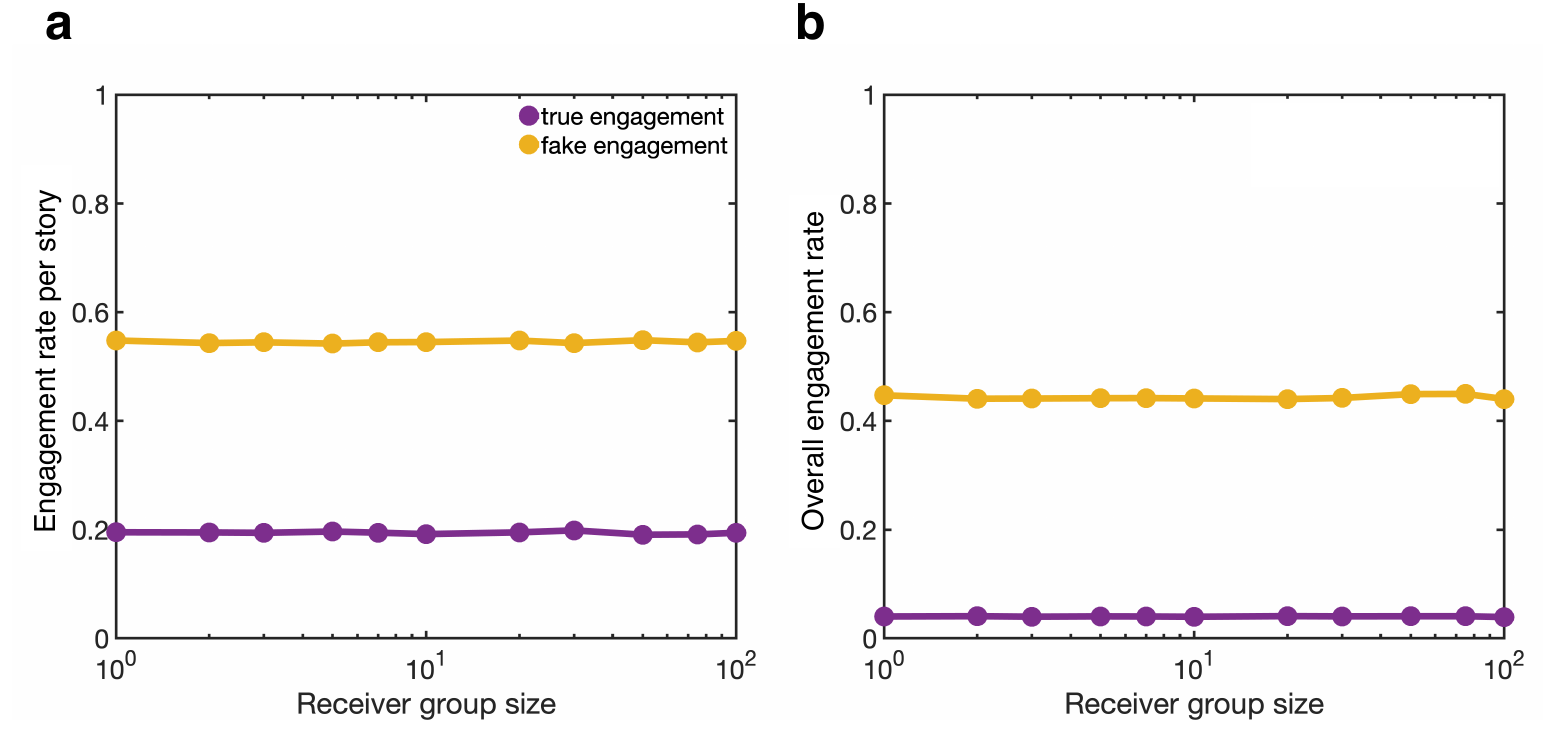}
\caption*{\small Figure S10: \textbf{Homogenous receiver groups --} a) A group of receivers optimizing in response to a single transmitter, under the same process as described for Figure S3, with the extension that that optimization occurs at the level of the group rather than the individual. We vary the size of the receiver group (x-axis) and calculate the equilibrium level of engagement with fake (orange) and true (purple) stories. b) We also calculate the equilibrium level of overall engagement. In all cases receiver mutations were local (see SI Section 1) and we set receiver payoffs $\pi_t=2$ and $\pi_f=-1$. Here misinformation sites use a fixed linear responsive strategy $\mathbf{r}=\{1.0,0.1,-0.9,-0.1\}$ see SI Section 1}
\end{figure}

Next we explored the effect of microtargetting on receiver engagement with misinformation in the face of a responsive transmitter. The engagement dynamics between a single transmitter and single receiver corresponds to the case of direct micro-targeting by a transmitter, in the sense that the transmitter makes their decision about what type of news to share (e.g.~by promoting stories on social media) with a receiver based solely on the engagement (or not) of that specific receiver. In reality, this degree of precision is unrealistic, but as we have shown above (Figure S9) the same effects also hold when a transmitter's audience is a group of receivers who employ similar behavioral strategies.

In order to explore how micro-targeting impacts the engagement per story and overall engagement habits of receivers interacting with a responsive misinformation strategy, we define the strength of micro-targeting as $M=1/G$, where $G$ is the number of independent individuals or groups to whom the transmitter promotes the same news. Here a ``group'' refers to a newsfeed, a social media page, or distribution list to which the transmitter promotes content. A group may contain only a single individual, or be composed of many people who share the same interest. When a transmitter promotes the same content to many such groups ($M\ll1$), each group updates their engagement independently in response, and so the strength of micro-targeting is low: a change in the behavior of any one group has a small impact on the overall engagement with the transmitter (Main Text Eq.~2), so the transmitter is not very responsive to a given receiver. When the strength of micro-targeting is high (e.g. when there is only one transmitter and one receiver, $M=1$) the transmitter is highly responsive to the engagement pattern of a given receiver.

Figure S11 shows the effect of varying the degree of micro-targeting available to the transmitter. We find that when micro-targeting is low the difference in engagement between false versus true stories declines, until the apparent attractiveness of misinformation disappears entirely. Thus the ability to target news stories at specific groups of receivers, either by news sites directly or by social media algorithms, can substantially contribute to the apparent attractiveness of false news stories.

\begin{figure}[th!] \centering \includegraphics[scale=0.6]{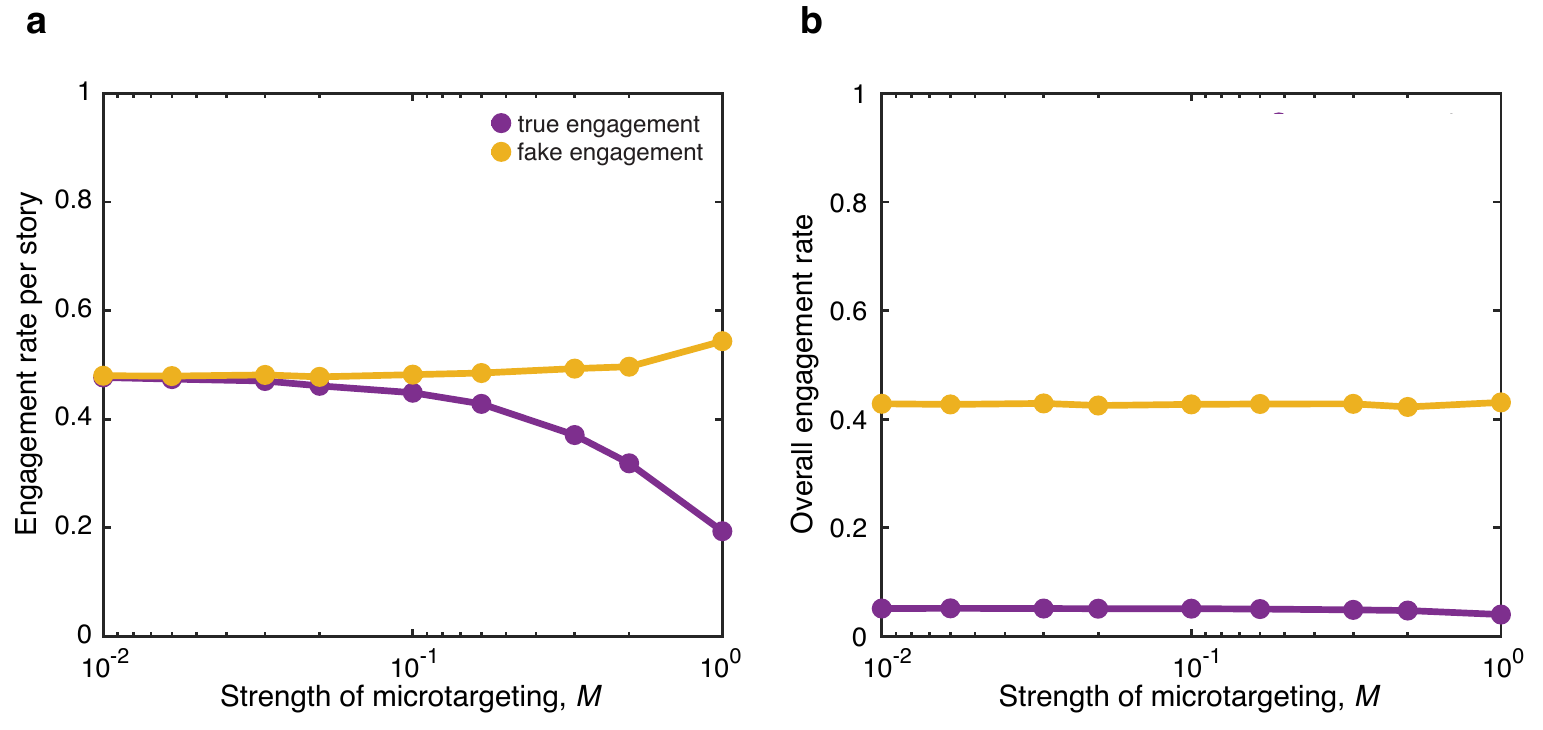}
\caption*{\small Figure S11: \textbf{Micro-targeting facilitates engagement with misinformation}  a) A single misinformation site micro-targeting multiple independent receivers. The strength of micro-targeting $M$ (x-axis) quantifies the responsiveness of the transmitter to each of the target groups of receivers (see main text). Each target receiver group is assumed to independently optimize their engagement strategy as described previously (Figure S3). The misinformation site uses the same transmission strategy for all the groups included in their targeting. The less precise the targeting of the transmitter (lower values of $M$), the smaller the difference in engagement between true and false news stories. b) However, the overall engagement probability remains unchanged with $M$.  In all cases receiver strategic mutations are local (see SI Section 1) and receiver payoffs are $\pi_t=2$ and $\pi_f=-1$. Other parameter choices are explored in the SI. Here the misinformation site uses a fixed linear responsive strategy $\mathbf{r}=\{1.0,0.1,-0.9,-0.1\}$ 
see SI Section 1}
\end{figure}

\clearpage

\subsection*{The effect of different types of receiver attention}

We studied the effects of varying different types of receiver attention to news content and to transmitter behavior. In the analysis presented in Figure S3 of the main text we assume that receivers are inattentive to both prior experience with a transmitter and prior information about the veracity of each headline, i.e. $a_0=a_1=0$ in Eq. 1. In Figure S12 we repeat the same analysis assuming receivers are attentive to past experience, i.e. $a_1=1$ in Eq. 1. We see that, as in Figure S3, the most successful misinformation strategies tend to produce greater overall engagement and engagement per story with fake stories than with true. Conversely, the most successful mainstream strategies produce more engagement  per story and overall engagement with true stories than false. 

\begin{figure}[th!] \centering \includegraphics[scale=0.25]{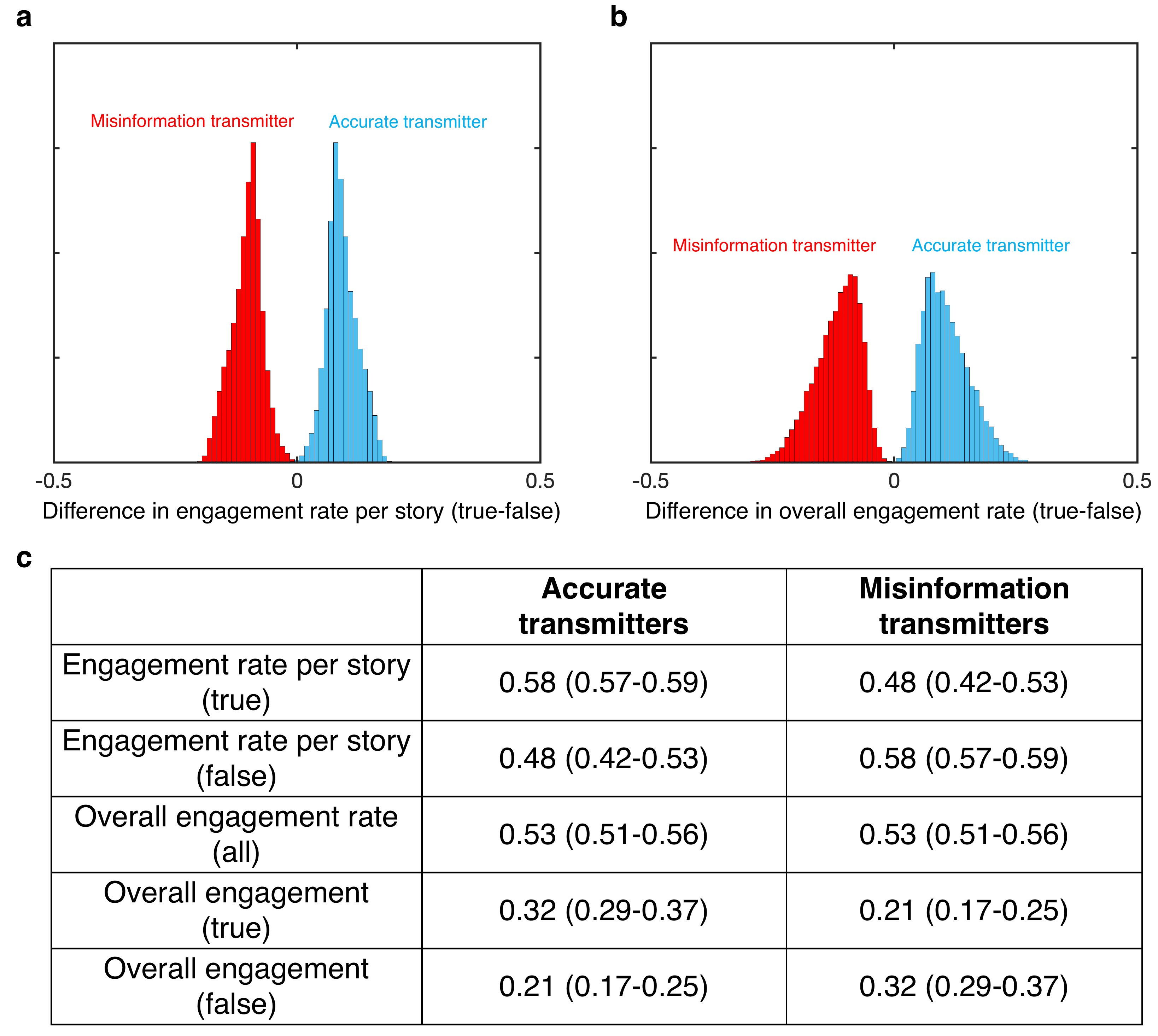}
\caption*{\small Figure S12: \textbf{Successful news transmission strategies with receiver memory --} To identify strategies that are successful at promoting engagement with true news stories (mainstream sites) or with misinformation stories (misinformation sites), we randomly drew $10^8$ transmitter strategies (Eq.~2) and allowed a single receiver, incentivized to engage with true news, to optimize their engagement strategy (see SI Section 1) over the course of $10^4$ interactions. The receiver was assumed to be inattentive to accuracy ($a_0=0$) but attentive to past stories ($a_1=1$) and optimization was assumed to be noisy ($\sigma=1$). a) We identified the transmitter strategies that  successfully promote engagement with true news stories (mainstream sites, blue), i.e those that produce a true news engagement probability within the 90th percentile from the $10^8$ transmitter strategies considered, as well as $v_f<0.5$. Similarly, we identified the transmitter strategies that successfully promote engagement with misinformation (misinformation sites, red), i.e those that produce a misinformation engagement probability within the 90th percentile, as well as $v_f>0.5$. We  plot the difference between engagement with true and fake stories, $v_{tc}/v_t-v_{fc}/v_f$, for all mainstream and misinformation sites. Mainstream site strategies induce engagement with accurate stories while misinformation site strategies induce the opposite effect. b) We also report the difference in overall engagement with true stories, $v_{tc}$, and overall engagement with fake stories $v_{fc}$. c) Summary of the engagement per story rates per story and overall engagement rates for mainstream and misinformation site transmission strategies. In all cases receiver mutations were local and we set receiver payoffs $\pi_t=2$ and $\pi_f=-1$. }
\end{figure}

\clearpage

We also calculate the regions FI, FII and FIII for the effect or transmitter preference on receiver engagement patterns in the case that receivers use memory-1, i.e. $a_1=1$ (Figure S13). We see qualitatively similar results to the regions in Figure 3. Surprisingly, we find that the region FII, in which receiver engagement reflects transmitter preference, despite receivers preferring accurate stories, is slightly larger than the case $a_1=0$.

\begin{figure}[th!] \centering \includegraphics[scale=0.33]{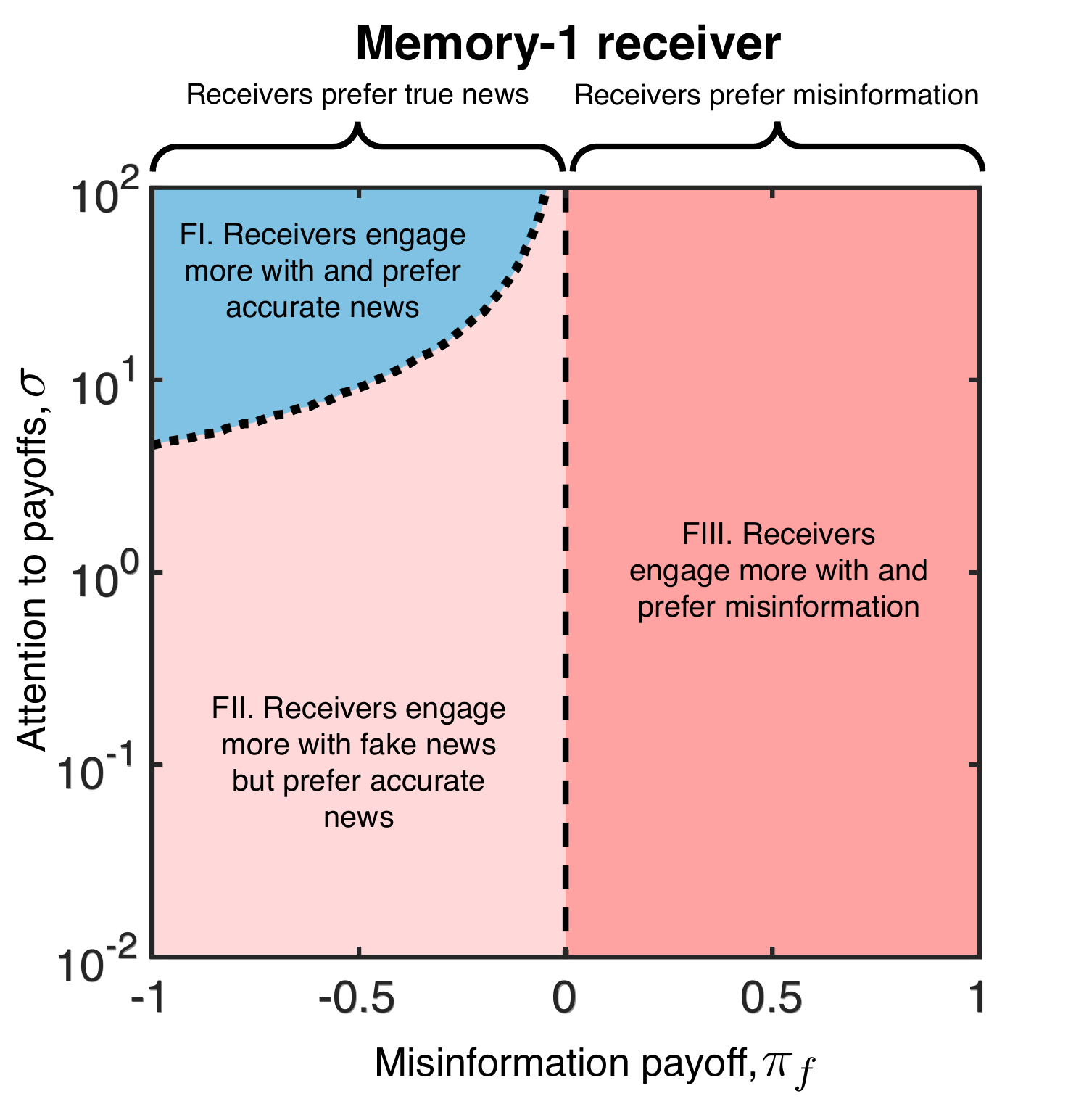}
\caption*{\small Figure S13: \textbf{Effect of memory, receiver preferences and attentiveness on engagement  --} 
We calculated the engagement patterns among receivers optimizing under different preferences (where we have set $\pi_t=-\pi_f$), and different levels of attention, $\sigma$ in the same way as for main text Figure 3, for a receiver with memory, $a_1=1$.  We see a qualitatively similar pattern to the case with no memory, $a_1=0$, with an increased region FII. In all cases receiver strategic exploration was local, with $a_0=0$, with transmitter attention set at $\sigma=100$. Optimization occurred over $10^4$ time-steps using ensembles of $10^3$ replicates for each value of $\{\pi_f,\sigma\}$ (see SI section 1).}
\end{figure}

The distribution of engagement per story and overall engagement for all $10^8$ transmitter strategies when receivers employ memory of their previous interaction ($a_1=1$) is shown in Figure S14. We see that the distribution of engagement is narrower than in Figure S3, when receivers are inattentive to memory, while the distribution of overall engagement follows a similar distribution in both cases.

\begin{figure}[th!] \centering \includegraphics[scale=0.2]{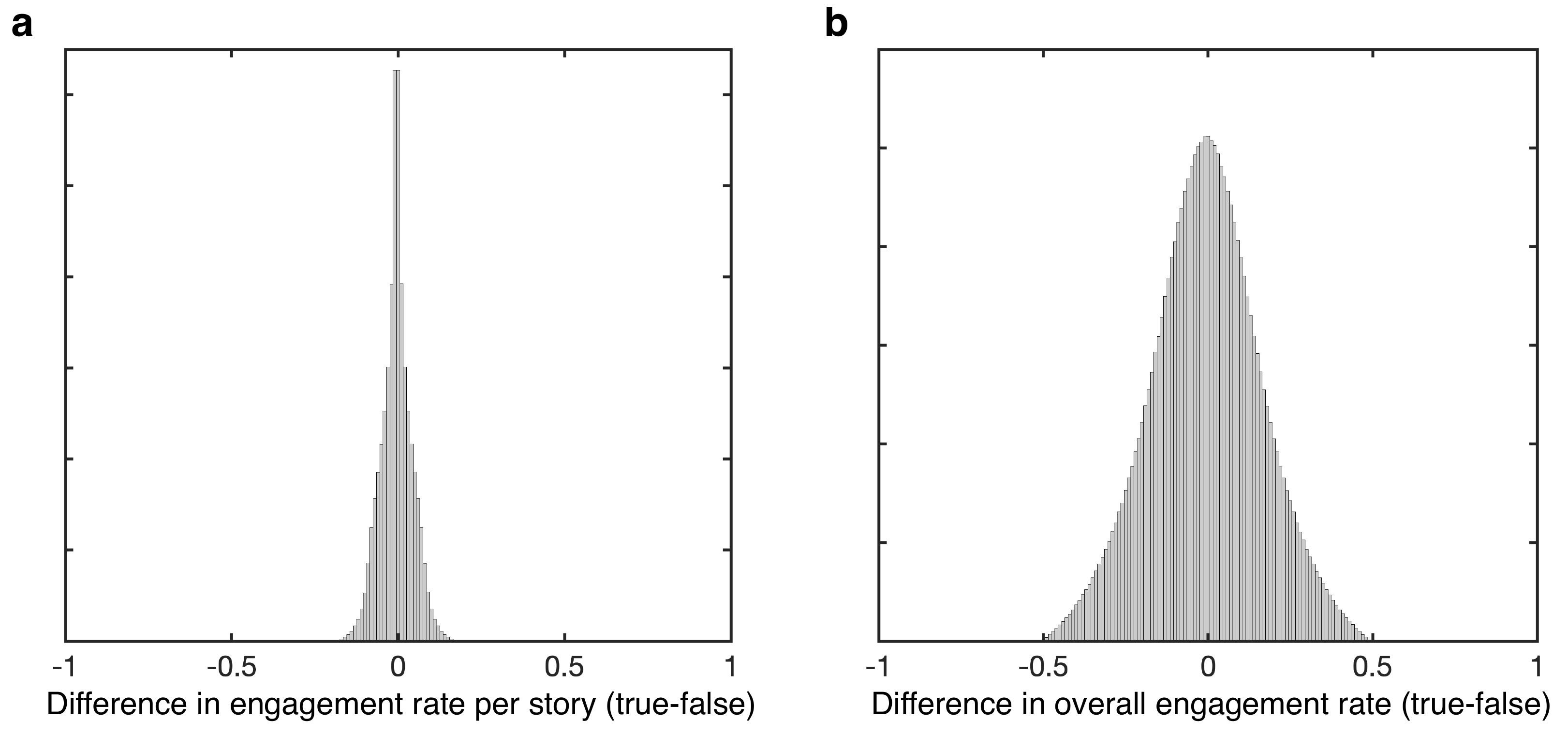}
\caption*{\small Figure S14: \textbf{Full distribution of engagement per story and overall engagement with receiver memory --} Shown are the distributions of the difference in engagement per story (a) and overall engagement (b) with true vs fake stories for $10^8$ transmitter strategies. We define successful misinformation strategies as those with $v_f>0.5$ and misinformation engagement, $v_{fc}/v_f$ in the 90th percentile among all observed strategies.  Similarly we define successful mainstream strategies as those with $v_t>0.5$ and misinformation engagement, $v_{tc}/v_t$ in the 90th percentile among all observed strategies.}
\end{figure}

We also considered (Figure S15) the effects of receivers conditioning their behavior on memory of previous interactions beyond simply the immediately preceding encounter. We assumed that receivers treated a transmitter as having shared misinformation with them when employing their strategy $\mathbf{p}$ (Eq. 1) if the transmitter has shared a fake story in \emph{any} of the last $m$ rounds. We show engagement per story and overall engagement as a function of such memory in Eq. S7, where we see that longer memory of this type has only a weak effect on engagement per story and overall engagement, and does not by itself reverse the trend of successful misinformation sites generating higher engagement per story and overall engagement with fake vs true stories among inattentive consumers.

\begin{figure}[th!] \centering \includegraphics[scale=0.5]{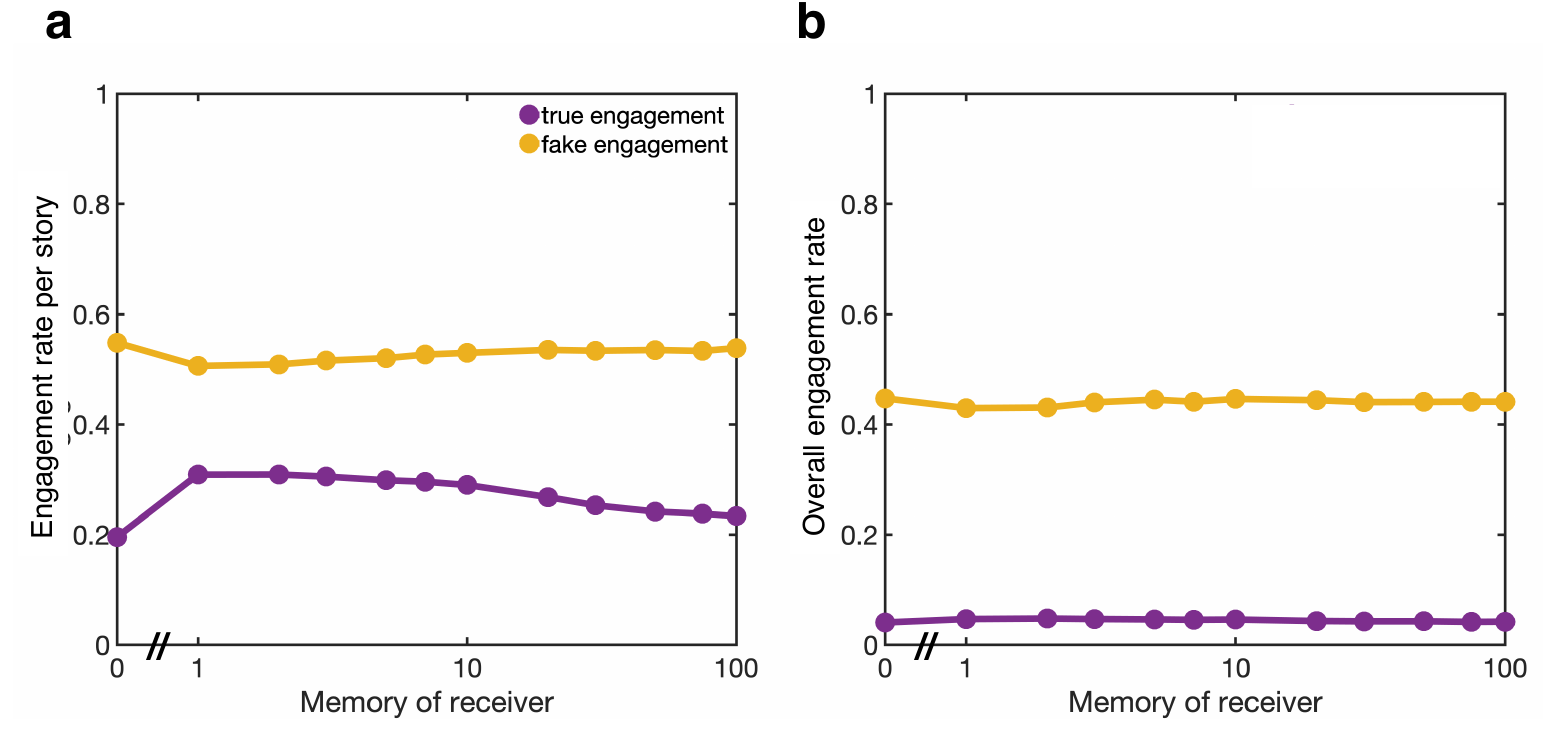}
\caption*{\small Figure S15: \textbf{Receiver memory length --} a) A single receiver optimizing in response to a single transmitter, under the same process as described for Figure S3. The receiver uses memory of preceding rounds to decide whether the transmitter shares true stories (and so is labelled $t$) or fake stories (and so is labelled $f$). We assume that if the receiver identifies any item of misinformation in their memory of the transmitter's shared stories, they label the transmitter as fake ($f$). We vary the length of this memory (x-axis) and calculate the equilibrium level of engagement with fake (orange) and true (purple) stories. b) We also calculate the equilibrium level of overall engagement. In all cases receiver mutations were local (see SI Section 1) and we set receiver payoffs $\pi_t=2$ and $\pi_f=-1$. Here misinformation sites use a fixed linear responsive strategy $\mathbf{r}=\{1.0,0.1,-0.9,-0.1\}$ see SI Section 1}
\end{figure}

We also studied the effect of varying attention to headline accuracy in Eq. 1, against successful misinformation promoting transmitter strategies. We varied attention to accuracy, $a_0$, memory of past interactions, $a_1$ and attention to payoffs, $\sigma$ among receivers. We also considered the case where receivers both prefer true news (as in the main text) and the case where they prefer misinformation, i.e. derive benefit $\pi_t$ from interacting with misinformation and pay cost $\pi_f$ for interacting with true news (Figure S16). 

\begin{figure}[th!] \centering \includegraphics[scale=0.2]{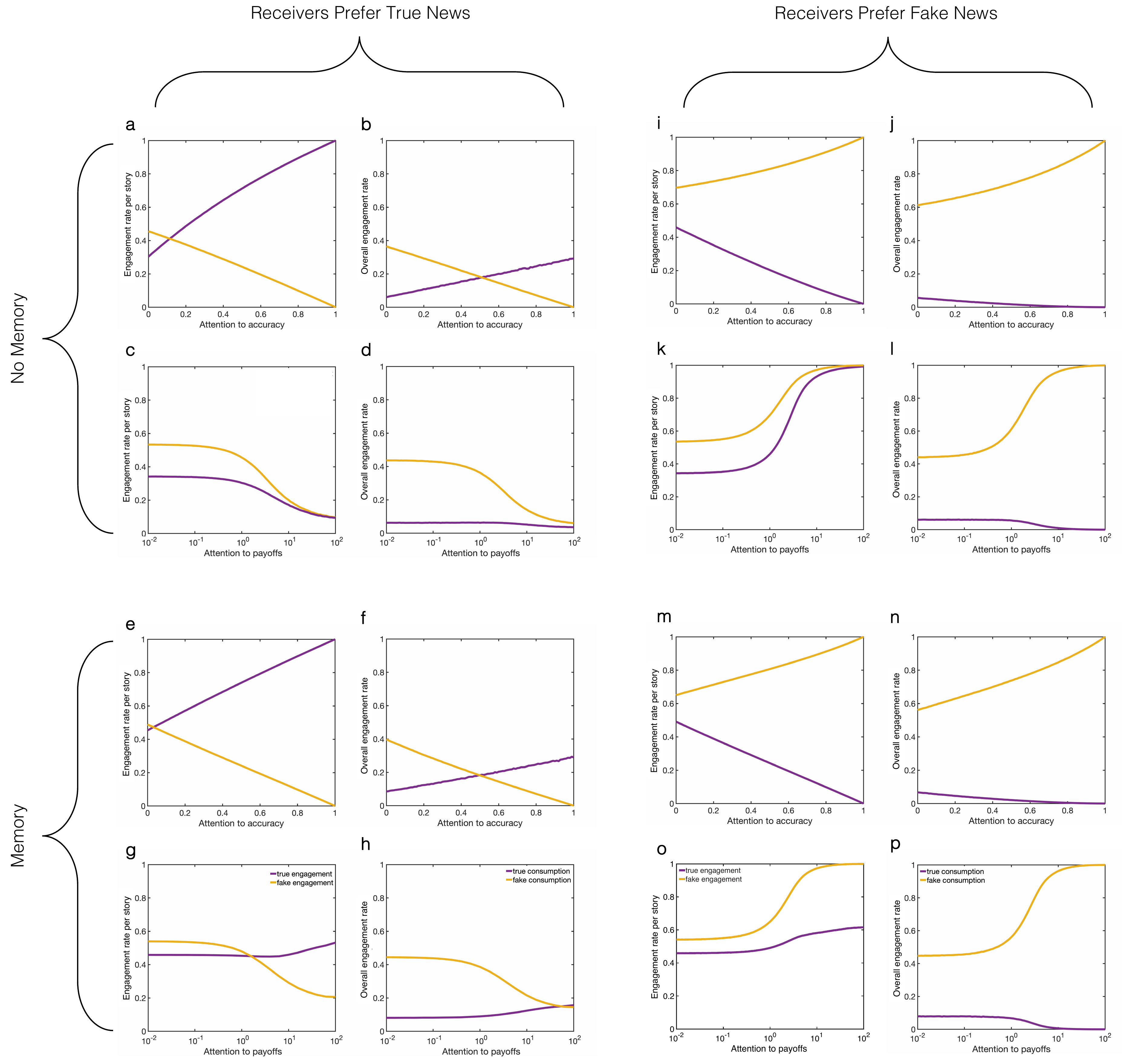}
\caption*{\small Figure S16: \textbf{Attention and news engagement --} We randomly drew $10^5$ misinformation strategies (see SI Section 1) and allowed receivers to optimize as described in Figure S3. (a-b) When receivers have no memory $a_1=0$, and pay little attention to long-term payoffs, $\sigma=1$, but prefer accurate information, we increased attention to accuracy in the optimization process ($a_0$) (x-axis), and calculated a) the average probability of engagement with true (purple) and fake (orange) stories. As attention to accuracy increases, overall engagement per story with fake stories decreases to zero, while engagement per story with true stories increases. b) Similarly, overall engagement probability with fake stories decreases with increasing attention to accuracy, while overall engagement with true stories increases. 
(c-d)  When receivers have no memory $a_1=0$, and no attention to accuracy, $a_0=0$ , but prefer accurate information, we increased attention to long-term payoffs in the optimization process ($\sigma$) (x-axis), and calculated c) the average probability of engagement with true (purple) and fake (orange) stories. As attention to payoffs increases, overall engagement with true and fake stories decreases to zero. d) Similarly, overall engagement probability with true and fake stories decreases with increasing attention to payoffs.
(e-h) Shows the same plots as in (a-d) except with memory of past interactions, i.e. $a_1=1$. (e-f) show similar patterns as (a-b) while (g-h) shows increasing engagement per story and overall engagement for true news (purple) with increasing attention to payoffs ($\sigma$).
(i-l) Shows the same plots as in (a-d) except with receivers preferring misinformation, i.e. gaining benefit $\pi_t$ from interacting with a fake story and paying a cost $\pi_f$ from interacting with a true story. (i-j) show opposite patterns to (a-b) while (k-l) show k) increasing engagement for both true and misinformation engagement with attention to payoffs, but l) declining overall engagement with true news.
(m-p) Shows the same plots as in (e-h) except with receivers preferring misinformation, i.e. gaining benefit $\pi_t$ from interacting with a fake story and paying a cost $\pi_f$ from interacting with a true story. (m-n) show opposite patterns to (e-f) while (o-p) show opposite patterns to (g-h). 
In all cases we set attention to payoffs $\sigma=1$ and receiver payoffs $\pi_t=2$ and $\pi_f=-1$ unless otherwise stated.}
\end{figure}

\clearpage

We see that in many cases increasing attention not only reduces engagement per story and overall engagement among receivers who prefer true news (Figure S16 panels a-h) it can also reverse the apparent preference for misinformation induced by responsive transmitter strategies. However increased attention tends to have the opposite effect on engagement per story and overall engagement when receivers prefer misinformation (Figure S16 panels i-p).

Finally we calculated the regions FI, FII and FIII for different rates of attention $a_0$ to headline veracity (Figure S7). As expected, increasing $a_0$ increases the size of region FI and decreases the size of region FII, i.e. receivers become increasingly likely to engage with news that reflects their own preference rather than that of the transmitter.

\begin{figure}[th!] \centering \includegraphics[scale=0.2]{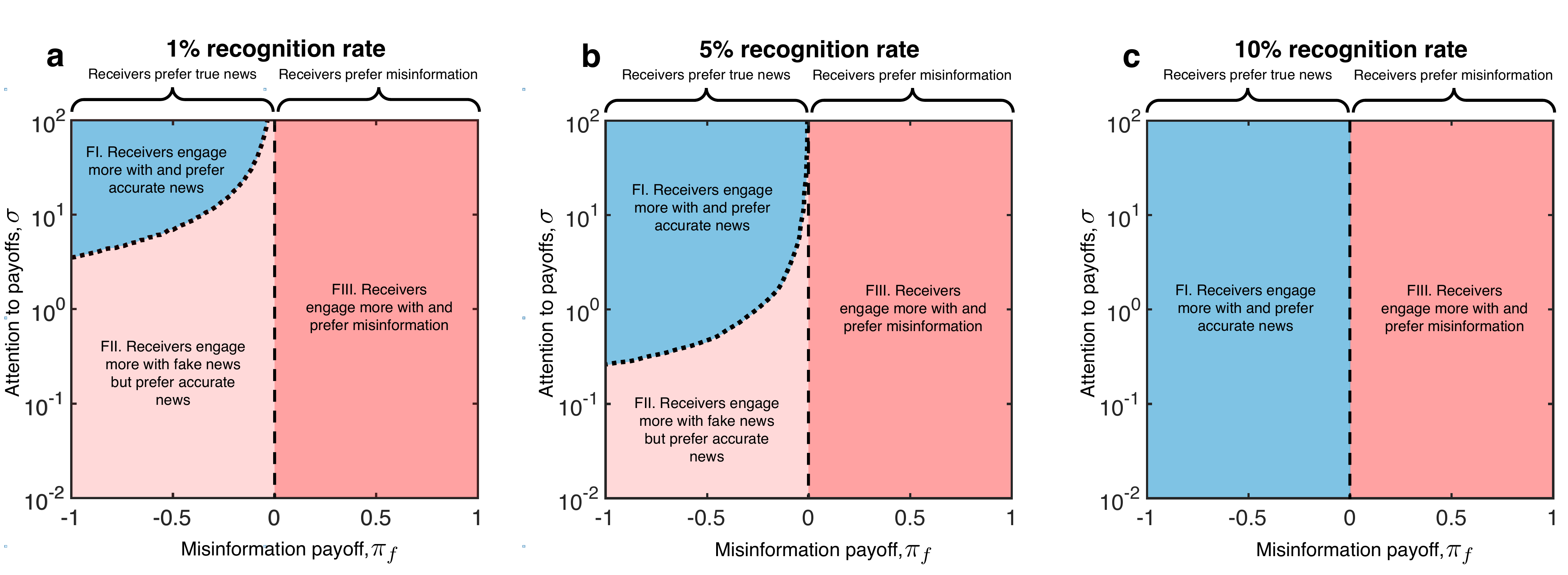}
\caption*{\small Figure S17: \textbf{Effect of prior knowledge, receiver preferences and attentiveness on engagement  --} 
We calculated the engagement patterns among receivers optimizing under different preferences (where we have set $\pi_t=-\pi_f$), and different levels of attention, $\sigma$ in the same way as for main text Figure 3, for different levels of attention to headline veracity, $a_0$. a) When $a_0=0.01$, meaning 1\% of false stories can be identified from their headline, we see little change in the size of region FII as compared to Figure S13 (for which $a_0=0$). b) When $a_0=0.05$, meaning 5\% of false stories can be identified from their headline, we see a significant increase in the size of region FII as compared to Figure S13. c) When $a_0=0.1$, meaning 10\% of false stories can be identified from their headline, region FII vanishes entirely, and receiver engagement patterns always match their preferences. In all cases receiver strategic exploration was local, with memory $a_1=1$, with transmitter attention set at $\sigma=100$. Optimization occurred over $10^4$ time-steps using ensembles of $10^3$ replicates for each value of $\{\pi_f,\sigma\}$ (see SI section 1).}
\end{figure}

\clearpage

\subsection*{The effect of competition between news sources}

We studied the effect of competition between news sources, with a receiver facing a single misinformation site within in a population of otherwise mainstream sites (Figure S18). We see that the receiver tends to engage with more misinformation from the misinformation site as the population size of mainstream sites grows. This makes intuitive sense, because the relative contribution of the misinformation site to the receiver's payoff necessarily declines as the population grows, and so avoiding misinformation becomes less important to receiver payoff.

\begin{figure}[th!] \centering \includegraphics[scale=0.5]{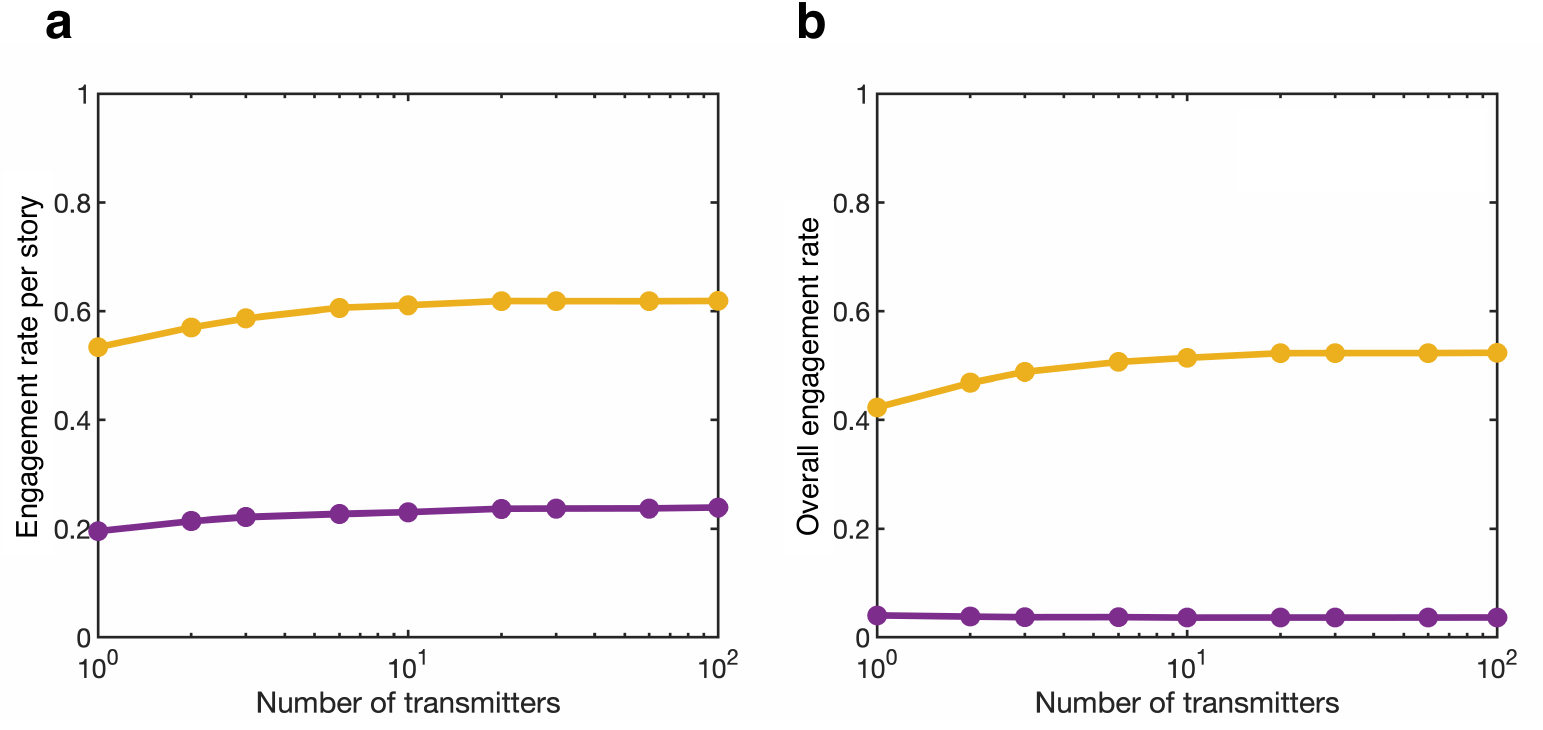}
\caption*{\small Figure S18 \textbf{Groups of transmitters --} a) A single receiver optimizing in the face of a population of transmitters, under the same process as described for Figure S3. The transmitter population consists of a single misinformation site and $N-1$ mainstream sites. Increasing the number of mainstream sites increases engagement with true (purple) and fake (orange) news stories from the misinformation transmitter as well as b) overall misinformation engagement. In all cases receiver mutations were local (see SI Section 1) and we set receiver payoffs $\pi_t=2$ and $\pi_f=-1$. Other parameter choices are explored in the SI. Here misinformation sites use a fixed linear responsive strategy $\mathbf{r}=\{1.0,0.1,-0.9,-0.1\}$ and mainstream sites use a fixed linear responsive strategy $\mathbf{r}=\{0.9,0.0,0.1,0.9\}$ see SI Section 1}
\end{figure}

\clearpage

\subsection*{The effect of supply and demand}

We studied the effects consumer demand on transmitter supply of misinformation. To do this we varied receiver payoff on overall engagement and engagement per story, shown in Figure S19 and Figure S20. In Figure S19 we fix the cost $\pi_f$ of consuming misinformation, and vary the benefit $\pi_t$ of consuming true news, for different levels of attention to payoff, $\sigma$. We see that increasing benefits of true news often has only weak effects on overall engagement and engagement per story, but do tend to increase overall levels of both once benefits become sufficiently large.

\begin{figure}[th!] \centering \includegraphics[scale=0.3]{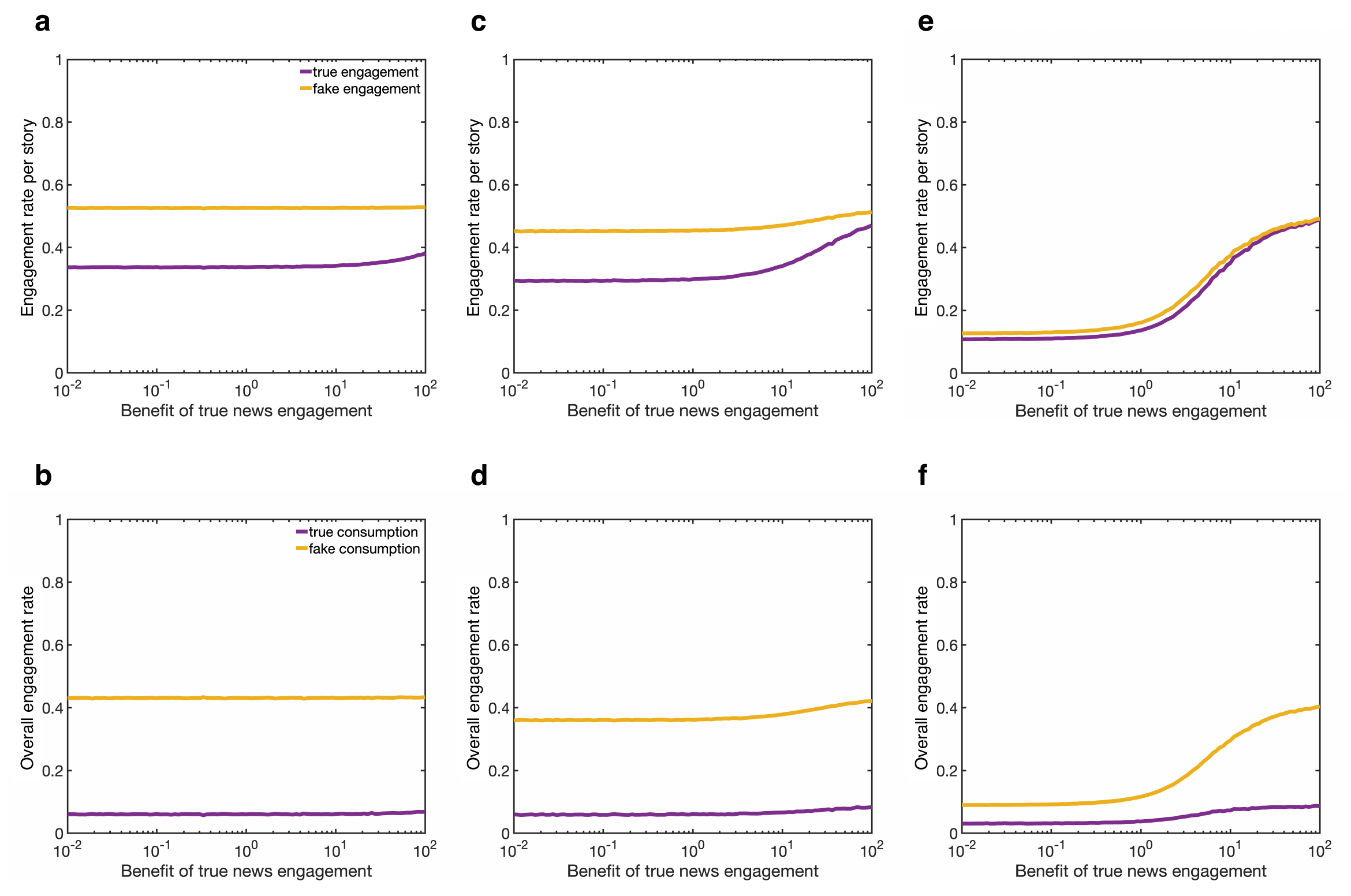}
\caption*{\small Figure S19: \textbf{Varying benefit of true news engagement --}  We randomly drew $10^5$ responsive misinformation strategies (see SI Section 1) and allowed receivers to optimize as described in Figure S3. a) When attention to payoff is low, ($\sigma=0.1$), cost of engaging with misinformation is fixed, ($\pi_f=-1$), and receivers have no memory and no attention to story accuracy, $a_0=a_1=0$, we increased the benefit of engaging with true news stories $\pi_t$ in the optimization process (x-axis), and calculated the average probability of engagement with true (purple) and fake (orange) stories. As attention to payoffs increases, engagement per story slowly increases and the difference in engagement with true and fake stories declines. b) Similarly, overall engagement when attention to payoffs is low ($\sigma=0.1$) remains close to constant. 
c) When attention to payoffs is moderate ($\sigma=1$) the same pattern holds for engagement per story while d) overall engagement slowly increases, and the difference between overall engagement with true and misinformation also increases. e) When attention to payoffs is strong ($\sigma=10$) there is little difference between true and misinformation engagement, and engagement per story increases with $\pi_t$ while f) overall engagement increases, and the difference between overall engagement with true and misinformation also increases. Unless otherwise stated parameters are the same as in Figure S3.}
\end{figure}

In Figure S20 we fix the benefit $\pi_t$ of consuming true news, and vary the cost $\pi_f$ of consuming misinformation, for different levels of attention to payoff, $\sigma$. We see that increasing costs of misinformation has a strong effect on overall engagement and engagement per story, reducing both close to zero once costs become sufficiently large. This result is consistent with our game-theoretic expectation that higher costs require a misinformation sharing strategy to include ever higher levels of true news to make engagement worthwhile for the receiver.

\begin{figure}[th!] \centering \includegraphics[scale=0.3]{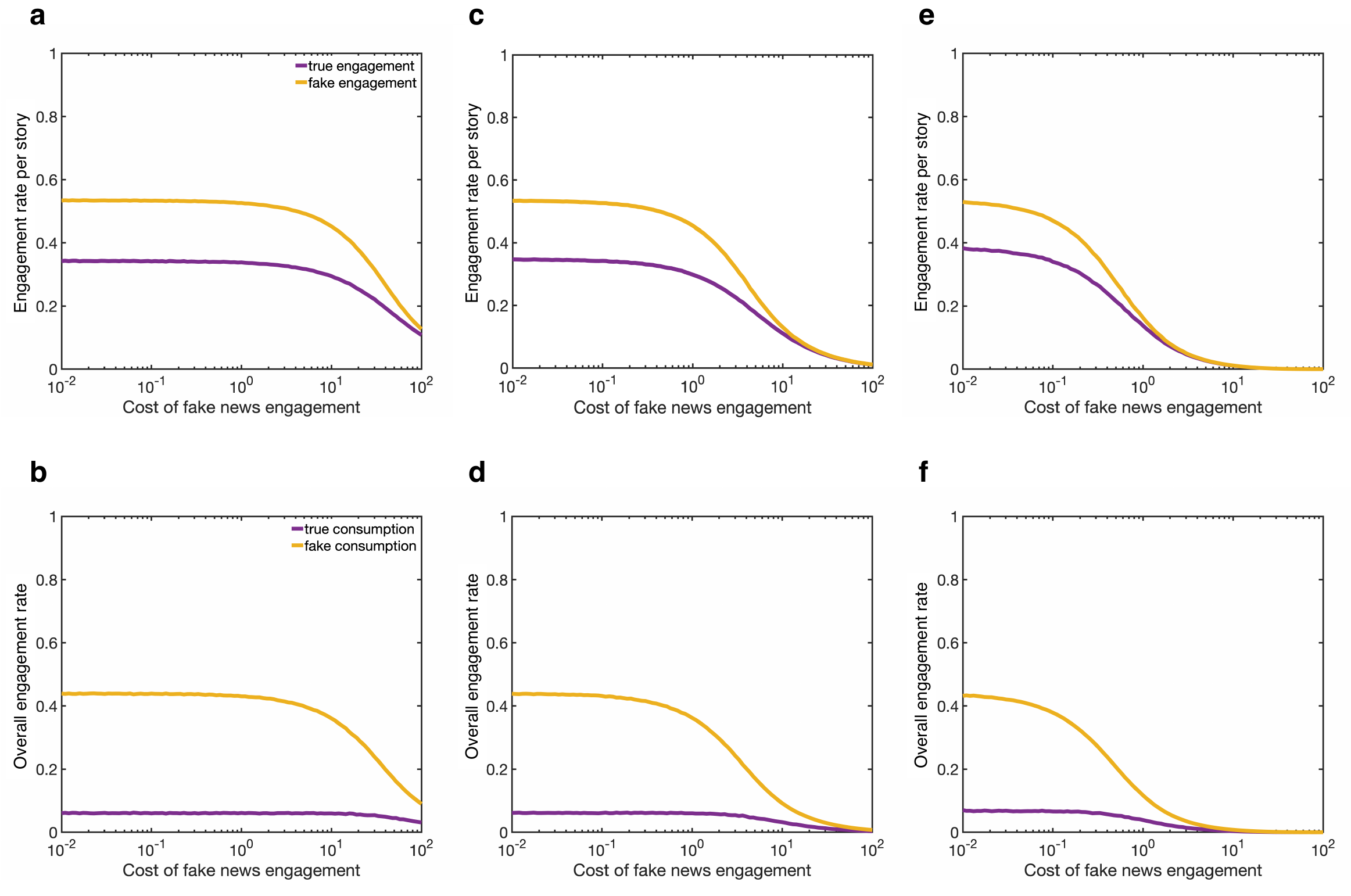}
\caption*{\small Figure S20: \textbf{Varying cost of true news engagement --}  We randomly drew $10^5$ responsive misinformation strategies (see SI Section 1) and allowed receivers to optimize as described in Figure S3. a) When attention to payoff is low, ($\sigma=0.1$), benefit of engaging with true news is fixed, ($\pi_t=1$), and receivers have no memory and no attention to story accuracy, $a_0=a_1=0$, we increased the cost of engaging with misinformation stories $\pi_f$ in the optimization process (x-axis), and calculated the average probability of engagement with true (purple) and fake (orange) stories. As attention to payoffs increases, engagement per story slowly decreases and the difference in engagement with true and fake stories declines. b) Similarly, overall engagement when attention to payoffs is low ($\sigma=0.1$) declines as $\pi_f$ increases. 
c) When attention to payoffs is moderate ($\sigma=1$) the same pattern holds for engagement per story and d) overall engagement. e) When attention to payoffs is strong ($\sigma=10$) both engagement per story and f) overall engagement decline to zero once $\pi_f<-1$. Unless otherwise stated parameters are the same as in Figure S3.}
\end{figure}

\clearpage

\section*{Supplementary information on experiments}

Here we provided additional details and analysis of our experimental findings.

\subsection*{Headline selection and engagement data}

We studied 20 mainstream and 20 misinformation sources, used in previous studies of misinformation online \cite{Pennycook2521}. We selected the most recent 20 unique news headlines with engagement data available for Facebook via CrowdTangle. We excluded stories whose headlines did not constitute news (e.g. a quiz). We also studied the relationship between source trust and patterns of engagement without a priori groupings of mainstream and misinformation sites (Figure S21). To do this we used trust ratings from \cite{DiasMisinfoRev}.

We gathered engagement data via CrowdTangle on the most recent 20 news headlines for each news source as available on November 22nd, 2020. This information was gathered by generating a historical report for each news site and gathering the 20 most recent news headlines in the database. The engagement information recorded for each story is the number of Facebook interactions reported by CrowdTangle at the time. This number accounts for likes, comments and shares of posts that contain the URL for the article across the CrowdTangle database.

The list of headlines for each source, the headline accuracy rating, the publication date of the story and the Facebook engagement information recovered via CrowdTangle is available via github. 

\subsection*{Meta-analysis}

We performed a meta-analysis on the results of linear regressions of engagement against accuracy, for the 20 misinformation and 20 mainstream sites. In order to determine whether there was a negative correlation between engagement and accuracy for misinformation sites, we performed Fisher's combined test, using the p-values resulting from a single-tailed t-test from a regression of engagement against accuracy. we used the same procedure to determine whether there was a positive correlation between engagement and accuracy for mainstream news sites. 

To determine whether there was significant heterogeneity among misinformation or mainstream sites we calculated the between study variance $\tau^2_{DL}$ using the  DerSimonian and Laird \cite{DERSIMONIAN1986177} moment-based estimate, and calculated the proportion of total variance attributable to between study effects, $I^2$, as well as the degree of homogeneity $Q_{RE}$ under a random-effects model \cite{https://doi.org/10.1002/sim.1186}. These results are summarized in Table S3 where we test regressions of $\log_{10}$-engagement vs accuracy, using both unstandardadized and standardized data, as well as regression using the rank of engagement and accuracy for each site. In all cases we find significant negative correlation for misinformation and significant positive correlation for mainstream sites with $p<0.01$. However we find no significant between site heterogeneity for either $I^2$ or $Q_{RE}$ with $p<0.05$.

\begin{figure}[th!] \centering \includegraphics[scale=0.25]{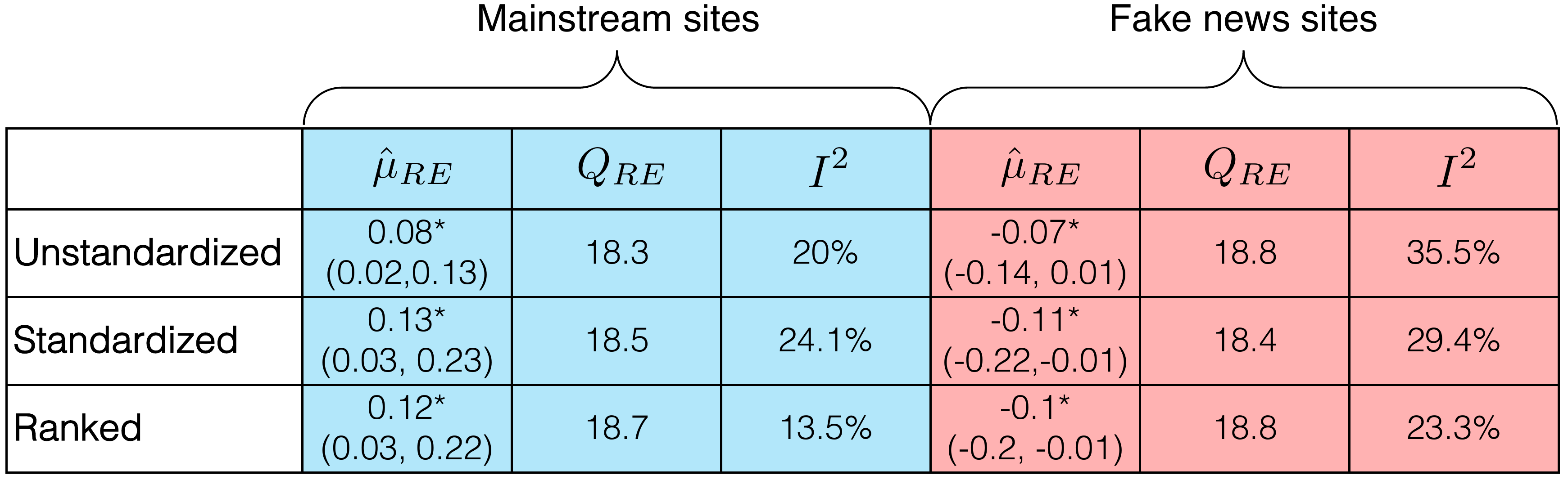}
\caption*{\small Table S3: \textbf{ Meta analysis --} The mean effect size $\hat{\mu}_{RE}$ for correlations with weights from a random effect model, as well as measures between study heterogeneity $I^2$ and $Q_{RE}$. Significance is indicated with an asterisk (*).}
\end{figure}

\clearpage

\subsection*{Experiment A: Empirical patterns of misinformation engagement (MTurk).}

In addition to the results for Experiment A presented in main text Figure 3, we also assessed each story's perceived accuracy by recruiting $1,000$ American participants from Amazon Mechanical Turk to rate the accuracy of $20$ headlines (yielding a total of $20,000$ accuracy ratings), which has been shown to produce good agreement with the ratings of professional fact-checkers via the wisdom of crowds \cite{doi:10.1126/sciadv.abf4393}. We see that mainstream news sites, which we assume seek to promote accurate information, tend to share headlines with higher accuracy ratings than fake news sites ($p<0.001$, Figure S21a). Importantly, however, both mainstream and fake news sites show wide variation in perceived headline accuracy. Thus, there is substantial overlap between the content produced by the two kinds of sites, with many articles from fake news sites being rated as more accurate than many articles from mainstream sites.

\begin{figure}[th!] \centering \includegraphics[scale=0.5]{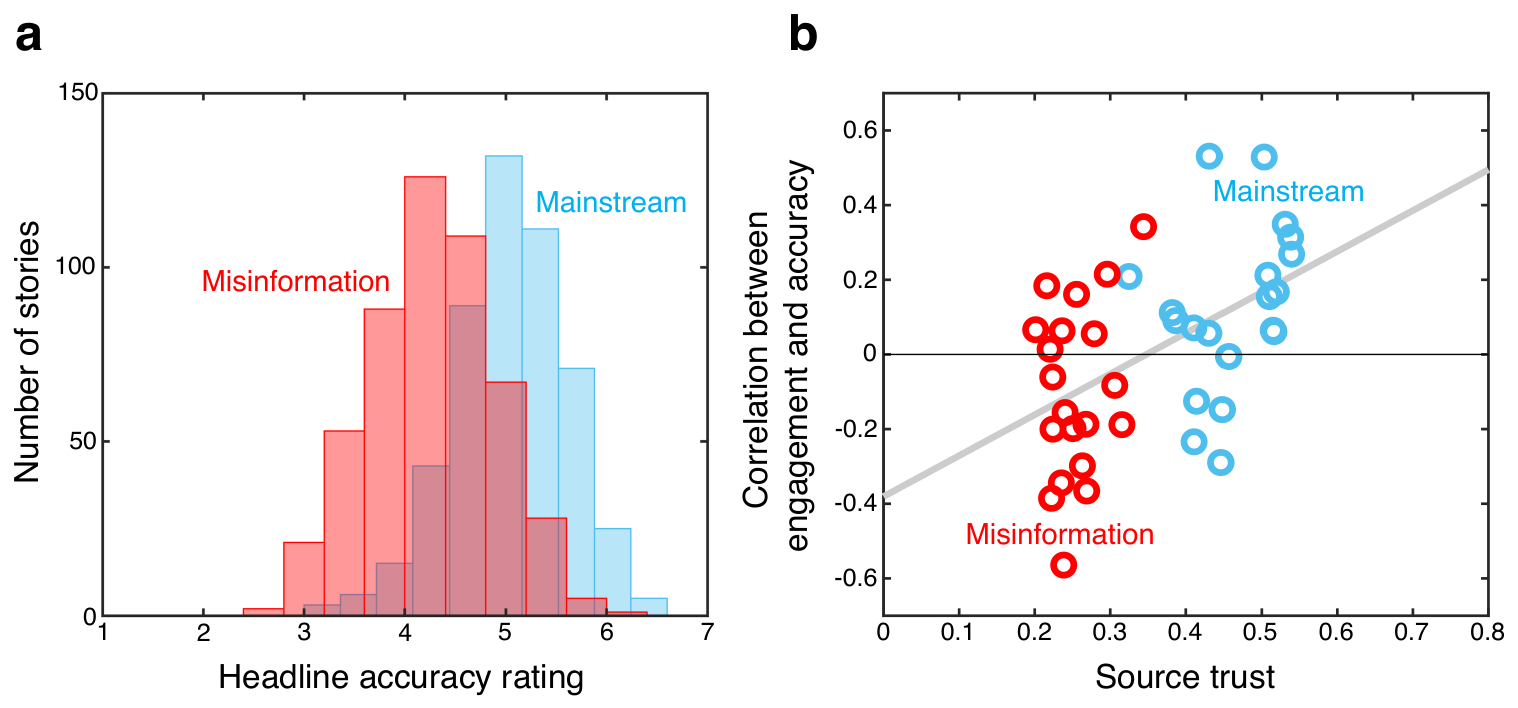}
\caption*{\small Figure S21: \textbf{Empirical patterns of news engagement and accuracy, for mainstream and misinformation sites --} We selected 20 mainstream and 20 fake news sites identified in previous studies of misinformation \cite{Pennycook2521}. Using Crowdtangle we selected the most recent 25 news stories for which engagement data on Facebook was available (see Methods). We then recruited American subjects from Amazon Mechanical Turk to assess headline accuracy ($20$ ratings per headline). a)  Distribution of accuracy ratings for mainstream sites (blue, mean=$5.05$, SD=$0.56$) and mainstream sites (red, mean=$4.27$, SD=$0.62$). As expected stories from misinformation sites are rated as significantly less accurate than stories from mainstream sites ($p<0.001$, two-sample t-test). b) Regression of individual site trust ratings \cite{DiasMisinfoRev} against standardized accuracy-engagement regression coefficient (see Methods). More trusted sites tend to produce higher engagement with accurate information, while less trusted sites tend to produce higher engagement with less accurate information ($p=0.0008$, $r^2=0.25$).}
\end{figure}

Table S4 shows the correlation between accuracy and engagement (as shown in main text Figure 4).

\begin{figure}[th!] \centering \includegraphics[scale=0.75]{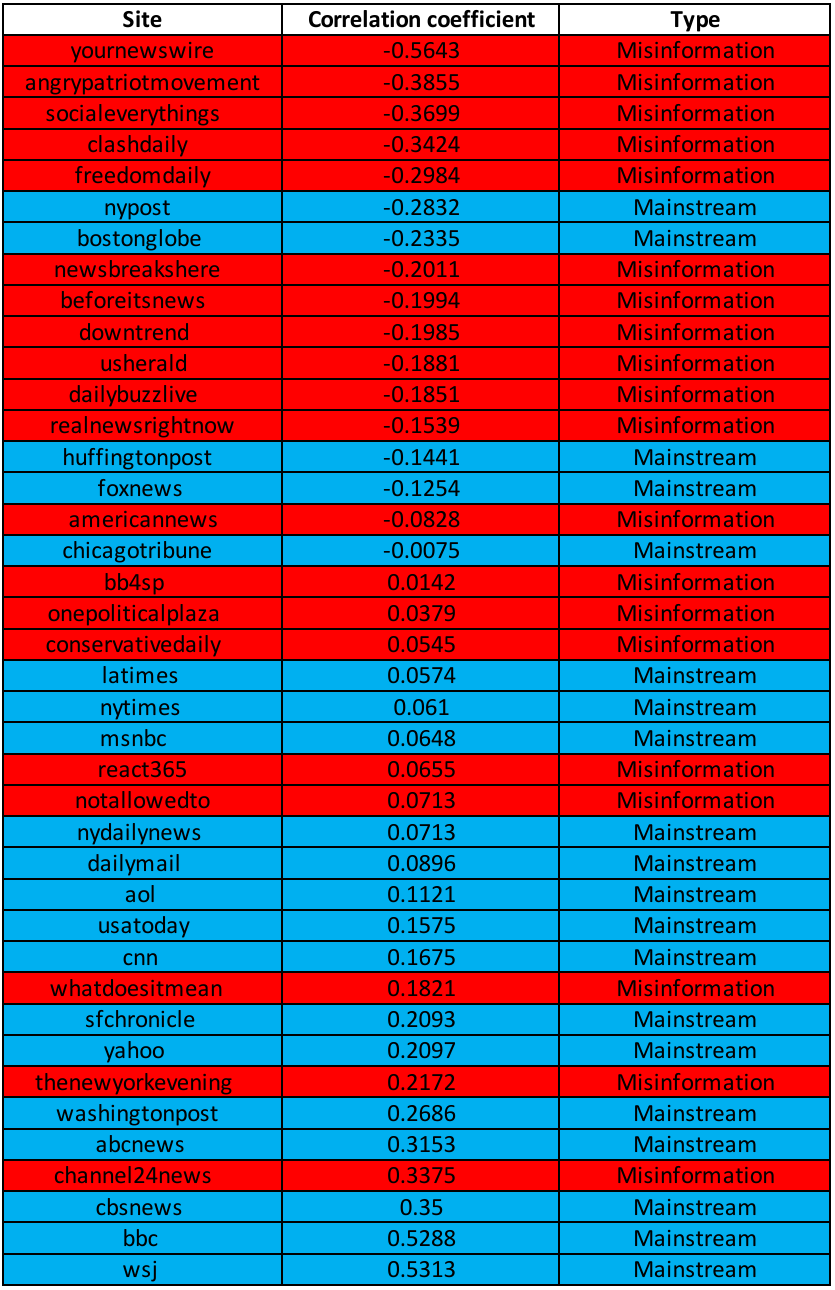}
\caption*{\small Table S4: \textbf{Correlation between accuracy and engagement}}
\end{figure}

\clearpage

\subsection*{Experiment B: Empirical patterns of receiver preference, with accuracy elicited (Lucid).}

Experiment B is detailed in the Methods section of the main text. We asked participants to assess the likelihood of sharing and the likelihood of clicking of 40 headlines. Unlike Experiment B, however, we did \textit{not} ask questions related to the accuracy of the displayed contents. Once participants completed the task, we asked them which domains they regularly used for news. The study was approved by MIT COUHES with the same protocol (1806400195).

The observed preferences for the general population sample (Figure S22) are consistent with those observed among Twitter users (Experiment C, see Figure 4). There is a strong positive correlation between perceived accuracy and willingness of readers to click or share an article ($p<0.001$). This holds for those recruited from the general population regardless of their self-reported use of misinformation sites. The results are qualitatively equivalent when using objective accuracy (as measured by professional fact-checkers; $p<0.001$ for both outcomes across both groups).

\begin{figure}[th!] \centering \includegraphics[scale=0.25]{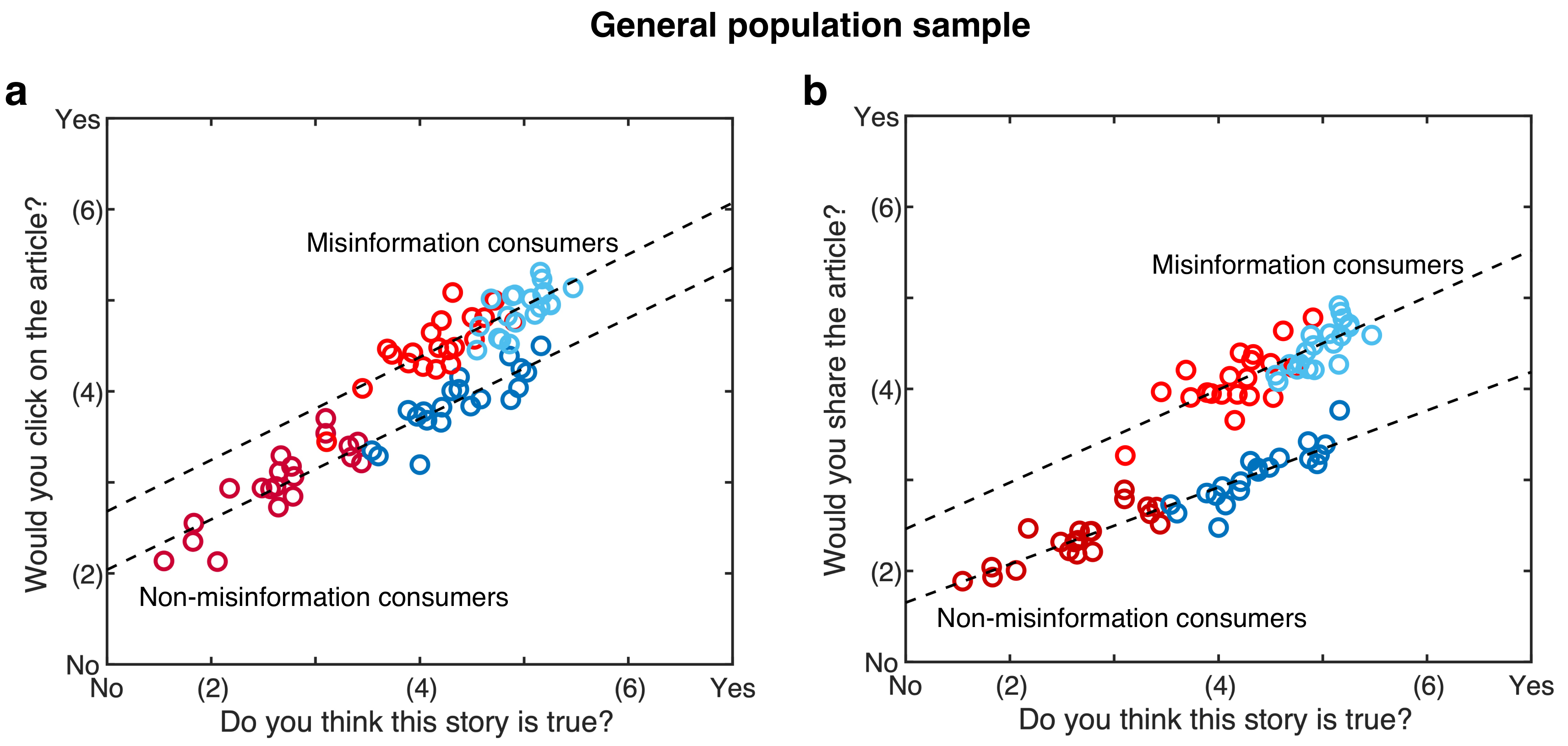}
\caption*{\small Figure S22: \textbf{Misinformation consumers prefer to engage with accuracy news.} We selected 20 mainstream (blue) and 20 misinformation (red) stories identified in previous studies of misinformation \cite{Pennycook2521}.  We then recruited American subjects from Lucid, to assess the accuracy of 10 headlines of each type. Each participant was then asked to rate their willingness to click on and share the article associated with the headline. a) A general population sample recruited from Lucid contained 124 participants who indicated engagement with misinformation sites (light colors) and 387 who indicated only engagement with mainstream sites (dark colors). Both fake and non-misinformation consumers show a strong positive correlation between perceived accuracy and willingness to click (misinformation consumers: $\beta=.662, z=18.13, p<0.001$; non-misinformation consumers: $\beta=.590, z=29.45, p<0.001$; difference between misinformation and non-misinformation consumers, $\beta=.072, z=1.83, p=0.068$), and between perceived accuracy and willingness to share (misinformation consumers: $\beta=.682, z=19.86, p<0.001$; non-misinformation consumers: $\beta=.585, z=22.75, p<0.001$; difference between misinformation and non-misinformation consumers, $\beta=-.097, z=-2.39, p=0.017$). 
}
\end{figure}

\subsection*{Supplemental Experiment: Empirical patterns of receiver preference, without accuracy elicited (Lucid).}

Similar to Experiment B detailed in the main text Methods, here we asked participants to assess the likelihood of sharing and the likelihood of clicking of 40 headlines. Unlike Experiment B, however, we did \textit{not} ask questions related to the accuracy of the displayed contents. Once participants completed the task, we asked them which domains they regularly used for news. The study was approved by MIT COUHES with the same protocol (1806400195).
\\
\\
\textbf{Participants.} From 27 April 2022 to 3 May 2022, we recruited 891 participants from Lucid but excluded: 270 who failed two trivial attention checks at the study outset, 75 who reported not having at least one social media account, and 30 who did not provide data on the use of at least one of the 60 listed domains for news. Thus, our final sample consists of 516 observations with valid records for the main analysis. The sample included 233 males and 258 females, with a mean age of 48.87 years (min. 18; max. 90). Median completion time was 10 minutes and 17 seconds.
\\
\\
\textbf{Materials.} We used the same set of 40 news "cards" and 60 domains as Experiment B. We then asked participants to : i) evaluate 20 cards, drawn randomly from the set of 40, on the likelihood of sharing (i.e. ``If you were to see the above headline online, would you share it?'', seven-point scale from ``Definitely NO'' to ``Definitely YES''), and likelihood of clicking (i.e. ``If you were to see the above headline online, would you click on it to read the article?'', seven-point scale from ``Definitely NO'' to ``Definitely YES'') of the information presented; ii) select which of the 60 domains they regularly use for news (with this information, we classified participants as misinformation media users if they selected at least one domain pre-identified as a misinformation outlet). Similar to Experiment B, the study concluded with a list of 20 exploratory items.
\\
\\
\textbf{Results} Our pre-registered main analysis was to restrict to data from participants who indicate regularly using one or more misinformation sites for news, and run  2 linear regressions with two-way clustered standard errors clustered on subject and headline, predicting sharing intentions or click intentions as the dependent variable and taking the average accuracy rating of the given headline collected in Experiment B as the independent variable. Consistent with the results of Experiment B, in the Supplemental Experiment we find that fake new site users show a significant positive relationship between out-of-sample perceived accuracy and intent to both share ($\beta=.048, t=2.96, p=0.003$) and click ($\beta=.047, t=2.26, p=0.024$). It is unsurprising that the strength of these correlations is substantially weaker than in Experiments B and C, because here we use population-average perceived accuracy from another study, rather than subject-specific perceived accuracy (and thus our predictor is measured with much more noise). Consistent with this interpretation, the associates are much higher stronger when averaging clicking and sharing intentions for each headline across subjects (thus matching the way perceived accuracy is calculated) and conducting a post hoc analysis at the headline level (see Figure S23). Be that as it may, the fact that we continue to observe significant positive associations in the Supplemental Experiment indicates that our key result from the main experiments is not simply an artifact of having asked participants to evaluate accuracy prior to making their sharing and clicking responses. 

We also pre-registered a secondary analysis comparing participants who engage with misinformation sites to participants who do not, by z-scoring each variable within user type and re-running the 2 main models including all subjects and adding a dummy for not visiting any misinformation sites, as well as the interaction between accuracy rating and this dummy. For sharing intentions, the association with out-of-sample perceived accuracy is marginally higher for non-fake-news-site users ($\beta=.028, z=1.82, p=0.069$); and significantly higher for clicking intentions ($\beta=.028, z=1.82, p=0.022$). The associations for both outcomes and both types of users are shown at the headline-level in Figure S23.

\begin{figure}[th!] \centering \includegraphics[scale=0.23]{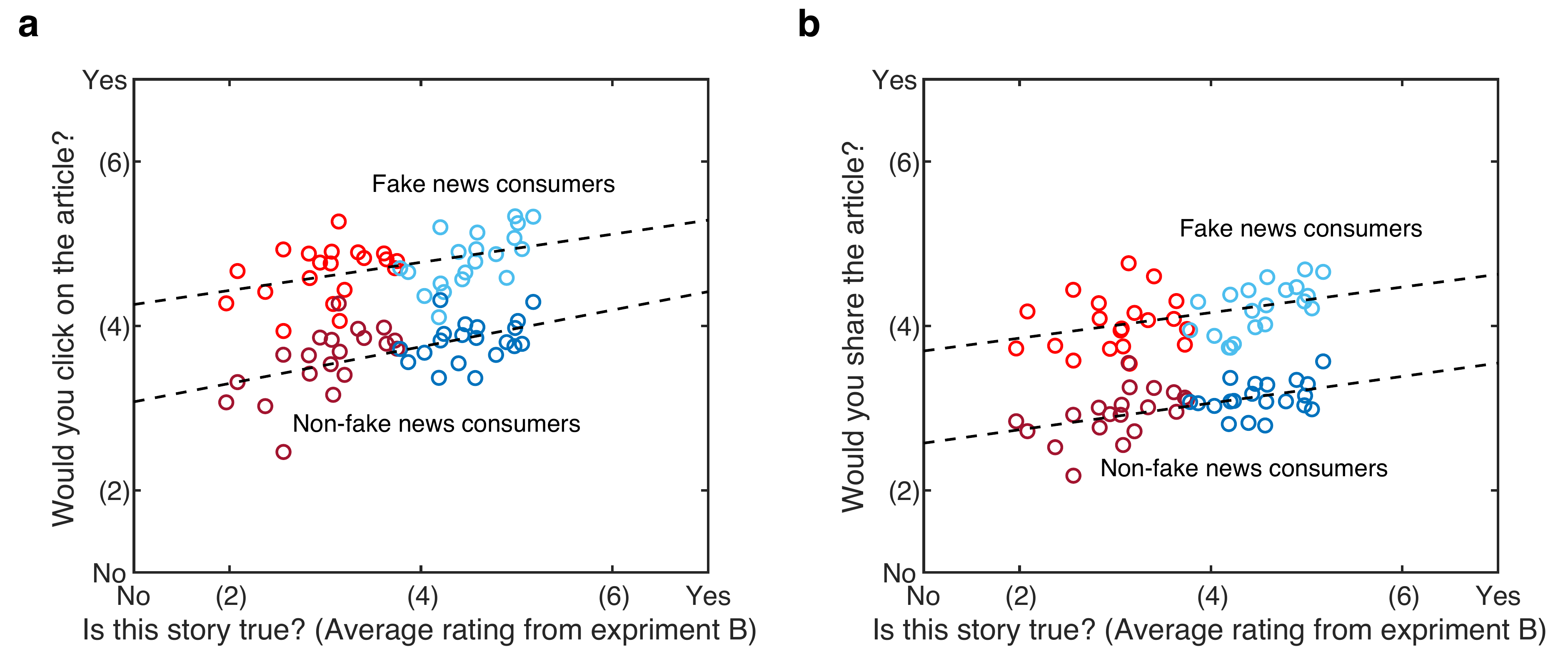}
\caption*{\footnotesize Figure S23: \textbf{The preferences of misinformation consumers.} We repeated Experiment B as described above, without prompting for accuracy. a) A general population sample recruited from Lucid contained 125 participants who indicated engagement with misinformation sites (light colors) and 421 who indicated only engagement with mainstream sites (dark colors). Both fake and non-misinformation consumers show a strong positive correlation between accuracy and willingness to click, (non-fake consumers: $r^2=0.31$, fake consumers:  $r^2=0.20$), and b) accuracy and willingness to share (non-fake consumers: $r^2=0.29$, fake consumers:  $r^2=0.17$), where accuracy ratings used are averaged from Experiment B.}
\end{figure}

\subsection*{Breakdown of the demographic and political leanings of our participants}

Here we present a table summarizing the age, sex, education, and political leaning of the participants in our four samples. Studies B and B bis match the expected quotas of a nationally-representative sample.

As can be seen, the study conducted on Twitter consists of a relatively older and more educated sample, with a slightly larger leaning towards the Republican party (6-point Likert scale; 1 - Strongly Democrat 6 - Strongly Republican). On the other hand, the study conducted on MTurk consists of a relatively younger and less educated sample, with a slightly larger leaning towards the Democratic party.

\begin{table}
\centering
\begin{tabular}{ |p{5cm}||p{1.5cm}|p{1.5cm}|p{1.5cm}|p{1.5cm}| }
 \hline
 \multicolumn{5}{|c|}{Study} \\
 \hline
  & A & B & B bis & C\\
 \hline
 
 Age bracket (Average)   &38.5 & 47.0 &48.9 & 57.8 \\
 \hspace{0.5cm}18 to 24   & 5.4\%    &10.7\% &   10.7\% & 0\% \\
 \hspace{0.5cm}25 to 34   & 42.8\%    &20.1\% &   18.8\% & 3.3\% \\
 \hspace{0.5cm}35 to 44   & 22.4\%    &16.6\% &   18.0\% & 12.2\% \\
 \hspace{0.5cm}45 to 54   & 16.5\%    &16.2\% &   15.9\% & 20.0\% \\
 \hspace{0.5cm}55 to 64   & 10.4\%    &15.8\% &   15.7\% & 32.2\% \\
 \hspace{0.5cm}Over 64   & 2.5\%    &20.7\% &   20.9\% & 32.2\% \\
Gender   &     & &    &  \\
 \hspace{0.5cm}Male   & 57.6\%    &47.3\% &   46.3\% & 50.4\% \\
 \hspace{0.5cm}Female   & 42.4\%    &51.7\% &   51.3\% & 44.4\% \\
Education   &     & &    &  \\
 \hspace{0.5cm}At least Bachelors Degree   & 21.0\%    &36.0\% &   31.0\% & 52.1\% \\
Political leaning (Average)   & 3.2    &3.4 &   3.4 & 3.8 \\
 \hspace{0.5cm}1-Strongly Democratic   & 17.3\% & 18.7\% & 15.7\% & 17.7\% \\
 \hspace{0.5cm}2-Democratic   & 26.0\% & 14.6\% & 15.3\% & 10.1\% \\
 \hspace{0.5cm}3-Lean Democratic   & 15.7\% & 17.1\% & 21.4\% & 7.6\% \\
 \hspace{0.5cm}4-Lean Republican   & 13.1\% & 21.7\% & 23.9\% & 26.9\% \\
 \hspace{0.5cm}5-Republican   & 17.8\% & 14.2\% & 11.6\% & 19.3\% \\
 \hspace{0.5cm}6-Strongly Republican   & 10.1\%    & 13.8\% & 12.2\% & 18.5\% \\
 \hline
\end{tabular}
\caption*{\small Table S5: \textbf{Demographic breakdown of participants.} }
\end{table}

\clearpage

\subsection*{Robustness of empirical results to partisanship}

In Experiment A our finding was that misinformation sites generate a negative correlation between accuracy and engagement while mainstream sites generate a positive correlation. To do this we compared engagement data from CrowdTangle with headline accuracy ratings from a survey experiment. Here we investigate whether the partisan preferences of those rating the accuracy of headlines impact our results. To do this we repeat the same analysis (calculating the regression coefficient of standardized engagement against accuracy for each news site) using accuracy ratings only from those participants who lean Democrat or Republican. We find no significant difference between the resulting correlation coefficients among headlines from mainstream sites ($p=0.23$, $t=1.22$) and no significant difference between the correlation coefficients among headlines from misinformation sites ($p=0.15$, $t=1.46$).

In Experiments B, C, and the Supplemental Experiment, our finding was that misinformation site users  prefer to engage with posts they believe are accurate. Here, we conduct a series of post hoc analyses to investigate whether there are partisan differences these preferences. To do so, we replicate the main analyses for each experiment (predicting sharing or clicking intentions using perceived accurate, using linear regression with robust standard errors clustered on subject and headline) while adding a user partisanship variable (preference for the Democratic versus Republican party on a 6-point Likert scale) and the interaction between user partisanship and perceived accuracy. This interaction indicates how the relationship between perceived accuracy and the outcome differs based on user partisanship. 

In Experiment B, we find no significant interaction between partisanship and perceived accuracy when predicting either sharing intentions ($\beta=-.050, z=-0.73, p=0.464$) or clicking intentions ($\beta=-.075, z=-1.01, p=0.311$).  

In Experiment C, we find no significant interaction between partisanship and perceived accuracy when predicting sharing intentions ($\beta=.000, z=0.01, p=0.992$). We do find a significant negative interaction when predicting clicking intentions ($\beta=-.052, z=-2.11, p=0.035$), such that the association is weaker for users who are more supportive of the Republican party. Nonetheless, even for maximally strong Republicans, we continue to find a strong positive association between perceived accuracy and clicking intention ($\beta=.318, z=3.50, p<0.001$).  

In the Supplemental Experimental, we find no significant interaction between partisanship and perceived accuracy when predicting either sharing intentions ($\beta=-.018, z=-1.38, p=0.167$) or clicking intentions ($\beta=-.011, z=-0.83, p=0.406$). 

Thus we do not find evidence that the positive association between perceived accuracy and engagement is unique to one political party or the other. 

\clearpage

\subsection*{Experimental interface}

\noindent\textbf{Experiment A.} In Experiment A we used Crowdtangle to gather the 25 most recently available news stories from 40 media outlets (i.e. 1,000 articles; 500 posted by 20 misinformation sites and 500 posted by 20 mainstream sites), along with the headline, lede, date of publication, link and level of engagement. See Supplementary Information section 3 for additional details. We then used this data to present the participants outlined above with 20 article headlines and ledes, drawn randomly from within one of the two media outlet subsets, and asked them to assess the accuracy of the information they were faced with. Specifically, they were asked ``Do you think this story is true?'', to which they responded on a seven-point scale from ``Definitely NO'' to ``Definitely YES''. The study concluded with seven demographic questions (age, gender, education, political conservativeness on social and economic issues, political position, and political preference) and a section to leave comments. An illustrative example is shown in Figure S24.

\begin{figure}[th!] \centering \includegraphics[scale=0.23]{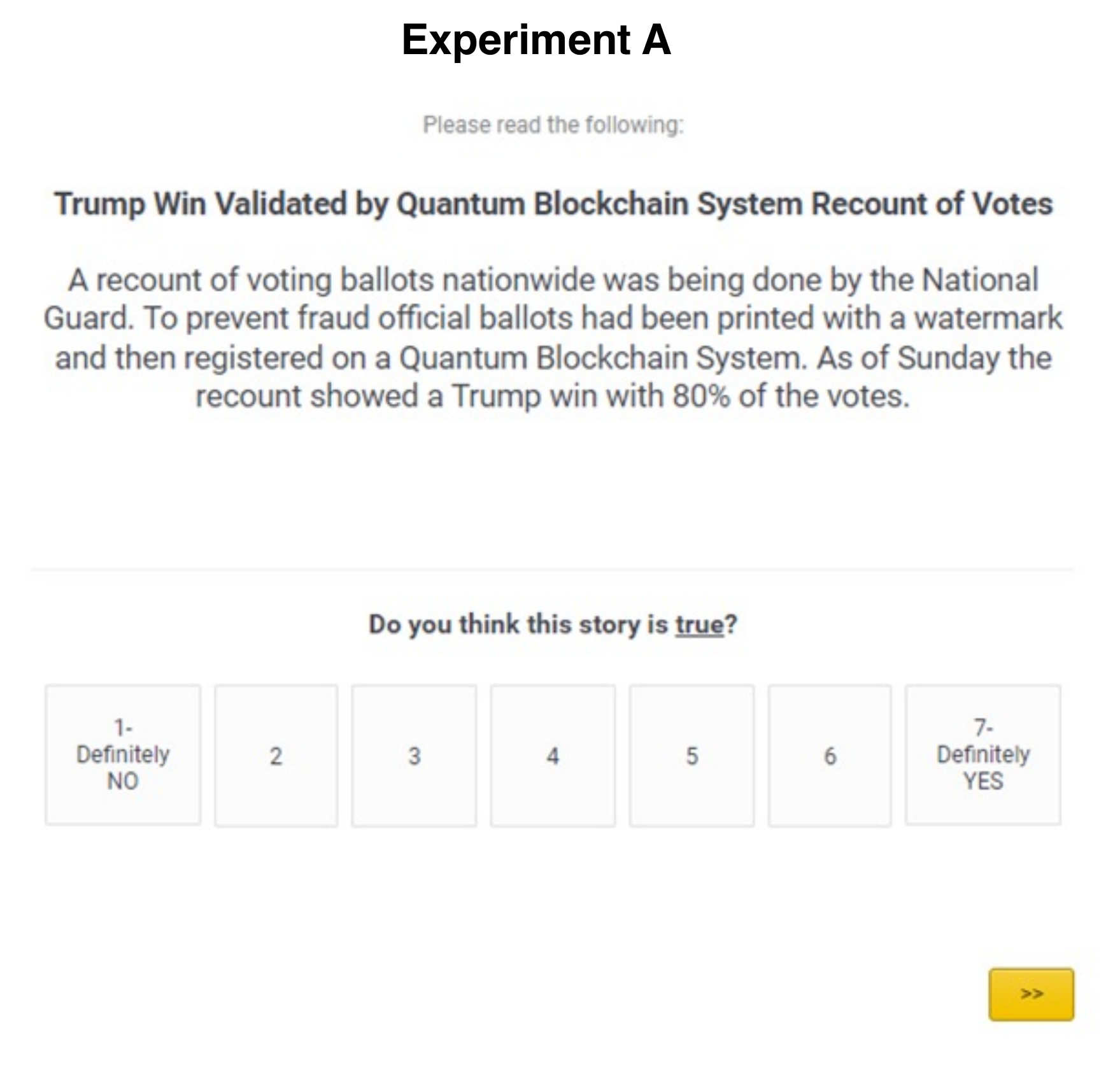}
\caption*{\footnotesize Figure S24: \textbf{Interface for Experiment A.}}
\end{figure}

\clearpage

\noindent\textbf{Experiment B \& C.} We identified a pool of 40 news "cards" (i.e. representations of Facebook posts with an image, headline, and a source; balanced on veracity and partisan lean) and a list of 60 domains regularly used for news (20 identified as mainstream, 20 as hyper-partisan, and as 20 fake by Pennycook and Rand, 2019  \cite{Pennycook2521}). We then asked participants to : i) evaluate 20 cards, drawn randomly from the set of 40, on the accuracy (i.e. ``Do you think this story is true?'', seven-point scale from ``Definitely NO'' to ``Definitely YES''), likelihood of sharing (i.e. ``If you were to see the above headline online, would you share it?'', seven-point scale from ``Definitely NO'' to ``Definitely YES''), and likelihood of clicking (i.e. ``If you were to see the above headline online, would you click on it to read the article?'', seven-point scale from ``Definitely NO'' to ``Definitely YES'') of the information presented; ii) select which of the 60 domains they regularly use for news (with this information, we classified participants as misinformation media users if they selected at least one domain pre-identified as a misinformation outlet). The study concluded with a list of 20 exploratory items. An illustrative example is shown in Figure S25.

\begin{figure}[th!] \centering \includegraphics[scale=0.23]{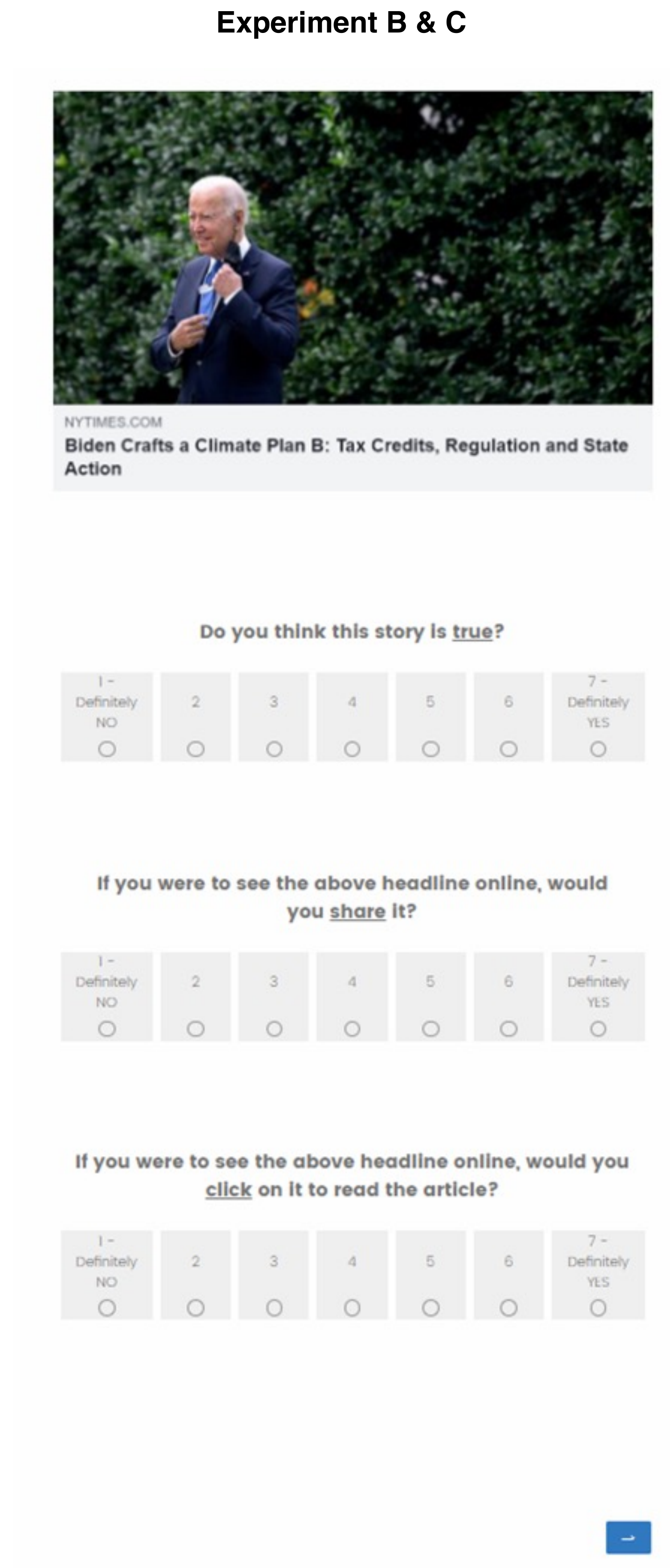}
\caption*{\footnotesize Figure S25: \textbf{Interface for Experiments B\&C}}
\end{figure}

\clearpage






\end{document}